\documentclass[aps,prb,twocolumn,superscriptaddress,longbibliography,floatfix]{revtex4-2}
\usepackage{amsmath,amssymb,bm,mathtools}
\usepackage{graphicx,booktabs,microtype}
\usepackage[hidelinks]{hyperref}
\newcommand{\EF}{E_F}
\newcommand{\kF}{k_F}
\newcommand{\vF}{v_F}
\newcommand{\aB}{a_B}
\newcommand{\Ost}{\Omega_*}
\newcommand{\lams}{\lambda_s}
\newcommand{\lampl}{\lambda_{\rm pl}}
\newcommand{\cG}{\mathcal G}
\newcommand{\cJ}{\mathcal J}
\newcommand{\cS}{\mathcal S}
\newcommand{\Ec}{E_{\rm C}}

\newcommand{\kappaA}{1.26}
\newcommand{\kappaB}{1.42}
\newcommand{\kappaC}{1.70}
\newcommand{\kappaD}{1.97}
\newcommand{\kappaE}{2.45}
\newcommand{\lamplErrA}{3\%}
\newcommand{\lamplErrB}{0.2\%}
\newcommand{\TcBareTwoFive}{-0.9}
\newcommand{\TcDynTwoFive}{-2.8}
\newcommand{\TcStatTwoFive}{-102}
\newcommand{\invlammin}{3}
\newcommand{\invlammax}{135}
\newcommand{\lnCmin}{$-1.6$}
\newcommand{\lnCmax}{$+6.6$}
\newcommand{\lnCtwoA}{$-1.29$}
\newcommand{\lnCtwoB}{$-1.34$}
\newcommand{\lnCtwoC}{$-1.37$}
\newcommand{\ndirect}{31}
\newcommand{\invlammaxdirect}{23}

\newcommand{\lnCminDirect}{$-1.6$}
\newcommand{\lnCmaxDirect}{$+0.2$}
\newcommand{\statSlopeErr}{0.08\%}
\newcommand{\statWinErr}{0.01\%}
\newcommand{\ladExcessCoef}{22.6}
\newcommand{\expErr}{13\%}
\newcommand{\slopemin}{0.97}
\newcommand{\slopemax}{1.08}
\newcommand{\slopeLowMin}{1.04}
\newcommand{\slopeLowMax}{1.06}
\newcommand{\muA}{0.28}
\newcommand{\muB}{0.39}
\newcommand{\ppaErr}{0.15}
\newcommand{\netfracmax}{8\%}
\newcommand{\netfracTwo}{1.2\%}
\newcommand{\netfracFive}{0.25\%}

\newcommand{\netslopeA}{$-4.0$}
\newcommand{\netslopeB}{$-4.2$}

\newcommand{\cutTen}{2.3}
\newcommand{\cutHundred}{8\%}
\newcommand{\gmbslope}{$0.98$}
\newcommand{\cgmbA}{0.9}
\newcommand{\cgmbB}{1.1}
\newcommand{\netrsA}{3.3}
\newcommand{\netrsB}{3.9}

\newcommand{\ladExcessOneTwo}{93}
\newcommand{\cGten}{35}
\newcommand{\Jw}{3.441\times10^{-4}}
\newcommand{\JwT}{0.101}
\newcommand{\ClowCoef}{0.1033}

\newcommand{\lowRatA}{1.06}
\newcommand{\lowRatB}{1.12}
\newcommand{\lowExtrapErr}{1.3\%}
\newcommand{\CfourSpread}{12\%}
\newcommand{\wedgeRatA}{1.114}
\newcommand{\wedgeRatB}{1.050}
\newcommand{\wedgeRatC}{1.023}
\newcommand{\wedgeRatD}{1.012}
\newcommand{\wedgeIntercept}{1.001}
\newcommand{\wedgeSlope}{3.3}
\newcommand{\RlowOne}{1.56}

\newcommand{\RlowFive}{0.56}

\begin{document}
\title{Exciton superfluidity in two-dimensional bilayers at zero and high magnetic fields}
\author{Sankar Das Sarma}
\author{Yang-Zhi~Chou}
\affiliation{Condensed Matter Theory Center and Joint Quantum Institute, Department of Physics, University of Maryland, College Park, Maryland 20742, USA}
\date{\today}
\begin{abstract}
We study how interlayer exciton superfluidity evolves with layer separation in two-dimensional bilayers, connecting the electron–hole system at zero magnetic field with the quantum Hall system at total filling factor $\nu_T=1$. A particle–hole transformation relates their order parameters, although the kinetic energy is quenched in the quantum Hall regime. Hartree–Fock theory with the unscreened Coulomb interaction predicts coherence at every finite separation, with an energy scale that decreases inversely with separation at large distances. We examine how screening, correlations, and spatial instabilities modify this prediction. At zero field, we solve the linearized gap equation with the full momentum and frequency dependence of the interaction screened within the random-phase approximation, without an imposed cutoff. The apparent plasmon enhancement of pairing in the ladder approximation is largely cancelled by the second-order crossed-rung correction. Static screening determines the leading pairing exponent in the controlled high-density limit at fixed separation. At large separation, the pairing scale falls exponentially with squared separation while the instability persists; extending this conclusion to fixed density assumes Fermi-liquid layers and higher-order dynamical contributions that are subleading to the static pairing interaction. At high field, we obtain an analytic uniform Hartree–Fock saddle point in the lowest Landau level that remains coherent at every finite separation. Within Hartree–Fock theory, its finite-wave-vector collective-mode instability favors a density-modulated coherent state, an excitonic supersolid. This distinguishes loss of spatial uniformity from loss of coherence. We conjecture that the quantum Hall bilayer evolves continuously from strong interlayer coherence to weak pairing of composite fermions, with a scale far below the mean-field prediction at large separation. Such suppression could in principle produce an apparent experimental critical separation at finite temperature without a ground-state quantum phase transition. The comparison separates the existence, energy scale, and spatial stability of interlayer coherence.
\end{abstract}
\maketitle

\section{Introduction}

An electron in one two-dimensional layer and a hole in a parallel layer attract each other, and at low temperature the resulting interlayer excitons can condense into a state with spontaneous coherence between the layers. Because the electron and the hole are in different layers the exciton cannot recombine, and the condensate is a superfluid that carries equal and opposite currents in the two layers \cite{Lozovik1976,Shevchenko1976}. It is the bilayer form of the excitonic instability of a semimetal \cite{Keldysh1965}. The subject has two standard settings. The first is the balanced electron-hole bilayer at zero magnetic field, realized in semiconductor and atomic double layers \cite{Croxall2008,Seamons2009,Ma2021,Nguyen2025,Qi2025}. The second is the quantum Hall bilayer at total filling factor $\nu_T=1$, in which each layer has a half-filled lowest Landau level and a particle-hole transformation of one layer turns the electron-electron bilayer into an electron-hole bilayer \cite{Eisenstein2004,Eisenstein2014}. It is in the second setting that the phenomenon was first observed and has been studied in the most detail.

The two settings are usually treated separately, with different languages: pairing and the Bardeen-Cooper-Schrieffer (BCS) instability in one, the pseudospin ferromagnet in the other. They are one problem. After a particle-hole transformation of one layer the Hamiltonian is the same apart from the kinetic energy and one-body terms, the broken symmetry is the same, the layer separation $d$ is the experimental tuning parameter in both, and the density enters through one further parameter, $r_s$ at zero field and the filling factor in a field. Here $r_s$ is the dimensionless Coulomb interaction parameter, defined as the mean interparticle spacing in units of the effective Bohr radius. In both, the analytic predictions for the dependence on $d$ come from mean-field theory, the Hartree-Fock (HF) approximation, and its extensions: the theory of Ref.~\cite{Zhu2024} and the Eliashberg theory with dynamical screening \cite{Sreejith2024} at zero field, and the theory of the coherent state and of its collective mode at $\nu_T=1$ \cite{Fertig1989,MacDonaldPlatzmanBoebinger1990,Moon1995}. And in both, mean-field theory gives coherence at every separation, with a scale that decays only as $1/d$. The questions are then the same too. Does the coherence at every $d$ survive beyond mean-field theory? What is the true dependence of its scale on $d$? Is the evolution from small to large $d$ a transition or a crossover? And what is the relation of the coherent state to the states of broken translation symmetry that appear on the way?

This paper treats both settings and compares them. The two parts are of unequal length, because the problems are unequal in what can be done. At zero field there are small parameters, and the theory can be carried beyond mean field in two limits. At $\nu_T=1$ the kinetic energy is quenched, there is no small parameter, and nothing is known in a controlled way beyond HF theory; what can be done there is to state that theory completely and to say exactly what it does and does not establish.

Three questions that are often run together must be kept apart, and their separation organizes the paper. The first is the existence of an interlayer-coherent solution. The second is the magnitude of its energy scale. The third is its stability against competing states. 

At zero field, screening qualitatively changes the unscreened mean-field scale, and the crossed-rung diagram largely cancels the enhancement obtained from dynamical screening in the ladder approximation. We analyze the resulting scale in two limits: small $r_s$ and large layer separation. Competing crystalline phases at low density lie outside our scope. At $\nu_T=1$, the uniform Hartree-Fock coherence scale is known in closed form, although its reliability at large separation remains unresolved. Within Hartree-Fock theory, the uniform state develops a finite-wave-vector instability beyond $d/\ell\simeq1.175$ (with $\ell$ being the magnetic length) toward a density-modulated state that retains interlayer coherence.

The difficulties in the subject arise where an answer to one of these questions is inferred from another: a large scale from the existence of the solution, or a loss of coherence from an instability of the uniform state. The results below are arranged so that each is assigned to the question it answers.

\emph{Zero field.} Whether the Coulomb interaction alone can produce a pairing instability of a metal at a useful temperature is an old question. For a single electron layer or a three-dimensional electron gas the bare interaction is repulsive, and $s$-wave pairing requires the dynamically screened interaction to be less repulsive at low frequency than at high frequency, the difference being supplied by the plasmon \cite{Takada1978,KMK1973,RietschelSham1983,GrabowskiSham1984}. A recent analysis of this problem \cite{DasSarma2026} emphasized that the standard Eliashberg route to a quantitative $T_c$ is uncontrolled there. Migdal's theorem \cite{Migdal1958,Eliashberg1960}, which justifies dropping vertex corrections for phonons because the Debye energy is much smaller than the Fermi energy, has no analog when the boson is the plasmon of the same electrons, because the plasmon energy is comparable to the Fermi energy. The calculated $T_c$ then depends on a frequency cutoff that the theory cannot fix, and for cutoffs small enough to make the neglect of vertex corrections plausible the $s$-wave instability disappears \cite{DasSarma2026}. A small parameter of the Migdal type can be restored in an all-electron bilayer by a large ratio of the layer masses \cite{Wang2026}.

The electron-hole bilayer is the natural system in which to revisit the question, because one essential ingredient is reversed. Electrons in one layer and holes in the other attract, so the bare interaction in the pairing channel is attractive at every wave vector \cite{Lozovik1976,Shevchenko1976,Keldysh1965}. Interlayer pairing, the condensation of interlayer excitons, is therefore present already at the HF level \cite{Zhu2024}, and screening can only reduce it. The role of dynamical screening was studied within Eliashberg theory in Ref.~\cite{Sreejith2024}, which found a finite $T_c$ over the range of density and separation studied, well below the unscreened HF value and above the statically screened one. Related questions arose earlier for graphene double layers, where static screening was argued to suppress the gap to unobservable values \cite{Kharitonov2008} and dynamical screening was found to restore much of it \cite{Sodemann2012,Lozovik2012}. A comparison of the energies of the exciton gas and of the electron-hole plasma found the outcome to depend strongly on how the screening of the attraction is treated \cite{Tu2024}. Quantum Monte Carlo calculations \cite{DePalo2002,LopezRios2018} and experiments on semiconductor and atomic double layers \cite{Croxall2008,Seamons2009,Ma2021,Nguyen2025,Qi2025} provide further motivation.

In the zero-field part of this paper we carry the program of Ref.~\cite{DasSarma2026} over to the bilayer. The plasmon of the single layer is replaced by the two plasmons of the double layer, superconductivity by interlayer superfluidity, and the ratio of the plasmon energy to the Fermi energy by a ratio involving the binding energy $E_B$ of the interlayer exciton. We ask three questions. What does the dynamically screened interaction look like, and where is it attractive? What transition temperature does it give when the equations are solved without ad hoc cutoffs? In which limits, if any, is the result controlled? The bilayer allows the third question to be answered. At large separation the plasmon energy that matters for interlayer pairing becomes small compared with the Fermi energy, so one can test whether that is what a theorem of the Migdal type requires.

\emph{High field.} The purpose of the high-field part of this paper is to set out in closed form, and in one place, what HF theory does and does not establish about the separation dependence of interlayer coherence in the problem projected onto the lowest Landau level (LLL).  Most of the ingredients are known: the collective mode and its finite-wave-vector softening \cite{Fertig1989,MacDonaldPlatzmanBoebinger1990,Brey1990}, the interlayer exchange integral and the stiffness \cite{Moon1995,JoglekarMacDonald2001}, the closed form of that integral in the closely related graphene double-layer problem \cite{FilKravchenko2009}, and the density-wave states of unrestricted HF \cite{CoteBreyMacDonald1992,BreyFertig2000,ChenQuinn1991}.  What we add is the explicit HF amplitude scale $T_c^{\rm HF}(d)$ and a clear statement of the logical content of these results.  The result is simple but consequential.  The interlayer exchange field is positive at every finite $d/\ell$, where $\ell$ is the magnetic length, and hence the $T=0$ HF state is coherent at every finite $d/\ell$.  The HF coherence scale has a $1/d$ tail.  There is  no zero-temperature critical separation in the projected LLL HF theory.

There is a separate finite-wave-vector instability.  The uniform coherent state develops a roton minimum which collapses at $d/\ell\simeq1.175$ in the zero-width model.  This instability has been widely identified, on heuristic grounds, with the disappearance of the $(111)$ state and the loss of interlayer coherence at $d/\ell\sim1$ \cite{MacDonaldPlatzmanBoebinger1990,JoglekarMacDonald2001}.  Within HF theory this identification is incorrect: the softening by itself does not imply a loss of coherence.  A finite-$q$ roton collapse destabilizes spatial uniformity.  It does not set the interlayer exchange field to zero.  The ensuing HF phase has coexisting interlayer coherence and layer-density modulation \cite{BreyFertig2000,JoglekarMacDonald2001}.  It is an excitonic supersolid.

Our main results are the following.

(i) Within the random-phase approximation (RPA) the screened interlayer interaction $W_{12}(q,i\nu)$ is attractive at every $q$ and every Matsubara frequency $\nu$, and its magnitude increases monotonically with $|\nu|$ from the statically screened value at $\nu=0$, approaching the magnitude of the bare, unscreened interaction as $|\nu|\to\infty$. On the real axis it is repulsive in the region between the acoustic and the optical plasmon (Sec.~\ref{sec:W}).

(ii) Within the ladder approximation, the leading gap function is nodeless in momentum and Matsubara frequency. The linearized gap equation is ultraviolet convergent and can be solved without an imposed cutoff. The bare interaction transfers only wave vectors $q\lesssim1/d$, which confines pairing to within $\hbar v_F/d$ of the Fermi surface and supplies an intrinsic cutoff (Secs.~\ref{sec:gap} and \ref{sec:numerics}).

(iii) For $d\gg a_B$ the pairing in the ladder approximation is dominated by a wedge-shaped region of frequency and momentum, in which $\nu$ exceeds the plasmon frequency $\omega_+(q)$ at the transferred wave vector and the interaction is therefore unscreened. In this region $\nu\gg v_Fq$. Virtual transitions through this region give a plasmon-induced coupling $\lambda_{\rm pl}$ that we obtain in closed form. It decays only as $d^{-1/2}$, whereas the static coupling decays as $d^{-2}$ (Sec.~\ref{sec:analytic}). The numerical solution confirms the analytic theory (Sec.~\ref{sec:numerics}).

(iv) The plasmon-induced coupling does not survive the leading vertex correction. In the wedge the carriers are slow compared with the plasmons, the opposite of Migdal's limit, and the crossed rung, which is of the same order in the screened interaction as $\lambda_{\rm pl}$, cancels it. We evaluate the crossed rung with the full dynamically screened interaction. The sum $\lambda_{\rm net}$ of the plasmon-induced and crossed terms is a small fraction of $\lambda_{\rm pl}$ at every separation examined. We obtain its limiting forms exactly, $C_3(d)\,r_s^3$ for $r_s\to0$ at fixed $d$ and $J_w\sqrt g\,r_s^4(a_B/d)^{7/2}$ for $d\to\infty$ at fixed $r_s$, and the numerical results reproduce both (Appendix~\ref{app:remainder}). The cancellation is exact at every order for carriers with linearized dispersion coupled to Gaussian density fluctuations, static carriers being the adiabatic limit; the remainder is an effect of the curvature of the Fermi surface, which we compute at second order (Sec.~\ref{sec:vertex}).

(v) The theory is controlled in two limits. For $r_s\to0$ at fixed $d$ the leading exponent of $T_c$ is $-1/\lams$; this is leading-order perturbation theory in the statically screened interaction. For large $d$ at any $r_s$ in the liquid phase we argue that it is $-(m^*/m)/\lams$; this rests on the cancellation just described, which is established at second order and in the linearized limit, and on Fermi-liquid theory. In both limits the answer is that of static screening, not that of the dynamically screened ladder, which at large $d$ overestimates $T_c$ by a factor that grows as $\exp({\rm const}\times d^2)$ (Sec.~\ref{sec:control}).

(vi) At $\nu_T=1$ the uniform HF theory of the LLL is analytic at every $d/\ell$. The interlayer exchange field $X_D$ is positive at every finite separation, so the uniform HF saddle at $T=0$ is interlayer coherent at every finite separation. The quasiparticle splitting is $\Delta_{\rm HF}=X_D$, the mean-field amplitude scale is $k_BT_c^{\rm HF}=X_D/4$, and $X_D\to e^2/\epsilon d$ at large $d$. The velocity of the collective mode and the phase stiffness follow in closed form from the same integral, and the dispersion of the mode from its Bessel transform (Sec.~\ref{sec:qh}).

(vii) The softening of the collective mode at $d/\ell\simeq1.175$ is an instability of the uniform state toward a coherent density wave, which within HF is an excitonic supersolid. It is not a loss of interlayer coherence, and the common identification of the two is not supported by the HF theory on which it rests (Secs.~\ref{sec:qh-mode} and \ref{sec:qh-supersolid}).

(viii) The two problems have the same mean-field structure and the same $1/d$ tail. At zero field the tail is replaced by a Gaussian decay in the two controlled limits, conditionally for $d\to\infty$, once the response of the layers is included. This is a precedent for one element of our expectation at $\nu_T=1$, the failure of the HF tail. The other element, a crossover from strong to weak coupling with no transition, is not derived in either problem. The expectation is stated as a conjecture, and the zero-field result is not evidence for it (Sec.~\ref{sec:compare}).

\emph{Scope.} The paper is a theory of mean-field coherence and of the weak-pairing side of the problem, in both settings. It is not a review of the experiments, of the quantum Monte Carlo and exact-diagonalization work, or of other filling factors, and the strongly paired regime at zero field enters only through the discussion of the crossovers in Appendix~\ref{app:long}. We distinguish throughout between statements of different kinds: a theorem within a stated approximation, an exact identity, a result in a controlled limit, a result that is conditional on a stated assumption, the output of an approximation with no estimate of its error, and a conjecture. Table~\ref{tab:status} in Sec.~\ref{sec:compare} assigns each of the main results to one of them.

\emph{Organization.} Section~\ref{sec:framework} sets out what the two problems have in common and fixes the notation. Sections~\ref{sec:model}--\ref{sec:discussion} contain the zero-field theory: the model, the screened interaction, the ladder theory and its analytic and numerical solution, the vertex corrections, and the controlled limits. Section~\ref{sec:qh} contains the theory at $\nu_T=1$. Section~\ref{sec:compare} compares the two, and Sec.~\ref{sec:conclusion} concludes. Appendices~\ref{app:pol}--\ref{app:num} contain the polarization function of a layer, the exciton binding energy, the solution of the unscreened gap equation, and the numerical details of the zero-field theory. Appendix~\ref{app:long} gives the complete set of pairing-channel diagrams through second order in the bare and in the screened interaction, states which of them our results contain, derives the two controlled limits diagram by diagram, and discusses the two crossovers to Bose condensation of excitons and the relation to the renormalization group. Appendix~\ref{app:remainder} contains the asymptotic evaluation of $\lambda_{\rm net}$. Appendices~\ref{app:qh-stiffness}--\ref{app:qh-ed} contain the response functions, the finite-temperature theory, the quantitative scales, and the derivations for $\nu_T=1$.

\section{One problem in two settings}
\label{sec:framework}

\subsection{Hamiltonian and symmetry}

Both systems consist of two parallel two-dimensional layers a distance $d$ apart, with a background dielectric constant $\epsilon$, Coulomb interactions, and no tunneling between the layers. Carriers in the same layer interact through $V_q=2\pi e^2/\epsilon q$, and carriers in different layers through an interaction of magnitude $U_q=V_qe^{-qd}$. Without tunneling the number of carriers is conserved in each layer separately, and the Hamiltonian is invariant under independent changes of phase in the two layers. The exciton condensate, in which an electron of one layer is coherent with a hole of the other, breaks the relative $U(1)$ symmetry and leaves the total one intact. This is the order of both problems. At zero field it appears as the pair amplitude of an electron and a hole, the gap function of Sec.~\ref{sec:gap}. At $\nu_T=1$ it appears as the interlayer coherence $\langle c^\dagger_{mT}c_{mB}\rangle$ of Sec.~\ref{sec:qh} between the two layers, labeled $T$ and $B$, at guiding center $m$; this is the in-plane component of the layer pseudospin.

At zero field the layers contain electrons and holes with kinetic energy $\hbar^2k^2/2m$, and the interlayer interaction is attractive [Eq.~\eqref{eq:H}]. At $\nu_T=1$ both layers contain electrons, each layer has a half-filled lowest Landau level, and the interlayer interaction is repulsive [Eq.~\eqref{qh:eq:H}]. A particle-hole transformation of one layer within the lowest Landau level turns the electrons of that layer at filling $1/2$ into holes at filling $1/2$ and reverses the sign of the interlayer interaction, apart from one-body terms \cite{Eisenstein2004}. In these variables the $\nu_T=1$ bilayer is a balanced electron-hole bilayer with an attractive interlayer interaction and no kinetic energy, and the Halperin $(111)$ state \cite{Halperin1983} is its exciton condensate. This is the sense in which the two problems are one.

\subsection{Parameters}

At zero field the model has two dimensionless parameters, the density parameter $r_s$ and the separation $d/\aB$, where $\aB$ is the Bohr radius (Sec.~\ref{sec:model}); the combinations that appear most often are $\kF d$ and the ratio $E_B/\EF$ of the exciton binding energy to the Fermi energy. At $\nu_T=1$, in the lowest Landau level and for layers of zero thickness, the only parameter is $d/\ell$. The two sets are related. A spin-polarized layer at filling $1/2$ has the density $n=1/4\pi\ell^2$, and a spin-polarized Fermi gas of that density at zero field would have $\kF=(4\pi n)^{1/2}=1/\ell$. In the notation of Sec.~\ref{sec:model} with $g=1$,
\begin{equation}
\frac{d}{\ell}=\kF d,\quad r_s=\frac{2\ell}{\aB}=2\kappa_{\rm LL},\quad\kappa_{\rm LL}\equiv\frac{e^2/\epsilon\ell}{\hbar\omega_c}=\frac{\ell}{\aB},
\label{eq:dictionary}
\end{equation}
where $\omega_c$ is the cyclotron frequency and $\kappa_{\rm LL}$ is the parameter of Landau-level mixing. The separation in units of the magnetic length is thus the same quantity as $\kF d$ at zero field: the layer separation measured against the distance between the carriers.

Projection onto the lowest Landau level is the limit $\kappa_{\rm LL}\to0$, that is, $r_s\to0$ at fixed $\kF d$. At zero field this is a limit of weak coupling, in which the interaction energy is small compared with the kinetic energy. In a strong field it is not, because the kinetic energy has been removed from the problem altogether and the Coulomb energy $e^2/\epsilon\ell$ is the only scale. This is the essential difference between the two problems. It is the reason the zero-field theory has controlled limits (Sec.~\ref{sec:control}), while at $\nu_T=1$ the only exact statement is the one at $d=0$.

\subsection{Mean-field theory and the diagrams}

In both problems the mean-field theory is the HF approximation with the bare Coulomb interaction, and in both the term that produces the coherence is the same diagram: the exchange self-energy with one interlayer interaction line. It is Fig.~\ref{fig:diag}(a), with the screened line replaced by the bare one, at zero field, and Fig.~\ref{qh:fig:diagrams}(b) at $\nu_T=1$. What differs is the set of states on which it acts. At zero field the carriers have a Fermi surface. The exchange self-energy is then anomalous, it satisfies a gap equation whose kernel contains the Cooper logarithm, multiplied for the unscreened interaction by the logarithm of the interaction itself, and the resulting scale is exponentially small, $\exp(-{\rm const}/\sqrt\alpha)$ [Eq.~\eqref{eq:Tc0}]. In the lowest Landau level all single-particle states are degenerate. There is no logarithm, the self-consistency condition is algebraic, and the gap is the exchange integral itself [Eq.~\eqref{qh:eq:XD}].

The corrections to mean-field theory are organized in the same way in the two problems. At zero field the particle-hole bubbles of each layer screen the interaction; their sum to all orders is the RPA of Sec.~\ref{sec:W}. The leading vertex correction is the crossed rung of Sec.~\ref{sec:vertex}, and the complete set of diagrams through second order is given in Appendix~\ref{app:long}. At $\nu_T=1$ the bubble within the filled pseudospin band vanishes, and the corrections come from particle-hole pairs between the filled and the empty pseudospin band [Figs.~\ref{qh:fig:diagrams}(c) and \ref{qh:fig:diagrams}(d)]. Their ladder sum is the time-dependent HF (TDHF) theory of the collective mode (Sec.~\ref{sec:qh-mode}), and their contribution to the self-energy is what uniform HF theory leaves out.

\subsection{The common questions}

Four questions are asked of both problems. (1) Does mean-field theory give interlayer coherence at every separation? (2) How does the scale of the coherence depend on $d$ at large separation, and does the mean-field answer survive the corrections? (3) Is the evolution from small to large $d$ a phase transition or a crossover from strong to weak coupling? (4) What is the relation of the coherent state to the states of broken translation symmetry? These questions distinguish the existence of a coherent solution from its energy scale and stability. In particular, an instability of the uniform state need not imply a loss of interlayer coherence. Table~\ref{tab:dictionary} lists the correspondence between the two problems and the answers obtained in Secs.~\ref{sec:model}--\ref{sec:qh}; they are compared in Sec.~\ref{sec:compare}.

\begin{table*}[t]
\caption{Correspondence between the two settings. The entries are established in the sections indicated; HF denotes the Hartree-Fock approximation with the bare interaction, LLL the lowest Landau level, and TDHF the time-dependent HF approximation.}
\label{tab:dictionary}
\begin{ruledtabular}
\begin{tabular}{p{0.20\textwidth}p{0.36\textwidth}p{0.38\textwidth}}
& Zero field (Secs.~\ref{sec:model}--\ref{sec:discussion}) & $\nu_T=1$ (Sec.~\ref{sec:qh})\\
\hline
Carriers & electrons in layer 1, holes in layer 2, spin degeneracy $g$ & electrons in both layers, spin polarized\\
Interlayer interaction & attractive, $-U_q$ & repulsive, $+U_q$; attractive after a particle-hole transformation\\
Kinetic energy & $\hbar^2k^2/2m$, Fermi surface & quenched\\
Units & $\aB$, ${\rm Ry}$, $\EF$ & $\ell$, $\Ec=e^2/\epsilon\ell$\\
Parameters & $r_s$, $d/\aB$ & $d/\ell$ ($\kappa_{\rm LL}\to0$)\\
Separation against carrier spacing & $\kF d$ & $d/\ell$\\
Mean-field theory & unscreened ladder (HF) & uniform LLL HF\\
Mean-field scale & $k_BT_c^{(0)}\propto(\hbar\vF/d)\,e^{-\pi^{3/2}/2\sqrt\alpha}$ & $k_BT_c^{\rm HF}=X_D/4\to e^2/4\epsilon d$\\
Origin of the scale & Cooper logarithm times the logarithm of the unscreened kernel & exchange integral, no logarithm\\
Collective modes & optical and acoustic plasmons & pseudospin Goldstone mode with a roton\\
Beyond the uniform mean-field state & RPA screening; crossed rung at second order & TDHF mode; density wave of unrestricted HF (literature); both within HF\\
Scale at large $d$ & $\ln T_c\propto-d^2$ (Sec.~\ref{sec:control}; conditional at fixed $r_s$) & HF: $\propto1/d$; beyond HF: not known\\
Broken translation symmetry & crystals at large $r_s$, outside the theory & coherent density wave for $d/\ell>1.175$ within HF\\
Where the theory is exact & leading exponent for $r_s\to0$ at fixed $d$ & $d=0$\\
\end{tabular}
\end{ruledtabular}
\end{table*}

\subsection{Conventions}

At zero field $T_c$ denotes the mean-field temperature at which the pairing instability occurs. In two dimensions the transition is of Kosterlitz-Thouless type \cite{KT1973}, with a transition temperature that approaches the mean-field one in the weak-coupling regime of interest there. At $\nu_T=1$ the corresponding mean-field quantity is written $T_c^{\rm HF}$. It is an amplitude scale, and it is far above the temperature of the Kosterlitz-Thouless transition (Appendix~\ref{qh:sec:finiteT}). The two are therefore kept apart in the notation.

Each part uses its natural units, $\aB$ and $\EF$ or ${\rm Ry}$ at zero field and $\ell$ and $\Ec=e^2/\epsilon\ell$ at $\nu_T=1$. Several symbols need care. $\Ec$ is the Coulomb unit of the high-field problem, while $E_c=\hbar^2/md^2$ is the curvature energy of the zero-field problem. The letter $\nu$ without a subscript is a Matsubara frequency in the zero-field sections; in Sec.~\ref{sec:qh} and its appendices it is a filling factor, in $\nu_T$ and in the $\nu=1$ ferromagnet of a single layer. $\Delta$, also written $\Delta_{\rm HF}$, is the HF quasiparticle splitting at $\nu_T=1$; at zero field $\Delta(\xi)$ is the gap function and $\bm\Delta$ in Sec.~\ref{sec:vertex} is a momentum transfer. $P$ is the pseudospin polarization in Sec.~\ref{sec:qh} and the dimensionless polarization function of a layer in the zero-field sections. The zero-field theory is written for spin degeneracy $g$, with numerical results for $g=2$, whereas the $\nu_T=1$ bilayer is spin polarized. The layers are labeled 1 and 2 at zero field and $T$ and $B$ at $\nu_T=1$. The cyclotron frequency $\omega_c$ appears only in Eq.~\eqref{eq:dictionary}; elsewhere $\omega_c$ is a frequency cutoff. The quantities $\delta=d/\ell$ and $Q=q\ell$, the function $F(\delta)$, and the kernels of Sec.~\ref{sec:qh-mode} are defined in Sec.~\ref{sec:qh} and have these meanings only there and in Appendices~\ref{app:qh-stiffness}--\ref{app:qh-ed}.

\section{Zero magnetic field: model and scales}
\label{sec:model}

We consider electrons of mass $m$ and density $n$ in layer 1 and holes of the same mass and density in layer 2, separated by a distance $d$, with a background dielectric constant $\epsilon$ and no interlayer tunneling,
\begin{align}
H={}&\sum_{l\mathbf k\sigma}\xi_{\mathbf k}\,\psi^\dagger_{l\mathbf k\sigma}\psi_{l\mathbf k\sigma}
+\frac{1}{2A}\sum_{l,\mathbf q}V_q:\!\rho_l(\mathbf q)\rho_l(-\mathbf q)\!:\nonumber\\
&-\frac1A\sum_{\mathbf q}U_q\,\rho_1(\mathbf q)\rho_2(-\mathbf q),
\label{eq:H}
\end{align}
with $\psi_{l=1\mathbf k\sigma}$ ($\psi_{l=2\mathbf k\sigma}$) being the annihilation operator of the electron in layer 1 (hole in layer 2) carrying momentum $\mathbf k$ and spin $\sigma$, $\xi_{\mathbf k}=\hbar^2k^2/2m-\EF$, $\rho_l(\mathbf q)=\sum_{\mathbf k\sigma}\psi^\dagger_{l\mathbf k\sigma}\psi_{l\mathbf k+\mathbf q\sigma}$, and
\begin{equation}
V_q=\frac{2\pi e^2}{\epsilon q},\qquad U_q=V_q\,e^{-qd}.
\label{eq:VU}
\end{equation}
Each layer has spin degeneracy $g$. All formulas are given for general $g$; numerical results are for $g=2$, and the spinless model of Ref.~\cite{Sreejith2024} is recovered by setting $g=1$. The interlayer attraction is independent of spin, so pairing of an electron with a hole of either spin is degenerate and we consider one spin pair.

The model has two dimensionless parameters, $r_s$ and $d/\aB$, with $\aB=\epsilon\hbar^2/me^2$ and $\pi r_s^2\aB^2n=1$. Table~\ref{tab:scales} collects the derived quantities. The total density of states of a layer is $N_0=gm/2\pi\hbar^2$, and that of one spin species, which enters the pairing problem, is $N_0/g$. We use
\begin{equation}
N_0V_q=\frac{c\,\kF}{q},\qquad c=g\alpha,\qquad \alpha=\frac{e^2}{\epsilon\hbar\vF}=\frac{\sqrt g\,r_s}{2},
\end{equation}
so that $\alpha$ is the fine-structure constant of the layer.

\begin{table}[tb]
\caption{Scales of the balanced electron-hole bilayer with spin degeneracy $g$ per layer. ${\rm Ry}=e^2/2\epsilon\aB$.}
\label{tab:scales}
\begin{ruledtabular}
\begin{tabular}{ll}
Fermi wave vector & $\kF\aB=2/(\sqrt g\,r_s)$\\
Fermi energy & $\EF=4\,{\rm Ry}/(g r_s^2)$\\
Coupling & $\alpha=e^2/\epsilon\hbar\vF=\sqrt g\,r_s/2$\\
Separation & $\kF d=2d/(\sqrt g\,r_s\aB)$\\
Capacitive parameter & $\cG=N_0(V_q\!-\!U_q)_{q\to0}=gd/\aB$\\
Classical binding energy & $E_d=e^2/\epsilon d$\\
 & $E_d/\EF=g r_s^2\aB/2d$\\
Kinematic cutoff & $\hbar\vF/d=2\EF/\kF d$\\
Plasmon energy at $q=1/d$ & $\hbar\Ost=(2g\EF E_d)^{1/2}$\\
\end{tabular}
\end{ruledtabular}
\end{table}

\subsection{The interlayer exciton}

\begin{figure*}[t]
\includegraphics[width=0.86\textwidth]{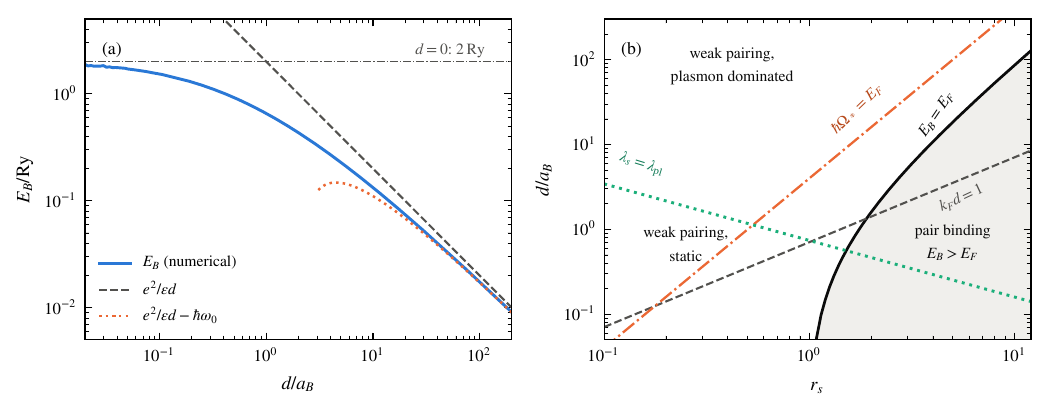}
\caption{(a) Binding energy of the interlayer exciton, compared with $E_d=e^2/\epsilon d$ and with Eq.~\eqref{eq:EB}. (b) Regimes of the bilayer for $g=2$. Shaded: $E_B>\EF$, where pairs are bound on the scale of the Fermi energy. Dashed: $\kF d=1$. Dash-dotted: $\hbar\Ost=\EF$. Dotted: $\lams=\lampl$, Eq.~\eqref{eq:crossover}; the plasmon-induced coupling dominates above this line.}
\label{fig:scales}
\end{figure*}

An isolated electron-hole pair is bound at every $d$. The binding energy $E_B$ follows from the relative-motion problem with reduced mass $m/2$ and potential $-e^2/\epsilon\sqrt{r^2+d^2}$, which we solve numerically (Appendix~\ref{app:exciton}). The result, Fig.~\ref{fig:scales}(a), interpolates between $E_B=2\,{\rm Ry}$ at $d=0$ and
\begin{equation}
E_B\simeq\frac{e^2}{\epsilon d}\Bigl[1-\sqrt{2\aB/d}\,\Bigr]\qquad(d\gg\aB),
\label{eq:EB}
\end{equation}
where the correction is the zero-point energy of the in-plane oscillation. The approach to $E_d=e^2/\epsilon d$ is slow: $E_B/E_d=0.33$, $0.57$, and $0.74$ at $d/\aB=1$, $5$, and $20$.

The ratio $E_B/\EF$ separates the regime of tightly bound excitons from that of weakly paired Fermi liquids \cite{Randeria1989,Nozieres1985}. The zero-field theory of this paper concerns the second regime, $E_B\lesssim\EF$, where the Fermi surfaces exist and the pairing is of the BCS type. The linearized gap equation at fixed chemical potential is not valid on the other side, where the chemical potential is pulled below the band edge. Figure~\ref{fig:scales}(b) shows the regimes in the $(r_s,d/\aB)$ plane together with three other lines that will emerge below: $\kF d=1$, $\hbar\Ost=\EF$, and the line on which the static and the plasmon-induced couplings are equal.

\section{Zero magnetic field: Dynamically screened interlayer interaction}
\label{sec:W}

\subsection{Random-phase approximation}

In the RPA the screened interaction is the $2\times2$ matrix $\hat W=[1-\hat V\Pi]^{-1}\hat V$ in the layer index, with $\hat V_{11}=\hat V_{22}=V_q$, $\hat V_{12}=-U_q$, and $\Pi(q,\omega)$ the noninteracting polarization of one layer \cite{Stern1967}, given in Appendix~\ref{app:pol}. The matrix is diagonal in the channels in which the two number densities oscillate out of phase and in phase. Because the carriers have opposite charge, the first is a charge oscillation with interaction $V_q+U_q$ and the second is charge neutral with interaction $V_q-U_q$. With
\begin{equation}
\varepsilon_\pm(q,\omega)=1-(V_q\pm U_q)\Pi(q,\omega),
\end{equation}
the intralayer and interlayer interactions are
\begin{align}
W_{11}&=\frac12\left[\frac{V_q+U_q}{\varepsilon_+}+\frac{V_q-U_q}{\varepsilon_-}\right],\\
W_{12}&=\frac12\left[\frac{V_q-U_q}{\varepsilon_-}-\frac{V_q+U_q}{\varepsilon_+}\right]=-\frac{U_q}{\varepsilon_+\varepsilon_-}.
\label{eq:W12}
\end{align}
A negative $W_{12}$ is attractive.

\subsection{Collective modes}

\begin{figure*}[t]
\includegraphics[width=0.86\textwidth]{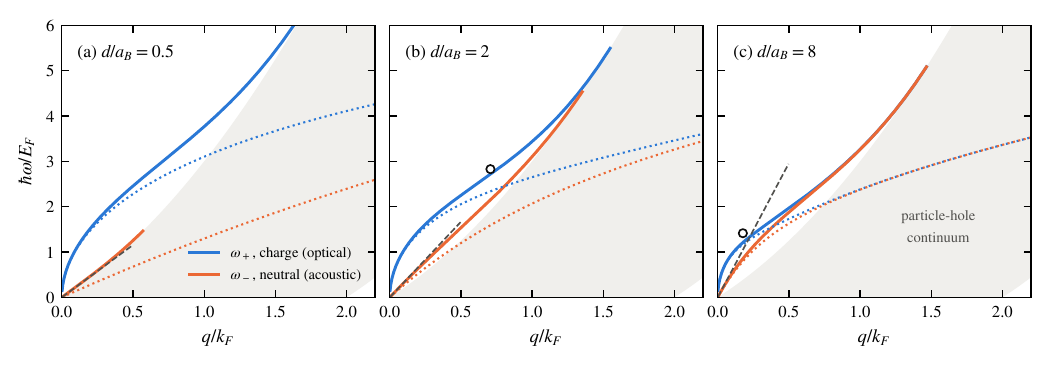}
\caption{Plasmons of the bilayer in the RPA for $r_s=2$. Solid: undamped modes from the zeros of $\varepsilon_\pm$. Dotted: long-wavelength forms, Eq.~\eqref{eq:plasmons}. Dashed: acoustic velocity, Eq.~\eqref{eq:vminus}. Shaded: particle-hole continuum. The circle marks $(q,\omega)=(1/d,\Ost)$.}
\label{fig:modes}
\end{figure*}

The zeros of $\varepsilon_\pm$ give the two plasmons of the double layer \cite{DasSarmaMadhukar1981,SantoroGiuliani1988}. With the high-frequency form of the polarization, valid for $\omega\gg\vF q$,
\begin{equation}
\omega_\pm^2(q)=\frac{2\pi ne^2}{\epsilon m}\,q\,\bigl(1\pm e^{-qd}\bigr),
\label{eq:plasmons}
\end{equation}
so that the charge mode is optical, $\omega_+\propto\sqrt q$, and the neutral mode is acoustic. For the optical mode the condition $\omega\gg\vF q$ holds at long wavelength for any $d$. For the acoustic mode it requires $\cG\gg1$, and Eq.~\eqref{eq:plasmons} then gives $v_-\simeq\vF\sqrt{\cG/2}$. At arbitrary separation the RPA velocity of the acoustic mode is
\begin{equation}
\frac{v_-}{\vF}=\frac{1+\cG}{\sqrt{1+2\cG}},\qquad\cG=\frac{g\,d}{\aB},
\label{eq:vminus}
\end{equation}
which exceeds $\vF$ for every $d$, so the mode is undamped at long wavelength. Figure~\ref{fig:modes} shows the dispersions.

The plasmon energy at the wave vector $q=1/d$ that characterizes the interlayer interaction will play a central role. We define it from Eq.~\eqref{eq:plasmons} at $q=1/d$ with the factor $1+e^{-qd}$ replaced by its long-wavelength value 2,
\begin{equation}
\hbar\Ost\equiv\Bigl(\frac{4\pi ne^2\hbar^2}{\epsilon md}\Bigr)^{1/2}=\sqrt{2g\,\EF E_d}.
\label{eq:Ostar}
\end{equation}
For $g=2$ this is $2\sqrt{\EF E_d}$. It is a characteristic scale, not a cutoff: $\omega_+(q)$ vanishes as $q\to0$, and the relevant comparison at each wave vector is with the local frequency $\omega_+(q)$. The plasmon energy relevant to interlayer pairing is thus proportional to the geometric mean of the Fermi energy and $E_d$, the exciton binding energy at large separation. This is the sense in which $E_B/\EF$ replaces the ratio $\hbar\omega_p/\EF$ of the single-layer problem.

\subsection{Real frequencies}

\begin{figure*}[t]
\includegraphics[width=0.92\textwidth]{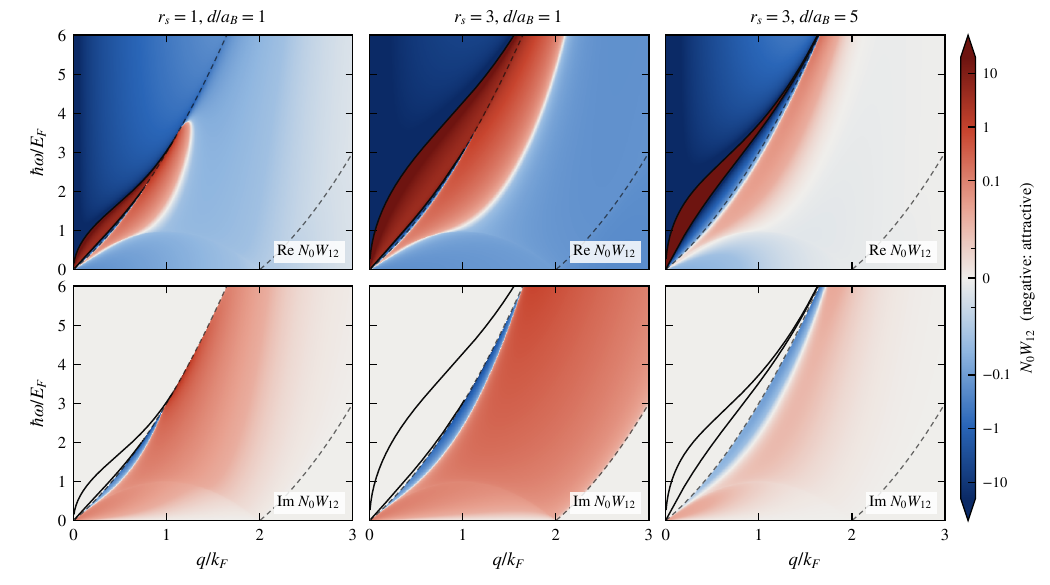}
\caption{Real (top) and imaginary (bottom) parts of the dynamically screened interlayer interaction $N_0W_{12}(q,\omega)$ in the RPA. Blue is attractive, the sign of the bare interaction, and red is repulsive; the color scale is logarithmic. Solid lines: undamped plasmons. Dashed lines: edges of the particle-hole continuum.}
\label{fig:real}
\end{figure*}

Figure~\ref{fig:real} shows the real and imaginary parts of $W_{12}(q,\omega+i0)$. The structure follows from Eq.~\eqref{eq:W12}. Outside the particle-hole continuum $\varepsilon_\pm$ are real. Above both plasmons they are positive and the interaction is attractive, tending to $-U_q$. Between the plasmons $\varepsilon_+<0<\varepsilon_-$ and the interaction is repulsive. Between the acoustic plasmon and the continuum both are negative and it is attractive again. At $\omega=0$ it is attractive at every $q$. The repulsive band continues into the continuum, where the imaginary part is of the same size as the real part. The imaginary part has both signs: the charge mode contributes ${\rm Im}\,W_{12}>0$ and the neutral mode ${\rm Im}\,W_{12}<0$.

The fraction of the $(q,\omega)$ plane in which the interaction is repulsive grows with $r_s$ as the optical plasmon moves up. In the low-energy window $q<2\kF$, $\hbar\omega<\EF$ it is between 6\% and 16\% for $0.5\le r_s\le10$ and $0.25\le d/\aB\le5$, and it decreases with $d$. This is the counterpart of Fig.~1 and Fig.~5 of Ref.~\cite{DasSarma2026}, where the bare interaction is repulsive and the plasmon produces an attractive region.

\subsection{Imaginary frequencies}

The pairing problem at $T_c$ involves the interaction at Matsubara frequencies. There the polarization is real and negative, $\Pi(q,i\nu)=-N_0P(q,\nu)$ with $0<P\le1$ decreasing in $|\nu|$ (Appendix~\ref{app:pol}), so that $\varepsilon_\pm(q,i\nu)=1+(V_q\pm U_q)N_0P\ge1$ and
\begin{equation}
-U_q\le W_{12}(q,i\nu)=-\frac{U_q}{\varepsilon_+(q,i\nu)\,\varepsilon_-(q,i\nu)}<0 .
\label{eq:sign}
\end{equation}
\emph{Within the RPA the interlayer interaction on the imaginary axis is attractive at every wave vector and frequency, and its magnitude increases monotonically with $|\nu|$ from the statically screened value to the bare value.} Screening cannot turn the attraction into a repulsion. The repulsive region on the real axis and Eq.~\eqref{eq:sign} are reconciled by the spectral representation
\begin{equation}
W_{12}(q,i\nu)=-U_q+\frac2\pi\int_0^\infty d\omega\,\frac{\omega\,{\rm Im}\,W_{12}(q,\omega)}{\omega^2+\nu^2},
\label{eq:spectral}
\end{equation}
in which the positive spectral weight of the charge mode screens and the negative weight of the neutral mode antiscreens, the net being a reduction of the attraction.

\begin{figure*}[t]
\includegraphics[width=0.86\textwidth]{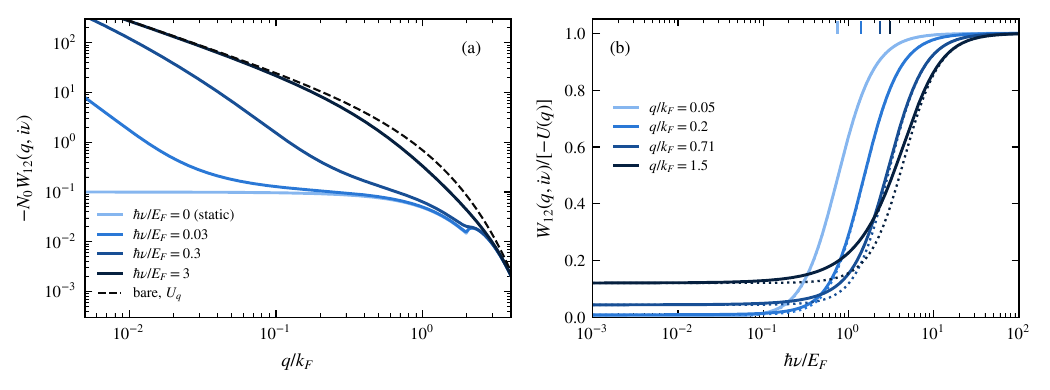}
\caption{Interlayer interaction at Matsubara frequencies for $r_s=2$, $d/\aB=2$. (a) $-N_0W_{12}(q,i\nu)$ versus $q$ for several $\hbar\nu/\EF$; dashed: bare interaction. (b) Ratio to the bare interaction versus $\nu$ for several $q$; dotted: two-pole plasmon-pole approximation; colored ticks: $\omega_+(q)$.}
\label{fig:imag}
\end{figure*}

Two limits of Eq.~\eqref{eq:sign} organize everything that follows. For $\nu\ll\vF q$ the screening is static. For $\nu\gg\vF q$ the polarization takes its high-frequency form $N_0P=nq^2/m\nu^2$, which is fixed by the $f$-sum rule, and
\begin{equation}
W_{12}(q,i\nu)\simeq\frac{-U_q\,\nu^4}{[\nu^2+\omega_+^2(q)][\nu^2+\omega_-^2(q)]}\quad(\nu\gg\vF q).
\label{eq:ppa}
\end{equation}
The interaction is unscreened for $\nu$ above the plasmon frequencies. Since $\omega_+\propto\sqrt q$ vanishes at long wavelength, at any nonzero $\nu$ there is a range of small $q$ in which the full $1/q$ attraction survives. Figure~\ref{fig:imag} illustrates this.

Interpolating between the two limits gives a two-pole plasmon-pole approximation, the bilayer version of the one used in Ref.~\cite{DasSarma2026}: in each channel $1/\varepsilon_\pm(q,i\nu)=1-\Omega_\pm^2/(\nu^2+\tilde\omega_\pm^2)$, with $\Omega_\pm^2=(V_q\pm U_q)nq^2/m$ fixed by the $f$-sum rule and $\tilde\omega_\pm^2=\Omega_\pm^2/[1-1/\varepsilon_\pm(q,0)]$ by the static limit. It is exact in both limits and is compared with the RPA in Fig.~\ref{fig:imag}(b).

\section{Zero magnetic field: Pairing theory}
\label{sec:gap}

\subsection{Linearized gap equation}

\begin{figure}[t]
\centering
\includegraphics{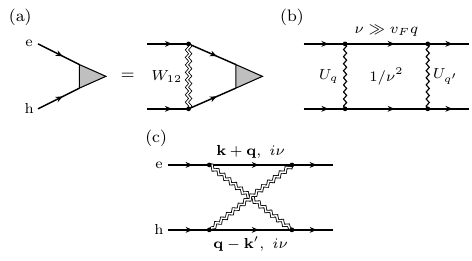}
\caption{(a) Linearized gap equation, Eq.~\eqref{eq:gap}: the anomalous self-energy (triangle) with the dynamically screened interlayer interaction (double wavy line). (b) The process responsible for the plasmon-induced coupling: a low-energy pair scatters by the unscreened interaction (wavy lines) into a virtual pair at small momentum transfer $q$ and at a frequency $\nu$ above the local plasmon frequency $\omega_+(q)$, so that $\nu\gg\vF q$, and back. (c) The crossed rung, the leading vertex correction. In the wedge the two propagators carry the same frequency, and the diagram equals minus the process (b).}
\label{fig:diag}
\end{figure}

With the anomalous Green function $\mathcal F(\mathbf p,i\omega)=-\langle\psi_{1,-\mathbf p}\psi_{2,\mathbf p}\rangle$ and the anomalous self-energy $\phi$, the linearized equation that determines $T_c$ is [Fig.~\ref{fig:diag}(a)]
\begin{align}
\phi(\mathbf k,i\omega_n)=-T\sum_m\!\int\!\!\frac{d^2p}{(2\pi)^2}\,
&W_{12}(\mathbf k-\mathbf p,i\omega_n-i\omega_m)\nonumber\\
&\times\frac{\phi(\mathbf p,i\omega_m)}{\omega_m^2+\xi_p^2}.
\label{eq:gap}
\end{align}
For $s$-wave pairing only the angular average of the interaction enters,
\begin{equation}
K(k,p;\nu)=N_0\!\int_0^{2\pi}\!\frac{d\theta}{2\pi}\,W_{12}\bigl(|\mathbf k-\mathbf p|,i\nu\bigr),
\label{eq:K}
\end{equation}
and with energies in units of $\EF$ Eq.~\eqref{eq:gap} becomes
\begin{equation}
\phi(k,\omega_n)=-\frac Tg\sum_m\int d\xi_p\,\frac{K(k,p;\omega_n-\omega_m)}{\omega_m^2+\xi_p^2}\,\phi(p,\omega_m).
\label{eq:gapK}
\end{equation}
The factor $1/g$ appears because the pairing equation involves the density of states per spin, $N_0/g$, whereas $N_0=gm/(2\pi\hbar^2)$ includes the spin degeneracy $g$. $T_c$ is the temperature at which the largest eigenvalue $\eta(T)$ of the kernel reaches unity. We consider three versions: the unscreened interaction, $W_{12}\to-U_q$, which is the linearized HF theory without the normal exchange self-energy that Ref.~\cite{Zhu2024} includes (Appendix~\ref{app:long}); the statically screened interaction $W_{12}(q,0)$; and the dynamically screened interaction $W_{12}(q,i\nu)$. In the first two the gap is independent of frequency and the Matsubara sum gives $\tanh(\xi_p/2T)/2\xi_p$.

By Eq.~\eqref{eq:sign}, $-K>0$ for all arguments. The kernel of Eq.~\eqref{eq:gapK} is therefore positive, the eigenfunction is nodeless, and $\eta(T)$ increases without bound as $T\to0$ because of the logarithm at the Fermi surface, indicating a pairing instability for arbitrarily weak attraction. \emph{Within the RPA there is a pairing instability at every $r_s$ and $d$}, in agreement with Ref.~\cite{Sreejith2024}. This is the essential difference from the single layer, where the kernel is repulsive and the instability depends on a delicate balance and on a cutoff \cite{DasSarma2026}.

\subsection{Normal self-energy and vertex}
\label{sec:ward}

Equation~\eqref{eq:gap} contains bare propagators and bare vertices. Eliashberg theory as usually applied, and as applied to this problem in Ref.~\cite{Sreejith2024}, dresses the propagators with the normal self-energy $\Sigma=-\mathcal G\,W_{11}$ (where $\mathcal{G}$ denotes the dressed normal-state one-particle Green function), in particular with the factor $Z=1-\partial\Sigma/\partial(i\omega)$ (the Eliashberg renormalization factor, with quasiparticle weight $Z^{-1}$), and keeps the vertices bare. For phonons this is justified by Migdal's theorem. Here it is not, and the self-energy and the vertex are tied together by the Ward identity for the density vertex $\Lambda$ \cite{Nozieres}. At $q=0$ and frequency transfer $\nu$ the identity is
\begin{equation}
\Lambda(\mathbf p;i\omega+i\nu,i\omega)=1-\frac{\Sigma(\mathbf p,i\omega+i\nu)-\Sigma(\mathbf p,i\omega)}{i\nu},
\label{eq:ward}
\end{equation}
which is equivalent to $\mathcal G(i\omega+i\nu)\,\Lambda\,\mathcal G(i\omega)=[\mathcal G(i\omega)-\mathcal G(i\omega+i\nu)]/i\nu$, the relation that the bare propagators satisfy with $\Lambda=1$. When the self-energy is linear in frequency over the interval, the right-hand side of Eq.~\eqref{eq:ward} is $1-\partial\Sigma/\partial(i\omega)=Z$, and a propagator near the Fermi surface is $\mathcal G=Z^{-1}/(i\omega-\xi^*_p)$. In the process of Fig.~\ref{fig:diag}(b) each of the two lines then contributes two vertices and the propagator of the virtual state between them, $Z\times Z^{-1}\times Z$, to the effective interaction between low-energy pairs, and the pair propagator that follows carries $Z^{-2}$. The quasiparticle weights cancel, and the only renormalization that remains is that of the density of states of the low-energy pair. The self-energy and vertex corrections must therefore be treated consistently with the Ward identity. Retaining $Z$ while leaving the density vertex bare ($\Lambda=1$) instead produces a factor $Z^{-4}$ in this limit.

The cancellation with vertex corrections and the $Z^{-4}$ factor without them are both estimates based on counting powers of $Z$. Both assume a self-energy that is linear in frequency over the full range of virtual states and independent of momentum.
The first condition requires $\hbar\Ost\ll\EF$ and therefore large $d$. In the opposite limit, virtual states at frequencies so high that the propagator and the vertex have both returned to their bare values, omitting $\Lambda$ makes no error, and the factor $Z^{-2}$ of the low-energy pair propagator is then a genuine reduction of the coupling and not an artifact. Where a given $r_s$ and $d$ lies between these limits depends on the frequency and momentum dependence of $\Sigma$ and of $\Lambda$, which the identity at $q=0$ does not determine and which we have not computed \cite{Chubukov2005}. The estimates are therefore not bounds. At general $r_s$ and $d$ the Ward identity therefore does not make the dressed ladder equal to the bare one, and we do not claim that it does.

We use bare propagators and bare vertices in Eq.~\eqref{eq:gap} for a different reason: this is the complete theory at first order in the screened interaction (Appendix~\ref{app:long}). The self-energy and the vertex corrections enter together at the next order, and including one without the other is not an improvement. We call this the ladder approximation. It is the bilayer counterpart of the theory of Ref.~\cite{DasSarma2026}. The remaining term of second order in the interlayer interaction, the crossed rung of Fig.~\ref{fig:diag}(c), is the subject of Sec.~\ref{sec:vertex}.

\subsection{Numerical method}
\label{sec:method}

Equation~\eqref{eq:gapK} is solved as an eigenvalue problem in the two variables $\xi_p$ and $\omega_m$. The kernel $K(k,p;\nu)$ has a logarithmic singularity at $p=k$ for $\nu\neq0$, a consequence of the unscreened $1/q$ interaction at small $q$, and the factor $1/(\omega_m^2+\xi_p^2)$ varies on the scale of the temperature, so both variables must be resolved over many decades. The momentum variable is divided into cells that are geometric in $|\xi_p|$ between $10^{-13}\EF$ and $63\,\EF$. The kernel is averaged over each cell, which integrates its singularity, and $1/(\omega_m^2+\xi_p^2)$ is integrated over each cell in closed form. The angular integral in Eq.~\eqref{eq:K} uses a substitution that resolves the forward direction. The Matsubara frequencies up to $10^5\EF$ are grouped into blocks of geometrically increasing size, with the kernel averaged over each block. The discretization errors are of second order in the logarithmic cell size and are removed by Richardson extrapolation in both variables. No momentum or frequency cutoff is imposed. Details and tests, including the reproduction of the BCS result for a model kernel to $0.2\%$, are given in Appendix~\ref{app:num}.

\section{Zero magnetic field: Analytic results at large separation}
\label{sec:analytic}

For $\kF d\gg1$ the bare interaction transfers only $q\lesssim1/d\ll\kF$. Near the Fermi surface $|\mathbf k-\mathbf p|^2\simeq(\xi_k-\xi_p)^2/\hbar^2\vF^2+\kF^2\theta^2$, the kernel depends only on $\xi_k-\xi_p$, and three results follow.

\subsection{Unscreened interaction}

For $W_{12}=-U_q$ the angular average is a modified Bessel function,
\begin{equation}
-\frac{K}{g}=A\,K_0\!\Bigl(\frac{|\xi_k-\xi_p|\,d}{\hbar\vF}\Bigr),\qquad A=\frac{c}{\pi g}=\frac{\alpha}{\pi}.
\label{eq:K0}
\end{equation}
It is logarithmic for $|\xi_k-\xi_p|\ll\hbar\vF/d$ and exponentially small beyond. The interaction cannot scatter a pair further than $\hbar\vF/d$ from the Fermi surface, which is the intrinsic cutoff of the problem. With $K_0(z)\simeq\ln(2e^{-\gamma}/z)$ the gap equation becomes
\begin{equation}
\Delta(\xi)=A\!\int_0^{\Omega_b}\!\frac{d\xi'}{\xi'}\tanh\frac{\xi'}{2T}\,\ln\frac{\Omega_b}{\max(\xi,\xi')}\,\Delta(\xi'),
\label{eq:son}
\end{equation}
with $\Omega_b=2e^{-\gamma}\hbar\vF/d$. The logarithm of the kernel multiplies the BCS logarithm, and the instability condition is $\sqrt A\,\ln(\Omega_b/T)\sim1$ in place of $\lambda\ln(\Omega/T)=1$. Equation~\eqref{eq:son} has the same form as the gap equation of color superconductivity with unscreened magnetic gluons \cite{Son1999}. It is solved in Appendix~\ref{app:son}, with the result
\begin{equation}
k_BT_c^{(0)}=\frac4\pi\,\kappa_0\,\frac{\hbar\vF}{d}\,\exp\!\Bigl[-\frac{\pi}{2\sqrt A}\Bigr],\qquad A=\frac\alpha\pi,
\label{eq:Tc0}
\end{equation}
where $\kappa_0\to1$ for $\alpha\to0$. The unscreened HF transition temperature thus decays only as $1/d$, and its dependence on the coupling is $\exp(-{\rm const}/\sqrt{r_s})$. The numerical solution confirms the $1/d$ law to one part in $10^3$ for $\kF d$ between 5 and 500 and gives $\kappa_0=\kappaA$, $\kappaB$, $\kappaC$, $\kappaD$, and $\kappaE$ for $r_s=0.5$, 1, 2, 3, and 5.

\subsection{Static screening}

For $\kF d\gg1$ the statically screened interaction depends on $q$ only through $y=qd$, and its Fermi-surface average, the BCS coupling constant, is
\begin{align}
\lams&\equiv\frac{N_0}{g}\bigl\langle|W_{12}(q,0)|\bigr\rangle_{\rm FS}=\frac{\cS(\cG)}{\pi g\,\kF d},\label{eq:lams}\\
\cS(\cG)&=\int_0^\infty\!dy\,\frac{\cG\,y\,e^{-y}}{[y+\cG(1+e^{-y})][y+\cG(1-e^{-y})]},\nonumber
\end{align}
where $\cG=gd/a_B$ is the capacitive parameter (not the Green function).
The function $\cS$ has a maximum of $0.25$ at $\cG\simeq0.8$ and falls as $\pi^2/8\cG$ for $\cG\gg1$, so that
\begin{equation}
\lams\simeq\frac{\pi}{16\,g^{3/2}}\,r_s\Bigl(\frac{\aB}{d}\Bigr)^{2}\qquad(d\gg\aB).
\label{eq:lamsasym}
\end{equation}
Both plasmon channels screen statically: the charge channel removes the long-range part and the neutral channel adds the factor $\aB/d$. The transition temperature $T_c^{(s)}\propto\exp(-1/\lams)$ is exponentially small in $d^2$. For $g=2$ and $r_s=2$, Eq.~\eqref{eq:lams} gives $1/\lams=55$ at $d=2\aB$ and $245$ at $d=5\aB$; the full Fermi-surface average of the static interaction, which is used in Table~\ref{tab:vertex}, gives $49$ and $234$.

\subsection{Dynamic screening: the plasmon-induced coupling}
\label{sec:lampl}

\begin{figure*}[t]
\includegraphics[width=0.88\textwidth]{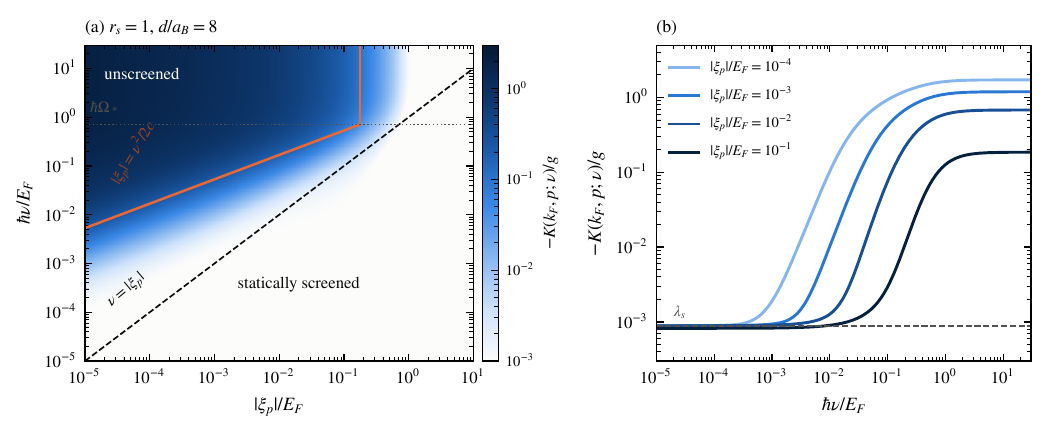}
\caption{(a) Pairing kernel $-K(\kF,p;\nu)/g$ of Eq.~\eqref{eq:K} for $r_s=1$, $d/\aB=8$, as a function of the energy $\xi_p$ of the scattered state and the Matsubara frequency transfer $\nu$. Below the dashed line $\nu=|\xi_p|$ the kernel has its static value $\lams\simeq9\times10^{-4}$. Inside the wedge bounded by the solid line, Eq.~\eqref{eq:wedge}, the interaction is unscreened and the kernel is of order unity. (b) Cuts at fixed $\xi_p$.}
\label{fig:wedge}
\end{figure*}

The dynamically screened kernel is shown in Fig.~\ref{fig:wedge} as a function of the two variables on which it depends near the Fermi surface. It takes the small static value $\lams$ when $\nu\lesssim|\xi_p|$, which is the region reached by on-shell scattering. In the wedge
\begin{equation}
|\xi_p|\lesssim\frac{\hbar^2\nu^2}{2c\,\EF}\quad(\nu<\Ost),\quad|\xi_p|\lesssim\frac{\hbar\vF}{d}\quad(\nu>\Ost),
\label{eq:wedge}
\end{equation}
it is larger by orders of magnitude, because there Eq.~\eqref{eq:ppa} applies with $\nu>\omega_+(q)$ and the interaction is unscreened. The first boundary is the condition $\nu=\omega_+(q)$ at $q=|\xi_p|/\hbar\vF$, and the second is the range of the bare interaction.

We use a two-step theory. First, we eliminate virtual states in the wedge to second order in the kernel to obtain an effective low-energy pairing coupling. We then solve the resulting BCS gap equation to determine $T_c$. States in the wedge are far off the energy shell, $\nu\gg|\xi_p|$, and their pair propagator is $1/\nu^2$ with no logarithm. They contribute through the second-order process of Fig.~\ref{fig:diag}(b): a pair at the Fermi surface is scattered into the wedge and back. Eliminating the wedge to second order in the kernel gives an effective coupling in the low-energy sector,
\begin{align}
\lambda_{\rm eff}&=\lams+\lampl,\label{eq:lameff}\\
\lampl&=\frac1{g^2}\int\!d\xi\!\int_0^\infty\!\frac{d\nu}{\pi}\,\frac{K(\xi;\nu)^2-K(\xi;0)^2}{\nu^2+\xi^2},\label{eq:lampl}
\end{align}
where $K(\xi;\nu)=K(\kF,p;\nu)$ at $\xi_p=\xi$. The subtraction removes the iteration of the static kernel, which is part of the BCS logarithm. The sign of $\lampl$ is positive whatever the sign of $K$: the gap function at high frequency is $-K$ times the low-energy pair amplitude, and it feeds back with a second factor of $-K$. This is the mechanism of plasmon-induced pairing in the repulsive single-layer problem. Here it adds to an interaction that is already attractive.

In the wedge, with $y=qd$ and $w=\nu/\Ost$, Eq.~\eqref{eq:ppa} becomes $W_{12}=-U_q\,s(y;w)/e^{-y}$ with
\begin{equation}
s(y;w)=\frac{e^{-y}}{\bigl[1+\frac{y(1+e^{-y})}{2w^2}\bigr]\bigl[1+\frac{y(1-e^{-y})}{2w^2}\bigr]},
\end{equation}
and $K=-(c/\pi)\,k(\eta;w)$ with $k(\eta;w)=\int_\eta^\infty dy\,s(y;w)/\sqrt{y^2-\eta^2}$ and $\eta=|\xi|d/\hbar\vF$. Equation~\eqref{eq:lampl} then reduces to a pure number times $c\,\hbar\Ost/\EF$:
\begin{align}
\lampl&=\frac{\cJ}{\pi^3g}\,\alpha\,\frac{\hbar\Ost}{\EF}
=\frac{\cJ}{\sqrt2\,\pi^3}\;r_s\sqrt{\frac{E_d}{\EF}}
=\frac{\cJ\sqrt g}{2\pi^3}\;r_s^2\sqrt{\frac{\aB}{d}},\label{eq:lamplasym}\\
\cJ&=\int_0^\infty\!\frac{dw}{w^2}\int_0^\infty\!d\eta\;k(\eta;w)^2=4.777.\nonumber
\end{align}
Numerically $\lampl=0.109\,r_s\sqrt{E_d/\EF}$, which for $g=2$ is $0.109\,r_s^2\sqrt{\aB/d}$. Equation~\eqref{eq:lamplasym} is the central analytic result of the ladder theory. It holds for $\cG\gg1$ and $\kF d\gg1$; the direct numerical evaluation of Eq.~\eqref{eq:lampl} with the full RPA kernel approaches it from above and is within \lamplErrA\ of it at $d=5\aB$ and within \lamplErrB\ at $d=20\aB$ for $r_s=1$ and 2.

The plasmon-induced coupling is the product of the Coulomb coupling $\alpha$ and the ratio of the plasmon energy at $q=1/d$ to the Fermi energy. It decays as $d^{-1/2}$, against $d^{-2}$ for $\lams$. The two are equal at
\begin{equation}
\frac{d_\times}{\aB}=\Bigl(\frac{\pi^4}{8\cJ g^2}\Bigr)^{2/3}r_s^{-2/3}\simeq0.74\,r_s^{-2/3}\quad(g=2),
\label{eq:crossover}
\end{equation}
and beyond $d_\times$ the plasmon-induced coupling dominates.

The low-energy sector that remains after the elimination is a BCS problem with coupling $\lambda_{\rm eff}$. Its logarithm is cut off at $\hbar\vF/d$: by Eq.~\eqref{eq:wedge} a state at energy $\xi$ couples only to the part of the wedge with $\nu>\omega_+(|\xi|/\hbar\vF)$, so the effective coupling falls off for $|\xi|\gtrsim\hbar\vF/d$, as the static one does. Hence
\begin{equation}
k_BT_c\simeq C\,\frac{\hbar\vF}{d}\,\exp\!\Bigl[-\frac{1}{\lams+\lampl}\Bigr],
\label{eq:Tcdyn}
\end{equation}
with $C$ of order unity. The elimination also gives the eigenvalue away from $T_c$. The wedge enters at second order, so in the eigenvalue equation its contribution carries a factor $1/\eta$, and
\begin{equation}
\eta^2=\bigl(\lams\,\eta+\lampl\bigr)\ln\frac{C\hbar\vF}{k_BT\,d}.
\label{eq:etaT}
\end{equation}
For $\lampl=0$ this is the BCS line $\eta=\lams\ln(1/T)$. When the plasmon-induced term dominates, $\eta$ grows as the square root of the logarithm. Equation~\eqref{eq:etaT} reduces to Eq.~\eqref{eq:Tcdyn} at $\eta=1$, and it provides a test of the theory that does not involve the exponentially small $T_c$ itself.

Terms of third and higher order in the kernel, in which the virtual pair scatters within the wedge before it returns, all have the same sign and raise $T_c$. The third-order term is smaller than $\lampl$ by a factor of order $\hbar\Ost/\EF$. In the exponent it is to be compared with $\lampl^2$, so it changes $\ln T_c$ by an amount of order $(\hbar\Ost/\EF)/\lampl\sim1/\alpha$, which does not depend on $d$. It belongs to the prefactor: $C$ is a function of $r_s$ and is not universal. At large $d$,
\begin{equation}
\ln\frac{k_BT_c\,d}{\hbar\vF}\simeq-\frac{2\pi^3}{\cJ\sqrt g}\,\frac{1}{r_s^{2}}\sqrt{\frac{d}{\aB}}=-\frac{9.2}{r_s^{2}}\sqrt{\frac{d}{\aB}}\quad(g=2).
\label{eq:stretched}
\end{equation}
The dynamically screened $T_c$ is a stretched exponential in $d$. It lies between the power law of the unscreened theory and the Gaussian of the statically screened one, and it is far closer to the former.

\section{Zero magnetic field: Numerical results within ladder approximation}
\label{sec:numerics}

\subsection{Transition temperature}

\begin{figure*}[t]
\includegraphics[width=0.95\textwidth]{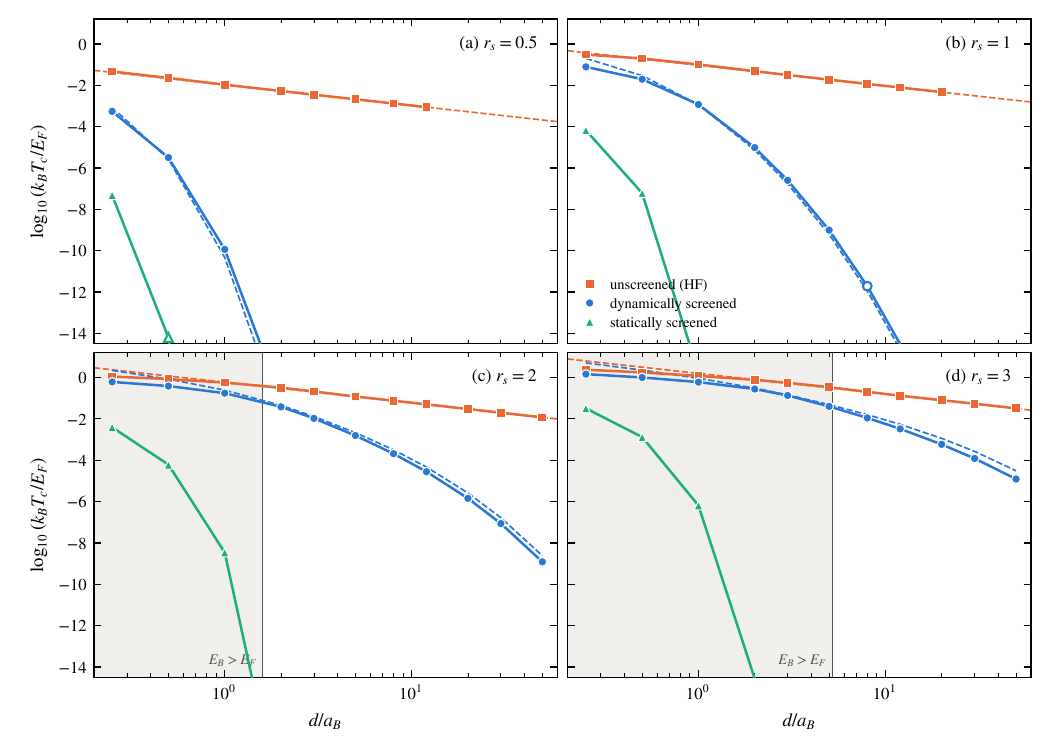}
\caption{Mean-field transition temperature versus layer separation for the unscreened (HF), dynamically screened, and statically screened interlayer interaction. Symbols: full numerical solution of Eq.~\eqref{eq:gapK}. Open symbols: values obtained by extrapolating $\eta(T)$ below the lowest temperature of the calculation. Dashed lines: Eq.~\eqref{eq:Tc0} for the unscreened interaction and Eq.~\eqref{eq:Tcdyn} with $C=0.5$ for the dynamically screened one. The statically screened $T_c$ leaves the frame at small $d$. Shaded: $E_B>\EF$, where the weak-pairing theory does not apply.}
\label{fig:tcd}
\end{figure*}

All results in this section are for the ladder approximation, Eq.~\eqref{eq:gapK}. Figure~\ref{fig:tcd} shows the mean-field $T_c$ from its full solution for the three interactions, and Table~\ref{tab:tc} lists representative values. The ordering $T_c^{(s)}<T_c<T_c^{(0)}$ is a direct consequence of Eq.~\eqref{eq:sign}, since the three kernels are ordered pointwise. The separations between them are very large. At $r_s=2$ and $d=5\aB$, for example, the unscreened, dynamically screened, and statically screened transition temperatures are $10^{\TcBareTwoFive}$, $10^{\TcDynTwoFive}$, and $10^{\TcStatTwoFive}$ in units of $\EF/k_B$. Within the ladder approximation, static screening is not an approximation to dynamic screening in any part of the parameter space we have examined.

\begin{figure}[t]
\includegraphics[width=\columnwidth]{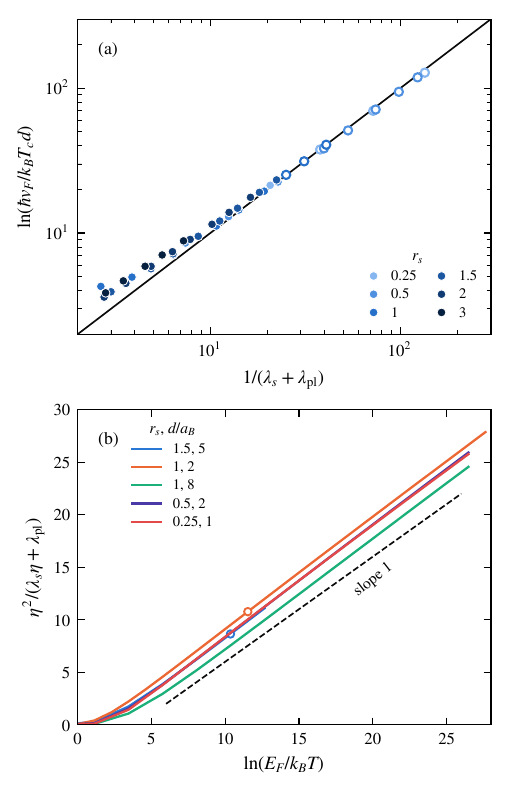}
\caption{Tests of the two-step theory of Sec.~\ref{sec:lampl}. (a) $\ln(\hbar\vF/k_BT_cd)$ from the numerical solution with dynamic screening versus $1/(\lams+\lampl)$ from Eqs.~\eqref{eq:lams} and \eqref{eq:lampl}, for all computed points with $E_B<\EF$. Line: unit slope, $C=1$ in Eq.~\eqref{eq:Tcdyn}. Open symbols: $T_c$ extrapolated below the lowest temperature of the calculation. (b) Test of Eq.~\eqref{eq:etaT}: $\eta^2/(\lams\eta+\lampl)$ from the numerical eigenvalue versus $\ln(\EF/k_BT)$. Circles mark $T_c$.}
\label{fig:scaling}
\end{figure}

The unscreened result follows the $1/d$ law of Eq.~\eqref{eq:Tc0}. The dynamically screened result decays faster than any power but much more slowly than an exponential in $d^2$. Figure~\ref{fig:scaling}(a) tests Eq.~\eqref{eq:Tcdyn}. It shows $\ln(\hbar\vF/k_BT_cd)$ from the full numerical solution against $1/(\lams+\lampl)$, with the two couplings computed independently from Eqs.~\eqref{eq:lams} and \eqref{eq:lampl}. All points with $E_B<\EF$, for $r_s$ between 0.25 and 3 and for exponents between \invlammin\ and \invlammax, lie on the line of unit slope to within an additive constant: $\ln C$ is between \lnCmin\ and \lnCmax, while the exponent varies by a factor of more than forty. For the points with $1/(\lams+\lampl)>10$ the numerical exponent and the two-step estimate differ by at most \expErr. The test rests on the \ndirect\ points at which $T_c$ is computed directly, which reach exponents up to \invlammaxdirect\ and have $\ln C$ between \lnCminDirect\ and \lnCmaxDirect. The points at larger exponents are extrapolated with the variable of Eq.~\eqref{eq:etaT} (Appendix~\ref{app:num}); the slope and the intercept of the extrapolation are taken from the numerical $\eta(T)$ and not from the theory, but these points assume the linearity that Fig.~\ref{fig:scaling}(b) shows over the computed range, and they are a weaker test. The constant is not universal. At fixed $r_s$ it tends to a limit at large $d$: for $r_s=2$, $\ln C$ is \lnCtwoA, \lnCtwoB, and \lnCtwoC\ at $d/\aB=20$, 30, and 50. It increases as $r_s$ decreases, as expected from the third-order term discussed in Sec.~\ref{sec:lampl}.

Figure~\ref{fig:scaling}(b) tests Eq.~\eqref{eq:etaT}. The combination $\eta^2/(\lams\eta+\lampl)$ formed from the numerical eigenvalue is a linear function of $\ln(1/T)$ below $k_BT\approx3\times10^{-3}\EF$, both above and below $T_c$, with a slope between \slopemin\ and \slopemax, and between \slopeLowMin\ and \slopeLowMax\ at the lowest temperatures. The excess over unity decreases slowly with decreasing temperature and is the signature of the third-order term.

\begin{figure}[t]
\includegraphics[width=\columnwidth]{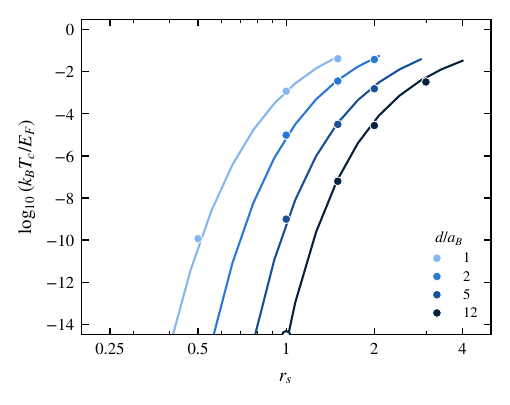}
\caption{Dynamically screened mean-field $T_c$ versus $r_s$ at fixed separation. Lines: Eq.~\eqref{eq:Tcdyn} with $C=0.5$ and the couplings of Eqs.~\eqref{eq:lams} and \eqref{eq:lampl}. Open symbols: extrapolated. Points with $E_B>\EF$ are omitted.}
\label{fig:tcrs}
\end{figure}

The dependence on $r_s$ at fixed $d$ is shown in Fig.~\ref{fig:tcrs}. $T_c/\EF$ increases rapidly with $r_s$, as $\exp(-{\rm const}/r_s^2)$ at large $d$ according to Eq.~\eqref{eq:stretched}, until $E_B$ becomes comparable to $\EF$, where the mean-field $T_c$ at fixed chemical potential ceases to be meaningful.

\begin{table*}[t]
\caption{Representative results for $g=2$. $T_c^{(0)}$, $T_c$, and $T_c^{(s)}$ are the mean-field transition temperatures for the unscreened, dynamically screened, and statically screened interlayer interaction, and $T_c^{\rm PPA}$ is the result of the two-pole plasmon-pole approximation; all are given as $\log_{10}(k_BT_c/\EF)$. The couplings $\lams$ and $\lampl$ are evaluated from Eqs.~\eqref{eq:lams} and \eqref{eq:lampl} with the full RPA kernel. The last two columns compare the two sides of Eq.~\eqref{eq:Tcdyn}.}
\label{tab:tc}
\begin{ruledtabular}
\begin{tabular}{ccccccccccccc}
$r_s$ & $d/\aB$ & $\kF d$ & $E_B/\EF$ & $\hbar\Ost/\EF$ & $\lams$ & $\lampl$ & $T_c^{(0)}$ & $T_c$ & $T_c^{\rm PPA}$ & $T_c^{(s)}$ & $\dfrac{1}{\lams+\lampl}$ & $\ln\dfrac{\hbar\vF}{k_BT_cd}$\\
\hline
0.25 & 0.25 & 1.41 & 0.04 & 1.00 & 0.0294 & 0.0190 & $-2.52$ & $-9.13$ & --- & $-14.84$$^{\rm a}$ & 20.7 & 21.4\\
 & 1 & 5.66 & 0.02 & 0.50 & 0.0061 & 0.0077 & $-3.15$ & $-31$$^{\rm a}$ & --- & $-71$$^{\rm a}$ & 72.1 & 69.8\\
0.5 & 0.25 & 0.71 & 0.15 & 2.00 & 0.0581 & 0.0770 & $-1.33$ & $-3.26$ & --- & $-7.32$ & 7.4 & 8.5\\
 & 1 & 2.83 & 0.08 & 1.00 & 0.0127 & 0.0313 & $-1.97$ & $-9.93$ & --- & $-34$$^{\rm a}$ & 22.7 & 22.5\\
 & 5 & 14.14 & 0.03 & 0.45 & 0.0010 & 0.0124 & $-2.68$ & $-32$$^{\rm a}$ & --- & $-426$$^{\rm a}$ & 74.2 & 71.3\\
1 & 0.5 & 0.71 & 0.46 & 2.83 & 0.0579 & 0.2005 & $-0.70$ & $-1.71$ & --- & $-7.21$ & 3.9 & 5.0\\
 & 1 & 1.41 & 0.33 & 2.00 & 0.0271 & 0.1290 & $-0.99$ & $-2.93$ & $-3.07$ & $-15.89$$^{\rm a}$ & 6.4 & 7.1\\
 & 2 & 2.83 & 0.22 & 1.41 & 0.0097 & 0.0832 & $-1.32$ & $-5.01$ & $-5.15$ & $-45$$^{\rm a}$ & 10.8 & 11.2\\
 & 5 & 7.07 & 0.11 & 0.89 & 0.0021 & 0.0499 & $-1.73$ & $-9.00$ & --- & $-211$$^{\rm a}$ & 19.2 & 19.5\\
 & 12 & 16.97 & 0.06 & 0.58 & 0.0004 & 0.0316 & $-2.11$ & $-14.52$$^{\rm a}$ & --- & $-1037$$^{\rm a}$ & 31.2 & 31.3\\
1.5 & 2 & 1.89 & 0.49 & 2.12 & 0.0152 & 0.1898 & $-0.82$ & $-2.44$ & --- & $-28.46$$^{\rm a}$ & 4.9 & 5.7\\
 & 5 & 4.71 & 0.26 & 1.34 & 0.0031 & 0.1125 & $-1.24$ & $-4.50$ & --- & $-139$$^{\rm a}$ & 8.6 & 9.5\\
 & 12 & 11.31 & 0.13 & 0.87 & 0.0006 & 0.0712 & $-1.62$ & $-7.20$ & --- & $-690$$^{\rm a}$ & 13.9 & 14.9\\
2 & 2 & 1.41 & 0.87 & 2.83 & 0.0205 & 0.3416 & $-0.49$ & $-1.42$ & $-1.48$ & $-20.95$$^{\rm a}$ & 2.8 & 3.6\\
 & 5 & 3.54 & 0.46 & 1.79 & 0.0043 & 0.2007 & $-0.92$ & $-2.81$ & $-2.85$ & $-102$$^{\rm a}$ & 4.9 & 5.9\\
 & 12 & 8.49 & 0.23 & 1.15 & 0.0008 & 0.1267 & $-1.30$ & $-4.56$ & --- & $-516$$^{\rm a}$ & 7.8 & 9.0\\
 & 30 & 21.21 & 0.10 & 0.73 & 0.0001 & 0.0795 & $-1.70$ & $-7.06$ & --- & $-2986$$^{\rm a}$ & 12.6 & 13.9\\
3 & 8 & 3.77 & 0.72 & 2.12 & 0.0028 & 0.3528 & $-0.69$ & $-1.96$ & --- & $-157$$^{\rm a}$ & 2.8 & 3.9\\
 & 12 & 5.66 & 0.52 & 1.73 & 0.0013 & 0.2856 & $-0.88$ & $-2.48$ & --- & $-340$$^{\rm a}$ & 3.5 & 4.7\\
 & 30 & 14.14 & 0.24 & 1.10 & 0.0002 & 0.1790 & $-1.27$ & $-3.92$ & --- & $-1987$$^{\rm a}$ & 5.6 & 7.1\\
\end{tabular}
\end{ruledtabular}
\begin{flushleft}
$^{\rm a}$Below the lowest temperature of the calculation; obtained by extrapolating the eigenvalue $\eta(T)$.
\end{flushleft}
\end{table*}

\subsection{Eigenvalue and gap function}

\begin{figure*}[t]
\includegraphics[width=0.86\textwidth]{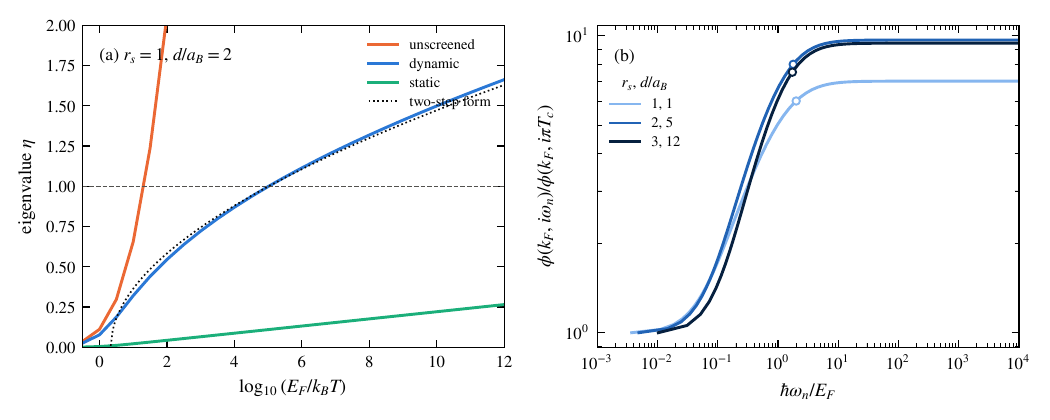}
\caption{(a) Largest eigenvalue of the linearized gap equation versus temperature for the three interactions at $r_s=1$, $d/\aB=2$. $T_c$ is where $\eta=1$. Dotted: Eq.~\eqref{eq:etaT} with $C$ fixed at $T_c$. (b) Gap function $\phi(\kF,i\omega_n)$ at $T_c$ for the dynamically screened interaction, normalized to its value at the lowest Matsubara frequency. Circles mark $\hbar\Ost$.}
\label{fig:eta}
\end{figure*}

Figure~\ref{fig:eta}(a) shows the eigenvalue $\eta(T)$ for the three interactions. For static screening it is a straight line in $\ln(1/T)$ with slope $\lams$, the BCS form. For the unscreened interaction it is quadratic, the signature of Eq.~\eqref{eq:son}. For dynamic screening it follows Eq.~\eqref{eq:etaT}: it rises steeply at temperatures comparable to the plasmon energy, where the wedge is populated directly, and then grows as the square root of the logarithm.

The gap function at $T_c$, Fig.~\ref{fig:eta}(b), is nodeless. It rises with frequency from its low-frequency value to a plateau that is larger by a factor of 7 to 10 in the examples shown, and the rise takes place at frequencies somewhat below $\Ost$, where the wedge opens. The plateau is the amplitude in the wedge that mediates $\lampl$. In the repulsive single-layer problem the corresponding gap function changes sign at the plasmon scale \cite{Takada1978,DasSarma2026}; here it does not.

\subsection{Frequency cutoff}

\begin{figure}[t]
\includegraphics[width=\columnwidth]{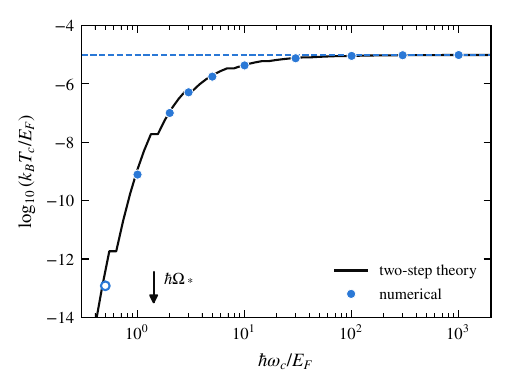}
\caption{Dependence of the dynamically screened $T_c$ on a Matsubara cutoff $\omega_c$ for $r_s=1$, $d=2\aB$. Symbols: numerical solution of Eq.~\eqref{eq:gapK}. Solid line: two-step theory, $T_c(\omega_c)=T_c(\infty)\exp[1/\lambda(\infty)-1/\lambda(\omega_c)]$ with $\lambda=\lams+\lampl$ and the frequency integral of Eq.~\eqref{eq:lampl} cut off at $\omega_c$. The small steps in the solid curve arise from the discrete frequency grid used to evaluate the cutoff integral. Dashed line: cutoff-free result. The arrow marks $\Ost$. }
\label{fig:cutoff}
\end{figure}

In the single-layer problem the calculated $T_c$ depends strongly on the Matsubara cutoff $\omega_c$, and no value of $\omega_c$ can be justified \cite{DasSarma2026}. Figure~\ref{fig:cutoff} shows the corresponding dependence here, obtained by restricting both frequencies in Eq.~\eqref{eq:gapK} to $|\omega|<\omega_c$, for $r_s=1$ and $d=2\aB$. $T_c$ rises by many orders of magnitude as $\omega_c$ is raised through $\Ost$ and the wedge is included, and it then converges to the cutoff-free value as $1/\omega_c$. It is lower than that value by a factor of \cutTen\ at $\hbar\omega_c=10\,\EF\approx7\hbar\Ost$ and by \cutHundred\ at $100\,\EF$. The whole dependence is reproduced by the two-step theory with the frequency integral of Eq.~\eqref{eq:lampl} cut off at $\omega_c$, which confirms that the effect of dynamic screening in the ladder is the coupling $\lampl$ and nothing else.

The convergence is the difference from the single layer. There the coupling keeps growing with $\omega_c$, and it competes with a static repulsion of the same size. Here the range of the interlayer interaction limits the wedge to $|\xi|\lesssim\hbar\vF/d$, the contribution of frequencies above $\Ost$ falls as $1/\nu^2$, and the static term has the same sign. The slow convergence nevertheless shows how much of the result comes from frequencies well above $\Ost$, where the interaction is the bare one.

The two-pole plasmon-pole approximation of Sec.~\ref{sec:W} reproduces the RPA result for $\log_{10}T_c$ to within \ppaErr\ at the points where we have compared them (Table~\ref{tab:tc}), as expected from the fact that it is exact in the wedge. It gives the lower value in each case.

\section{Zero magnetic field: Vertex corrections and the crossed rung}
\label{sec:vertex}

The results of Secs.~\ref{sec:analytic} and \ref{sec:numerics} are those of the ladder approximation. The large enhancement of $T_c$ over the statically screened value comes from one term, $\lampl$, which is of second order in the screened interaction. At that order the irreducible vertex has another term that contains only the interlayer interaction, the crossed rung of Fig.~\ref{fig:diag}(c). A theorem of the Migdal type would make it negligible. We show here that it is not negligible, that it can be computed, and that it cancels $\lampl$.

\subsection{Why Migdal's theorem does not apply}

Migdal's argument requires the boson to be slow, $\omega\ll\vF q$ at the wave vectors that matter. The vertex correction is then of order $\omega_D/\EF$ \cite{Migdal1958}. In the bilayer the wave vectors that matter are $q\lesssim1/d$, and both collective modes are fast there:
\begin{equation}
\frac{\Ost}{\vF/d}=\sqrt{\cG},\qquad\frac{v_-}{\vF}\simeq\sqrt{\cG/2},\qquad\cG=\frac{g\,d}{\aB}.
\label{eq:fast}
\end{equation}
The wedge of Eq.~\eqref{eq:wedge}, from which the whole of $\lampl$ comes, is by its definition the region $\nu>\omega_+(q)\gg\vF q$. The smallness of the plasmon energy compared with the Fermi energy, $\hbar\Ost/\EF=(2gE_d/\EF)^{1/2}\ll1$ at large $d$, does not help: Migdal's parameter is the ratio of the boson velocity to the Fermi velocity, and it is large. The same is true of the plasmon of a single layer \cite{DasSarma2026}.

\emph{A bilayer in which Migdal's theorem does hold.} The criterion is made explicit by comparison with the all-electron bilayer with unequal layer masses of Ref.~\cite{Wang2026}. There the electrons of a light layer, of mass $m_1$, are paired by the exchange of the acoustic plasmon of a heavy layer, of mass $m_2$, and the Coulomb interaction is purely repulsive. At small separation and comparable densities the velocity of that plasmon is of order $\hbar\kF/(m_1m_2)^{1/2}$, which is below the Fermi velocity of the light layer by a factor of order $(m_1/m_2)^{1/2}$. For $m_2/m_1\gg1$ the boson is therefore slow for the carriers that pair. The heavy layer plays the part of the ions and its acoustic plasmon that of the acoustic phonon, Migdal's theorem holds with the mass ratio as its small parameter, and Eliashberg theory is controlled although no part of the interaction is attractive. As the mass ratio approaches unity the control is lost and the result depends on a frequency cutoff \cite{Wang2026}. The equal-mass electron-hole bilayer is at the opposite end of the same axis. By Eqs.~\eqref{eq:vminus} and \eqref{eq:fast} both modes are faster than both carriers of the pair at every separation, there is no mass ratio to make them slow, and the theorem that applies is the adiabatic one of Sec.~\ref{sec:adiabatic}. The two problems are complementary. In one, the retardation of a slow mode produces an attraction from a repulsion, and the ladder is controlled. In the other, the mode is fast, retardation is ineffective, and the ladder overestimates the pairing. In an electron-hole bilayer with a large mass ratio and small separation, the acoustic mode can be slower than the light carrier but faster than the heavy carrier. Since an interlayer pair contains both carriers, neither limit discussed above directly applies. Whether the dynamical enhancement of pairing survives vertex corrections in this regime remains unresolved by the present analysis. At sufficiently large $d$, the acoustic mode is faster than both carriers for any finite mass ratio.

\subsection{Second-order vertex at the Fermi surface}

Consider an electron at $\mathbf k$ and a hole at $-\mathbf k$ scattered to $\mathbf k'$ and $-\mathbf k'$, all at the Fermi surface and at zero frequency, with $\bm\Delta=\mathbf k'-\mathbf k$. To second order in $W_{12}$ the amplitude is
\begin{align}
\Gamma^{(2)}={}&\int\!\frac{d\nu}{2\pi}\!\int\!\frac{d^2q}{(2\pi)^2}\,W_{12}(\mathbf q,i\nu)\,W_{12}(\bm\Delta-\mathbf q,i\nu)\nonumber\\
&\times\bigl[B_{\rm lad}+B_{\rm cr}\bigr],
\label{eq:Gamma2}
\end{align}
where $B_{\rm lad}$ and $B_{\rm cr}$ denote the electron–hole propagator factors for the second-order ladder and crossed-rung diagrams, respectively.
The first term is the second iteration of the ladder. The electron is at $\mathbf k+\mathbf q$ with frequency $\nu$ and the hole at $-\mathbf k-\mathbf q$ with frequency $-\nu$, so that $\mathcal G_e\mathcal G_h=1/(\nu^2+\xi_1^2)$ with $\xi_1=\xi_{\mathbf k+\mathbf q}$. The second term is the crossed rung. The electron is again at $\mathbf k+\mathbf q$, but the hole absorbs the two momentum transfers in the opposite order: it is at $\mathbf q-\mathbf k'$ and it carries the frequency $+\nu$. With $\xi_2=\xi_{\mathbf k'-\mathbf q}$ the product is $1/[(i\nu-\xi_1)(i\nu-\xi_2)]$. The interaction factors are symmetric under $\mathbf q\to\bm\Delta-\mathbf q$, which exchanges $\xi_1$ and $\xi_2$. After symmetrization
\begin{align}
B_{\rm lad}&=\frac12\Bigl[\frac1{\nu^2+\xi_1^2}+\frac1{\nu^2+\xi_2^2}\Bigr],\nonumber\\
B_{\rm cr}&=\frac{\xi_1\xi_2-\nu^2}{(\nu^2+\xi_1^2)(\nu^2+\xi_2^2)},\nonumber\\
B_{\rm lad}+B_{\rm cr}&=\frac{(\xi_1+\xi_2)^2}{2(\nu^2+\xi_1^2)(\nu^2+\xi_2^2)}.
\label{eq:Bnet}
\end{align}
This is an identity. For a parabolic band
\begin{equation}
\xi_1+\xi_2=\frac{\hbar^2}{m}\,\mathbf q\cdot(\mathbf q-\bm\Delta).
\label{eq:xisum}
\end{equation}
The terms of first order in $\mathbf q$ have cancelled. The electron and the hole of a pair move in opposite directions, so for scattering at small angle a momentum transfer that raises the energy of one lowers the energy of the other by the same amount, $\xi_2=-\xi_1$, and then $B_{\rm cr}=-B_{\rm lad}$ at every frequency. What is left is the curvature of the dispersion.

As in Eq.~\eqref{eq:lampl} we separate the product of the static interactions and define, for $X={\rm lad}$, cr,
\begin{align}
\lambda_X={}&\frac{N_0}{g}\Bigl\langle\int\!\frac{d\nu}{2\pi}\!\int\!\frac{d^2q}{(2\pi)^2}\bigl[W_{12}(\mathbf q,i\nu)W_{12}(\bm\Delta-\mathbf q,i\nu)\nonumber\\
&-W_{12}(\mathbf q,0)W_{12}(\bm\Delta-\mathbf q,0)\bigr]B_X\Bigr\rangle_{\mathbf k'},
\label{eq:lamX}
\end{align}
where the average is over the direction of $\mathbf k'$. The ladder term is the plasmon-induced coupling, $\lambda_{\rm lad}=\lampl$. The crossed term $\lambda_{\rm cr}$ is an addition to the pairing interaction between the low-energy states, of the same order, and the effective coupling of the low-energy sector becomes
\begin{equation}
\lambda_{\rm eff}=\lams+\lambda_{\rm net}+\lambda_{\rm GMB},\qquad\lambda_{\rm net}=\lampl+\lambda_{\rm cr}.
\label{eq:lamnet}
\end{equation}
The last term is the crossed rung with the two static interactions,
\begin{align}
\lambda_{\rm GMB}=-\frac{N_0}{g}\Bigl\langle\int\!\frac{d^2q}{(2\pi)^2}\,&W_{12}(\mathbf q,0)W_{12}(\bm\Delta-\mathbf q,0)\nonumber\\
&\times\frac{\theta(-\xi_1\xi_2)}{|\xi_1-\xi_2|}\Bigr\rangle_{\mathbf k'},
\label{eq:gmb}
\end{align}
the analog of the induced interaction of Gor'kov and Melik-Barkhudarov \cite{Gorkov1961}. It is repulsive and of order $\lams^2\ln\kF d$.

In the wedge $\nu\gg|\xi_{1,2}|$, so $B_{\rm lad}\simeq1/\nu^2$ and, by Eqs.~\eqref{eq:Bnet} and \eqref{eq:xisum}, $B_{\rm lad}+B_{\rm cr}\simeq(\xi_1+\xi_2)^2/2\nu^4$. The cancellation leaves a fraction $(\hbar^2q^2/m\hbar\nu)^2$, which at $q\sim1/d$ and $\nu\sim\Ost$ is $1/[\cG(\kF d)^2]$. We denote by $\lambda_w$ the residual contribution of the wedge to $\lambda_{\mathrm{net}}$ after the partial cancellation between the ladder and crossed-rung diagrams. It is of order $\lambda_{\mathrm{pl}}/[\cG(k_Fd)^2]$, with a coefficient evaluated in Appendix F:
\begin{equation}
\lambda_{w}=J_w\,\sqrt g\,r_s^{4}\Bigl(\frac{\aB}{d}\Bigr)^{7/2},\qquad J_w=\Jw,
\label{eq:lamnetscaling}
\end{equation}
to be compared with $\lampl\propto d^{-1/2}$ and $\lams\propto d^{-2}$. The wedge is not the only contribution to $\lambda_{\rm net}$. At frequencies and energies of the order of the curvature energy $\hbar^2/md^2$ the factor $B_{\rm lad}+B_{\rm cr}$ is not small, and the screened interaction differs from its static value by the term linear in $|\nu|/\vF q$ that comes from the particle-hole continuum. This region gives the leading term of $\lambda_{\rm net}$ at small $r_s$, which can be obtained in closed form (Appendix~\ref{app:remainder}):
\begin{align}
\lambda_{\rm net}&\to C_3(d)\,r_s^{3}\qquad(r_s\to0),\nonumber\\
C_3&\to\frac{7\pi\zeta(3)}{256\,g^{5/2}}\Bigl(\frac{\aB}{d}\Bigr)^{5}\qquad(d\gg\aB),
\label{eq:lamnetlow}
\end{align}
with $C_3(d)$ given for every $d$ by Eq.~\eqref{eq:C3}. The integrand of $\lambda_{\rm net}$ is nonnegative, so the two contributions add. The second is of lower order in $r_s$ and falls off faster with $d$, so that Eq.~\eqref{eq:lamnetlow} is the limit of $\lambda_{\rm net}$ for $r_s\to0$ at fixed $d$ and Eq.~\eqref{eq:lamnetscaling} its limit for $d\to\infty$ at fixed $r_s$.

\begin{figure*}[t]
\includegraphics[width=0.9\textwidth]{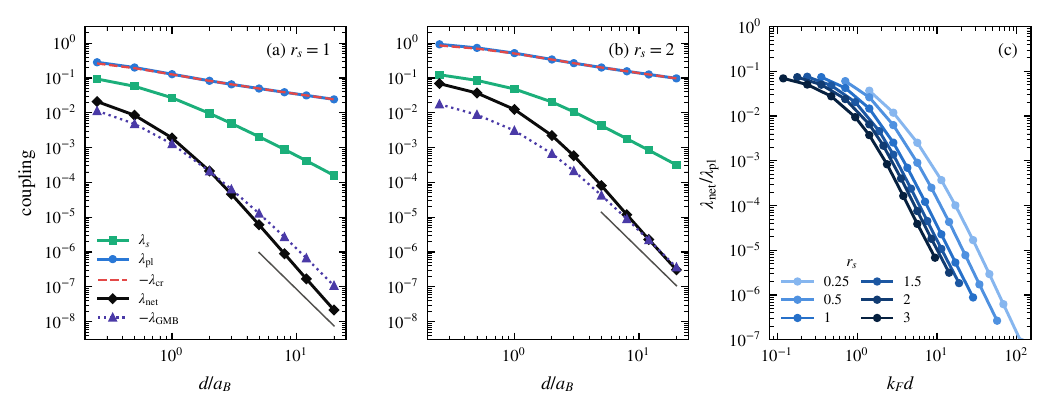}
\caption{Second-order couplings between Fermi-surface states. (a) $r_s=1$ and (b) $r_s=2$: static coupling $\lams$, plasmon-induced coupling of the ladder $\lampl$, crossed rung $-\lambda_{\rm cr}$, their sum $\lambda_{\rm net}$, and the static crossed term $-\lambda_{\rm GMB}$. The curves for $\lampl$ and $-\lambda_{\rm cr}$ are indistinguishable on this scale for $d\gtrsim\aB$. Thin line: $d^{-7/2}$. (c) Fraction of the plasmon-induced coupling that survives the crossed rung, versus $\kF d$.}
\label{fig:vertex}
\end{figure*}

We have evaluated Eqs.~\eqref{eq:lamX} and \eqref{eq:gmb} with the full RPA interaction by direct four-dimensional quadrature (Appendix~\ref{app:num}). The ladder term reproduces the independently computed $\lampl$ of Eq.~\eqref{eq:lampl} to better than $0.5\%$. The results are shown in Fig.~\ref{fig:vertex} and Table~\ref{tab:vertex}. The crossed rung is negative and nearly equal in magnitude to $\lampl$ at every separation, including $\kF d<1$, because the wedge is by construction the region in which the carriers are slow. The net coupling is at most \netfracmax\ of $\lampl$ in the range examined, at most \netfracTwo\ for $\kF d>2$ and \netfracFive\ for $\kF d>5$, and it falls off with $d$ approximately as $d^{-4}$: between $d=8\aB$ and $20\aB$ the local exponent is between \netslopeA\ and \netslopeB\ for the values of $r_s$ in Fig.~\ref{fig:vertex}. This is between the powers $7/2$ and $5$ of Eqs.~\eqref{eq:lamnetscaling} and \eqref{eq:lamnetlow}. The dependence on $r_s$ at fixed $d$ is likewise between $r_s^3$ and $r_s^4$: the local exponent is between \netrsA\ and \netrsB\ for $d\ge5\aB$ and $0.25\le r_s\le3$, and decreases toward 3 as $r_s$ decreases. Both limits are reproduced by the numerical results with no adjustable quantity [Appendix~\ref{app:remainder} and Fig.~\ref{fig:remainder}]. In additional calculations for $r_s$ down to $1/32$ the ratio $\lambda_{\rm net}/C_3r_s^3$ tends to 1 linearly in $r_s$, and its extrapolation to $r_s=0$ is within \lowExtrapErr\ of 1 at every separation examined. In calculations for $d$ up to $320\,\aB$ at $r_s=1$ the ratio $(\lambda_{\rm net}-C_3r_s^3)/\lambda_w$ decreases to \wedgeRatD\ and extrapolates to \wedgeIntercept\ at $d=\infty$. For $d\ge5\aB$ the computed values are described by $\lambda_{\rm net}\simeq C_3(d)r_s^3+C_4(d)r_s^4$, where $C_4$ depends on $r_s$ by at most \CfourSpread\ for $0.25\le r_s\le3$ and approaches $J_w\sqrt g\,(\aB/d)^{7/2}$ from above. For $\kF d\gtrsim2$ the net coupling is small compared with $\lams$, and it is partly offset by $\lambda_{\rm GMB}$.

\begin{table*}[t]
\caption{Second-order couplings between Fermi-surface states for $g=2$: static coupling $\lams$, plasmon-induced coupling of the ladder $\lampl$, crossed rung $\lambda_{\rm cr}$, their sum $\lambda_{\rm net}$, and the static crossed term $\lambda_{\rm GMB}$ [Eqs.~\eqref{eq:lamX}--\eqref{eq:gmb}]. The last two columns give the magnitude of the exponent of $T_c$ in the ladder approximation and with the crossed rung, $\lambda_{\rm eff}=\lams+\lambda_{\rm net}+\lambda_{\rm GMB}$.}
\label{tab:vertex}
\begin{ruledtabular}
\begin{tabular}{ccccccccccc}
$r_s$ & $d/\aB$ & $\kF d$ & $\lams$ & $\lampl$ & $-\lambda_{\rm cr}$ & $\lambda_{\rm net}$ & $\lambda_{\rm GMB}$ & $\lambda_{\rm net}/\lampl$ & $\dfrac{1}{\lams+\lampl}$ & $\dfrac{1}{\lambda_{\rm eff}}$\\
\hline
0.5 & 0.5 & 1.41 & 0.0305 & 0.0496 & 0.0483 & $1.3\times10^{-3}$ & $-1.8\times10^{-3}$ & $0.0264$ & 12.5 & 33.3\\
 & 2 & 5.66 & 0.0046 & 0.0206 & 0.0206 & $1.9\times10^{-5}$ & $-6.3\times10^{-5}$ & $9.1\times10^{-4}$ & 39.7 & 218.7\\
1 & 0.25 & 0.35 & 0.0938 & 0.2836 & 0.2625 & $0.0211$ & $-0.0114$ & $0.0744$ & 2.7 & 9.7\\
 & 1 & 1.41 & 0.0271 & 0.1290 & 0.1271 & $1.9\times10^{-3}$ & $-1.3\times10^{-3}$ & $0.0148$ & 6.4 & 36.0\\
 & 2 & 2.83 & 0.0097 & 0.0832 & 0.0830 & $2.1\times10^{-4}$ & $-2.1\times10^{-4}$ & $2.6\times10^{-3}$ & 10.8 & 103.5\\
 & 5 & 7.07 & 0.0021 & 0.0499 & 0.0499 & $6.1\times10^{-6}$ & $-1.3\times10^{-5}$ & $1.2\times10^{-4}$ & 19.2 & 487.0\\
 & 12 & 16.97 & 0.0004 & 0.0317 & 0.0317 & $1.7\times10^{-7}$ & $-6.7\times10^{-7}$ & $5.3\times10^{-6}$ & 31.2 & 2388.0\\
2 & 0.25 & 0.18 & 0.1239 & 0.9310 & 0.8619 & $0.0691$ & $-0.0177$ & $0.0742$ & 0.9 & 5.7\\
 & 1 & 0.71 & 0.0484 & 0.5200 & 0.5074 & $0.0125$ & $-3.1\times10^{-3}$ & $0.0241$ & 1.8 & 17.3\\
 & 2 & 1.41 & 0.0205 & 0.3417 & 0.3395 & $2.2\times10^{-3}$ & $-6.7\times10^{-4}$ & $6.5\times10^{-3}$ & 2.8 & 45.3\\
 & 5 & 3.54 & 0.0043 & 0.2007 & 0.2006 & $8.1\times10^{-5}$ & $-4.3\times10^{-5}$ & $4.0\times10^{-4}$ & 4.9 & 231.7\\
 & 12 & 8.49 & 0.0008 & 0.1268 & 0.1268 & $2.3\times10^{-6}$ & $-2.2\times10^{-6}$ & $1.8\times10^{-5}$ & 7.8 & 1185.5\\
3 & 0.5 & 0.24 & 0.0977 & 1.5006 & 1.4244 & $0.0761$ & $-0.0108$ & $0.0507$ & 0.6 & 6.1\\
 & 2 & 0.94 & 0.0288 & 0.7775 & 0.7702 & $7.4\times10^{-3}$ & $-1.1\times10^{-3}$ & $9.5\times10^{-3}$ & 1.2 & 28.6\\
 & 8 & 3.77 & 0.0028 & 0.3530 & 0.3529 & $5.8\times10^{-5}$ & $-1.8\times10^{-5}$ & $1.6\times10^{-4}$ & 2.8 & 355.2\\
\end{tabular}
\end{ruledtabular}
\end{table*}

\subsection{The adiabatic limit and the cancellation beyond second order}
\label{sec:adiabatic}

The cancellation is not special to second order. Its origin can be identified exactly in two limits, and we state separately what these limits prove and what they do not.

\emph{Static sources.} In the wedge the energies of the carriers change by much less than $\hbar\nu$ during the process, so the carriers act as static sources for the density fluctuations of the two layers, which in the RPA are Gaussian. For an electron and a hole held at in-plane separation $\mathbf r$ and coupled linearly to a Gaussian field, the pair propagator in imaginary time is known exactly. Apart from the self-energy factors of each charge,
\begin{align}
P(\tau)&\propto\exp\Bigl[-\frac1\hbar\int_0^\tau\!\!\int_0^\tau\!d\tau_1d\tau_2\,W_{12}(\mathbf r,\tau_1-\tau_2)\Bigr]\nonumber\\
&=\exp\Bigl[-\frac1\hbar\int\!\frac{d\nu}{2\pi}\,W_{12}(\mathbf r,i\nu)\,\frac{2-2\cos\nu\tau}{\nu^2}\Bigr].
\label{eq:static}
\end{align}
The exponent is linear in the interaction. At long times it is $-\tau\,W_{12}(\mathbf r,0)/\hbar$ plus terms that grow more slowly than $\tau$. The interaction energy of the two sources is the statically screened interaction. The slower terms contain the dynamical part of the interaction. Those from the plasmons tend to a constant. Those from the particle-hole continuum grow logarithmically and change the pair amplitude by a power law. Neither contributes to the energy. A ladder built from the same interaction contains at second order a term linear in $\tau$ and proportional to $\int d\nu\,W_{12}^2/\nu^2$, which is $\lampl$. The crossed diagram removes it. The static-source approximation applies to finite-mass carriers in the wedge, where $\nu\gg v_Fq$; it does not require taking $m\to\infty$. At fixed density and layer separation, the literal infinite-mass limit gives $E_B/E_F\to\infty$ and lies outside the weak-pairing regime considered here.

\emph{Carriers with linear dispersion.} The result does not require the sources to be at rest. For a fermion with linearized dispersion, $\xi_{\mathbf k+\mathbf q}=\hbar\mathbf v\cdot\mathbf q$, the propagator in an arbitrary potential that depends on space and time is known in closed form, and its average over a Gaussian field is again the exponential of a form bilinear in the sources \cite{DzyaloshinskiiLarkin1974,Kopietz1997,Metzner1998}. For an electron and a hole on opposite patches of the Fermi surface the ladder and all crossed diagrams, with any number of interlayer lines, then sum to an exponential whose exponent is of first order in $W_{12}$. In this limit the logarithmic renormalization of the interaction in the Cooper channel is cancelled by the crossed channel, and the coupling does not flow. This is a statement about the interaction and not about correlation functions: the pair susceptibility remains singular, and it is a power law of the energy, as in a one-dimensional system of counterpropagating fermions \cite{Metzner1998}. Equation~\eqref{eq:Bnet} with $\xi_2=-\xi_1$ is the second-order instance: $B_{\rm cr}=-B_{\rm lad}$ at every frequency.

\emph{What follows, and what does not.} The distinction between the irreducible vertex and the full amplitude matters here. In the linearized limit the crossed diagram, which belongs to the irreducible vertex $I$ of Eq.~\eqref{eq:BS}, does not vanish: it equals minus the iteration of the first-order rung, which is reducible. What vanishes, at every order and every frequency, is their sum, that is, the contribution of second and higher order to the effective interaction between low-energy pairs. A pairing instability at a finite temperature therefore requires what the linearization omits, the curvature of the dispersion, and every term of second and higher order in $W_{12}$ in that effective interaction is proportional to it. At second order we have evaluated these terms without approximation: they are $\lambda_{\rm net}$ and $\lambda_{\rm GMB}$. We have not evaluated the curvature terms of third and higher order. The exact cancellations for static sources and for carriers with linearized dispersion, both assuming Gaussian density fluctuations, suggest that the higher-order curvature terms are also suppressed. We proceed on this expectation, which remains unverified beyond second order for carriers with quadratic dispersion. Three points limit what the solvable limits establish. (i) They concern the pair propagator for prescribed kinematics, not the irreducible vertex of the full problem with curvature. (ii) They assume Gaussian density fluctuations, which is the RPA. In an interacting Fermi liquid the higher cumulants of the density are suppressed at $q\ll\kF$ by the cancellation of fermion loops \cite{Metzner1998}, which is the regime relevant at large $d$, but we have not shown this for arbitrary $r_s$. (iii) The criterion for slow carriers is local in $q$. The plasmon is gapless, and what is required is $\nu\gtrsim\omega_+(q)\gg\vF q$ at each wave vector. Since $\omega_+(q)/\vF q$ is of order $\sqrt{\cG}$ at $q=1/d$ and increases as $q\to0$, the criterion is met uniformly at large $d$.

We refer to the statement that carriers slow compared with the modes of the medium interact through its static response as the adiabatic theorem. It is the opposite of Migdal's limit. Migdal's theorem states that the vertex corrections are negligible when the boson is slow. Here the boson is fast, the vertex corrections are of order unity, and in the solvable limits their sum is known: the dynamical part of the interaction drops out of the pairing. For carriers with curvature the cancellation is incomplete at second order by the amounts given in Eqs.~\eqref{eq:lamnetscaling} and \eqref{eq:lamnetlow}. The intralayer vertex and self-energy corrections of Sec.~\ref{sec:ward}, which cancel in the same region by the Ward identity, are part of the same structure.

\subsection{Consequences}

With the crossed rung included the effective coupling is $\lams+\lambda_{\rm net}+\lambda_{\rm GMB}$, which for $\kF d\gtrsim2$ differs little from $\lams$ (Table~\ref{tab:vertex}). To the accuracy of the exponent the transition temperature is then that of the statically screened theory, and the ladder overestimates it by a factor $\exp[1/\lams-1/(\lams+\lampl)]$. At $r_s=1$ and $d=2a_B$, this factor is $\sim10^{40}$: the ladder estimate exceeds the statically screened estimate by about 40 orders of magnitude, to exponential accuracy.

Two remarks are in order. First, the identity of Eq.~\eqref{eq:Bnet} applies without change to the particle-particle channel of a single layer, where the plasmon-induced attraction of the ladder arises from the same second-order process through the region $\nu\gg\vF q$. This suggests that a large part of that attraction is removed by the crossed diagram there too. At $\kF d=0.18$, where the interlayer interaction is close to that of a single layer, we find that the crossed rung removes $93\%$ of $\lampl$. This is in line with the conclusion of Ref.~\cite{DasSarma2026} that the Eliashberg treatment of the plasmon mechanism is not reliable. It does not show that a single layer has no pairing instability. In the single layer the momentum transfers are not confined to small angles, we have not computed the remainder for that case, and other routes to pairing are not addressed by this argument. Second, the unscreened HF theory is not affected by this argument in the same way, because it has no small parameter at all: the adiabatic limit selects the statically screened interaction, not the bare one.

\section{What is controlled at zero field}
\label{sec:control}

This section states the conclusions; the derivations, diagram by diagram, are in Appendix~\ref{app:long}. It is natural to expect the theory to be exact in two limits: at small $r_s$ for every $d$, because the interaction is weak, and at large $d$ for every $r_s$, by an argument of the Migdal type, because the exciton binding energy is then small compared with the Fermi energy. Both limits are controlled, the first as a rigorous statement about the leading exponent and the second under an assumption that we state. In neither is the result that of the dynamically screened ladder.

\subsection{Small $r_s$}

At fixed $d/\aB>0$ and $r_s\to0$ one has $\kF d\to\infty$, so Eq.~\eqref{eq:lams} applies: $\lams=\sqrt g\,r_s\aB\,\cS(g d/\aB)/(2\pi g\,d)$ is of first order in $r_s$. All other terms are of higher order, and
\begin{equation}
\lim_{r_s\to0}\ \lams\,\ln\frac{\EF}{k_BT_c}=1\qquad(d/\aB\ \text{fixed}).
\label{eq:smallrs}
\end{equation}
Four qualifications apply.

First, only the leading exponent is exact, and the corrections to $\ln T_c$ are not bounded. They grow as $\ln(1/r_s)$, which is small compared with $1/\lams\propto1/r_s$ but is not a constant. Two sources are identified in Appendix~\ref{app:long}. The static crossed rung replaces the energy scale $\hbar\vF/d$ of the prefactor by $\hbar^2/md^2$, which is $2\EF/(\kF d)^2\propto r_s^2\EF$. And the intralayer terms are of relative order $\alpha\ln(1/\alpha)$, as in the effective mass of the two-dimensional electron gas at high density \cite{Galitski2004}, which changes $\ln T_c$ by an amount of order $\ln(1/\alpha)$. The prefactor of $T_c$ is therefore a power of $r_s$ times a function of $d/\aB$, and its determination requires the other terms of second order: $\lambda_{\rm GMB}$, the intralayer vertex and self-energy corrections, and the corrections to the polarization that the Ward identity requires together with them. The net dynamical interlayer term is not among them. It changes $\ln T_c$ by $\lambda_{\rm net}/\lams^2\to R(d)\,r_s$, which vanishes with $r_s$ [Eq.~\eqref{eq:Rd}]; $R=\RlowOne$ at $d=\aB$ and $\RlowFive$ at $d=5\aB$. The situation is that of the weakly interacting Fermi gas, where the induced interaction changes the prefactor of the BCS $T_c$ \cite{Gorkov1961}.

Second, the ladder reproduces the leading weak-coupling exponent at fixed $d/a_B$ because $\lambda_{\mathrm{pl}}/\lambda_s\propto r_s\to0$. However, this does not ensure an accurate $T_c$. Even when $\lambda_{\mathrm{pl}}\ll\lambda_s$, the ladder enhances the transition temperature relative to the statically screened estimate by the factor $\exp[1/\lams-1/(\lams+\lampl)]\simeq\exp(\lambda_{\mathrm{pl}}/\lambda_s^2).$
Since $\lambda_s\propto r_s$ and $\lambda_{\mathrm{pl}}\propto r_s^2$, this factor approaches a constant as $r_s\to0$, but that constant can be enormous. For $g=2$, the full expression gives an exponent of about 980 at $d=2a_B$; its large-separation asymptote is $22.6(d/a_B)^{7/2}$. Moreover, reaching $\lambda_{\mathrm{pl}}\ll\lambda_s$ requires $r_s\ll0.64(a_B/d)^{3/2}$ at large separation. The crossed rung is needed to remove this spurious dynamical enhancement. 
At $r_s=1$ and $d=2\aB$, where $\lampl$ exceeds $\lams$ and this expansion does not apply, the difference of the exponents is $1/\lams-1/(\lams+\lampl)=\ladExcessOneTwo$. For any fixed $r_s$, however small, the second-order term of the ladder exceeds the first-order one beyond $d_\times$ of Eq.~\eqref{eq:crossover}, and along any path on which $d$ grows faster than $r_s^{-2/3}$ the ladder has the wrong exponent. A second-order term that overtakes the first-order term is the sign that a contribution of the same order has been left out, and Sec.~\ref{sec:vertex} identifies it. With the crossed rung, $\lambda_{\rm net}/\lams$ is the sum of a term proportional to $r_s^3(\aB/d)^{3/2}$ and one proportional to $r_s^2(\aB/d)^{3}$ at large $d$, by Eqs.~\eqref{eq:lamnetscaling} and \eqref{eq:lamnetlow}, and it vanishes whether $r_s\to0$ or $d\to\infty$. The quantity that the ladder gives as $\lampl/\lams^2$, a constant at small $r_s$, becomes $\lambda_{\rm net}/\lams^2=R(d)\,r_s$. The ladder approximation alone reproduces the leading weak-coupling exponent in Eq. (43) at fixed $d/a_B$. Including the crossed rung is necessary to make this conclusion uniform in increasing separation, including paths on which $d/a_B$ grows faster than $r_s^{-2/3}$. 

Third, Eq.~\eqref{eq:lams} requires $\kF d\gg1$, and the form $\lams\propto r_s$ is not uniform as $d\to0$. For $\kF d\lesssim1$ the coupling is the full Fermi-surface average of the static interaction, which at $d=0$ is $(\alpha/\pi)\ln(2/g\alpha)$ and so behaves as $r_s\ln(1/r_s)$. Equation~\eqref{eq:smallrs} is expected to hold there with this average for $\lams$, but along paths on which $\kF d$ does not become large the small-angle expressions of this paper do not apply.

Finally, the controlled weak-coupling regime yields an unobservably small pairing scale. At $r_s=0.25$ and $d=a_B$, $1/\lambda_s=163$, giving a leading exponential factor $e^{-1/\lambda_s}\sim10^{-71}$ in $T_c$. This limit therefore serves primarily as a controlled benchmark for assessing the pairing theory.

\subsection{Large $d$}

Consider now $d\to\infty$ at fixed $r_s$, in the liquid phase. Then $\kF d\gg1$, $\cG\gg1$, and $E_B\simeq E_d\ll\EF$.

\emph{It is not Migdal's theorem.} By Eq.~\eqref{eq:fast} the plasmons are fast, and Eliashberg theory, which keeps the ladder and drops the vertex corrections, is not justified. It retains $\lampl$, which the full theory does not contain. Dressing the propagators with $Z$ while keeping the vertices bare (Sec.~\ref{sec:ward}) reduces $\lampl$ but does not remove it.

\emph{The control comes from the large energy denominators of the plasmons.} All interlayer momentum transfers are $q\lesssim1/d\ll\kF$, the carriers are slow compared with both plasmons, and by the argument of Sec.~\ref{sec:vertex} the pairing interaction is the static one. Because $q\ll\kF$ it can be written in terms of Fermi-liquid quantities. The statically screened interlayer interaction is $-U_q/\{[1+(V_q+U_q)\chi][1+(V_q-U_q)\chi]\}$ with $\chi=\partial n/\partial\mu$ the exact proper compressibility of a layer, and the quasiparticles couple to it through the static density vertex, which by the Ward identity is $\chi/N^*$ in units of the inverse quasiparticle weight, with $N^*=gm^*/2\pi\hbar^2$ and $m^*$ being the quasiparticle effective mass of an isolated layer \cite{Nozieres}. For $(V_q\pm U_q)|\chi|\gg1$, which holds for $d\gg\aB$, the compressibility cancels,
\begin{equation}
W_{12}^{\rm qp}(q)=-\frac{1}{N^{*2}}\,\frac{U_q}{V_q^2-U_q^2}=-\frac{1}{N^{*2}}\,\frac{\epsilon\,q}{4\pi e^2\sinh qd},
\end{equation}
which is the interaction of two quasiparticles through the geometric capacitance of the double layer. It does not depend on the sign of the compressibility. Averaging over the Fermi surface with the quasiparticle density of states gives $\lambda=(m/m^*)\lams$ with $\lams$ from Eq.~\eqref{eq:lamsasym}, and
\begin{equation}
\ln T_c=-\frac{m^*}{m}\,\frac{16\,g^{3/2}}{\pi\,r_s}\Bigl(\frac{d}{\aB}\Bigr)^{2}\bigl[1+o(1)\bigr].
\label{eq:exactlarged}
\end{equation}

\begin{figure*}[t]
\includegraphics[width=0.86\textwidth]{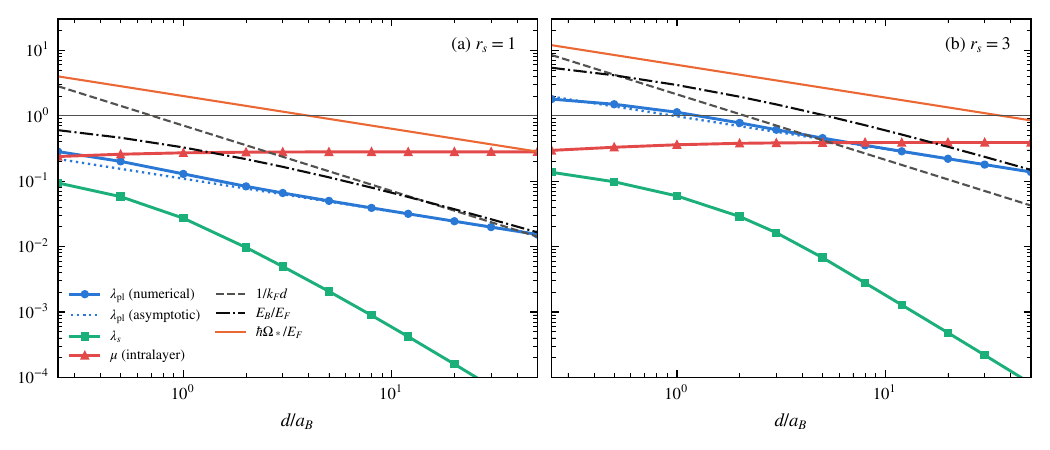}
\caption{Parameters of the pairing problem versus separation for (a) $r_s=1$ and (b) $r_s=3$: static coupling $\lams$, plasmon-induced coupling of the ladder $\lampl$ from Eq.~\eqref{eq:lampl} (solid) and its asymptotic form, Eq.~\eqref{eq:lamplasym} (dotted), the parameters $1/\kF d$, $E_B/\EF$, and $\hbar\Ost/\EF$, and the Fermi-surface average $\mu$ of the statically screened intralayer repulsion, which measures the intralayer vertex and self-energy corrections and does not decrease with $d$.}
\label{fig:params}
\end{figure*}

The corrections that have been computed, to $\cS(\cG)$, to the Fermi-surface average, and from the terms of second order, are of relative order $\aB/d$ and $1/\kF d$ or smaller. For the terms that have not been computed the assumption stated next gives only that their relative contribution vanishes as $d\to\infty$, without a rate. Equation~\eqref{eq:exactlarged} rests on two assumptions. The first is that each layer is a Fermi liquid. The second is that the cancellation of the dynamical terms, which is exact in the linearized limit at every order and which we have verified with curvature at second order, holds for the curvature terms of higher order as well. Precisely, the assumption is that the sum $\delta\lambda$ of all contributions to the pair coupling that are not contained in the static quasiparticle interaction satisfies $\delta\lambda/[(m/m^*)\lams]\to0$ as $d\to\infty$. A contribution proportional to $d^{-2}$ would not change the form of the result but would change its coefficient, and is excluded by this assumption and not by anything we have proven. At second order the assumption can be checked, because the remainder is known: for $d\to\infty$ the net dynamical term is $\lambda_w$ of Eq.~\eqref{eq:lamnetscaling}, which changes $\ln T_c$ by $\lambda_w/\lams^2=\JwT\,r_s^2(d/\aB)^{1/2}$ for $g=2$ [Eq.~\eqref{eq:dlnTcw}]. This increases with $d$, but it is smaller than the leading term by the factor $\lambda_w/\lams\propto r_s^3(\aB/d)^{3/2}$. We consider the second assumption very likely to be correct at higher order as well, but we have not proven it, and Eq.~\eqref{eq:exactlarged} should be read with that qualification. If it holds, the large-$d$ limit is controlled at any $r_s$, as anticipated. The conditions are $\kF d\gg1$ and $d\gg\aB$. Since $E_d/\EF=gr_s^2\aB/2d$, the condition $E_B\ll\EF$ implies both of them when $r_s\gtrsim1$, and in this sense the ratio of the exciton binding energy to the Fermi energy is the control parameter. But the mechanism is the adiabatic one and not Migdal's, and the theory it selects is static screening and not Eliashberg theory. The transition temperature in the controlled regime decreases as a Gaussian in $d$, Eq.~\eqref{eq:exactlarged}, and not as the stretched exponential of Eq.~\eqref{eq:stretched}. Figure~\ref{fig:params} shows the parameters.

\subsection{Between the limits}

For $r_s\gtrsim1$ and $d\lesssim r_s\aB$, which includes the experimental systems, $\kF d\lesssim1$ and no parameter is small. The crossed rung still removes most of the plasmon-induced coupling there (Table~\ref{tab:vertex}), but the remainder is no longer negligible, the couplings are not small, the intralayer vertex corrections are of order $\mu\approx\muA$ to $\muB$, and $E_B$ is comparable to $\EF$ or larger, so that the problem approaches that of bound pairs. The numbers in Fig.~\ref{fig:tcd} are then the output of a definite approximation, the RPA ladder, with no estimate of its error. The analysis of Sec.~\ref{sec:vertex} indicates the sign of the error: both corrections we have computed lower the coupling. What survives is qualitative. Within the RPA the interaction on the imaginary axis is attractive throughout, so the ladder has an instability at every $r_s$ and $d$; this does not fix the sign of the complete irreducible vertex at intermediate coupling. Within the ladder the unscreened and the statically screened results bracket the dynamically screened one. No such bound holds for the full problem: with the two corrections we have computed the effective coupling is already below $\lams$ at some points of Table~\ref{tab:vertex}, because $\lambda_{\rm GMB}$ is negative. And the true result is expected to lie much closer to the statically screened theory than the ladder suggests.

Table~\ref{tab:compare} summarizes the comparison with phonons and with the single-layer plasmon problem.

\begin{table*}[t]
\caption{Comparison of pairing by phonons, by the plasmon of a single electron layer \cite{DasSarma2026}, and by the screened interlayer attraction in the electron-hole bilayer at large separation.}
\label{tab:compare}
\begin{ruledtabular}
\begin{tabular}{llll}
& Phonons & Single-layer plasmon & Electron-hole bilayer, $d\gg\aB$, $\kF d\gg1$\\
\hline
Bare interaction in pairing channel & attractive (retarded) & repulsive & attractive\\
Boson energy$/\EF$ & $\omega_D/\EF\sim10^{-2}$ & $\hbar\omega_p/\EF\sim1$ & $\hbar\Ost/\EF=(2gE_d/\EF)^{1/2}$\\
Boson velocity$/\vF$ & $\ll1$ & $\gg1$ & $\sqrt{gd/\aB}\gg1$\\
Kernel at Matsubara frequencies & attractive below $\omega_D$ & repulsive, weaker at low $\nu$ & attractive at all $q$, $\nu$\\
Ladder (Eliashberg) result & $T_c\sim\omega_De^{-1/\lambda}$ & depends on cutoff $\omega_c$ & $\ln T_c\propto-\sqrt d$, no cutoff needed\\
Vertex corrections & negligible (Migdal) & of order unity & of order unity; cancel $\lampl$ (adiabatic)\\
Asymptotic result & Eliashberg theory & none & static screening, $\ln T_c\propto-(m^*/m)\,d^2$ \footnotemark[1]\\
\end{tabular}
\end{ruledtabular}
\footnotetext[1]{Conditional on the assumption stated after Eq.~\eqref{eq:exactlarged}.}
\end{table*}

\section{Discussion of the zero-field theory}
\label{sec:discussion}

\emph{Relation to earlier work.} The existence of the instability at all $d$ within the RPA, and the ordering of the three theories, agree with Ref.~\cite{Sreejith2024}. That work describes the decay of $T_c$ with $d$ as exponential over the range of separations it covers. Our Eq.~\eqref{eq:stretched} gives the asymptotic form of the ladder result as a stretched exponential, $\exp(-{\rm const}\sqrt d)$, which cannot be distinguished from an exponential over a short range. Reference~\cite{Sreejith2024} includes the normal self-energy without the vertex correction, which reduces the plasmon-induced coupling relative to the ladder with bare propagators (Sec.~\ref{sec:ward}). A quantitative comparison of the two calculations would require the frequency dependence of the self-energy and must allow for the different spin degeneracy, and we do not attempt it. The analyses of graphene double layers \cite{Sodemann2012,Lozovik2012} reached the same qualitative conclusion about dynamic against static screening. All of these calculations, like our Secs.~\ref{sec:analytic} and \ref{sec:numerics}, solve a gap equation with one screened interaction line, and the large enhancement over static screening that they find is the term $\lampl$. Section~\ref{sec:vertex} shows that at second order this term belongs to the ladder approximation and not to the problem: the crossed rung, which is of the same order, cancels it. The earlier conclusion that static screening suppresses the transition temperature to very small values at large separation \cite{Kharitonov2008} is, by this analysis, the correct one in the regime where the pairing is weak. Reference~\cite{Tu2024} compared the ground-state energies of the exciton gas and of the electron-hole plasma with the electron-hole attraction taken as unscreened, statically screened, or dynamically screened in a plasmon-pole form, and concluded that the phase diagram is very sensitive to this choice and, in the dynamical case, to the energy at which the screening is evaluated. The present analysis removes that ambiguity at the two ends. On the weak-pairing side the relevant energy is far below the plasmon energies and the screening is static. On the Bose side the pairs are neutral and the attraction is unscreened (Appendix~\ref{app:long}). This is consistent with the finding of Ref.~\cite{Tu2024} that the exciton gas has the lower energy at very low density in every approximation. The onset density for superfluidity found in mean-field theory when the screening is computed self-consistently in the paired state \cite{Perali2013}, and in quantum Monte Carlo calculations \cite{LopezRios2018}, concerns the strongly paired side and is outside the scope of the linearized theory.

\emph{Magnitudes.} The temperatures in the regime where the theory is controlled are extremely small. In the ladder approximation $T_c\approx10^{-9}\EF$ at $r_s=1$, $d=5\aB$ (Table~\ref{tab:tc}), and with the crossed rung the exponent there is $-1/\lams\approx-490$. The examples we have computed suggest that observable transition temperatures require $E_B$ comparable to $\EF$ or larger, where the weak-pairing theory does not apply and the relevant description approaches that of a Bose gas of dipoles; the crossover itself has not been calculated here. The interlayer coherence observed in the experiments is in that range of parameters.

\emph{Limitations.} The results are for the ideal model. (a) Any mismatch of the two Fermi contours, from a density imbalance or from different anisotropies of the two bands, cuts off the logarithm and removes the instability once it exceeds the gap. A difference of the isotropic masses alone does not: at equal densities the two Fermi circles coincide whatever the masses, and the logarithm survives with $\vF$ replaced by the mean of the two Fermi velocities. (b) Disorder that is not identical in the two layers breaks pairs \cite{Lozovik1976}. Both effects eliminate the exponentially small gaps of the controlled regime. (c) We assumed a Fermi liquid in each layer. At large $r_s$ the layers, or the excitons, may crystallize \cite{DePalo2002,Astrakharchik2007,Joglekar2006}, and the argument does not apply. (d) The sign property, Eq.~\eqref{eq:sign}, is a property of the RPA. It relies on $\Pi(q,i\nu)\le0$, which the exact proper polarization need not satisfy at $r_s\gtrsim2$, where the compressibility of a layer is negative. Equation~\eqref{eq:exactlarged} does not depend on it. (e) The crossed rung was evaluated between states at the Fermi surface and at zero frequency, which is what determines the exponent of a weak-coupling $T_c$. We have not solved a gap equation with the crossed rung in the kernel, and for $\kF d\lesssim1$ and $r_s\gtrsim2$, where the couplings are not small, the two-step estimate of Eq.~\eqref{eq:lamnet} is only indicative. (f) We took equal masses. The cancellation of Eq.~\eqref{eq:Bnet} does not depend on this: to first order in $\mathbf q$ it requires only that the hole energy be odd under $\mathbf q\to-\mathbf q$ at fixed $\mathbf k$, which holds for any dispersion.

\emph{Open problems.} The dynamical terms of third order with curvature have not been evaluated; their calculation would test the assumption on which Eq.~\eqref{eq:exactlarged} rests. The prefactor of $T_c$ in both limits requires the remaining terms of second order, the static crossed rung beyond its logarithm and the intralayer terms; its energy scale is $\hbar^2/md^2$ (Appendix~\ref{app:long}). The coefficient $C_4(d)$ of the next term of $\lambda_{\rm net}$ is known only numerically (Appendix~\ref{app:remainder}). The intermediate regime requires a method that does not rely on a small parameter. A solution of the gap equation with the crossed rung included in the kernel at all frequencies would give the corrected $T_c$ at $\kF d\lesssim1$.

\section{High magnetic field: the $\nu_T=1$ quantum Hall bilayer}
\label{sec:qh}

We now turn to the same problem in a strong magnetic field. By the particle-hole transformation of Sec.~\ref{sec:framework}, and apart from one-body terms, the quantum Hall bilayer at total filling factor one is the electron-hole bilayer of the preceding sections with the kinetic energy removed and the interaction projected onto the lowest Landau level, which brings in its form factor and the guiding-center algebra; by Eq.~\eqref{eq:dictionary} $d/\ell$ takes the place of $\kF d$. The controlled limits of Sec.~\ref{sec:control} are then not available. In exchange the mean-field theory, which at zero field requires the numerical solution of a gap equation, becomes analytic.

The $\nu_T=1$ bilayer quantum Hall state is the simplest realization of spontaneous interlayer coherence in a two-component electron system.  Two half-filled layers form the Halperin $(111)$ state \cite{Halperin1983}, an incompressible integer quantum Hall state with a neutral interlayer phase mode.  In the pseudospin representation the layer index is a spin-$1/2$ label and the ground state is an easy-plane quantum Hall ferromagnet \cite{Fertig1989,MacDonaldPlatzmanBoebinger1990,Yang1994,Moon1995}.  In the electron-hole representation it is an exciton superfluid.  The experimental development includes the original transport observation of the bilayer quantum Hall state at $\nu_T=1$ and of its disappearance at large layer separation \cite{Murphy1994}, the resonantly enhanced zero-bias tunneling peak \cite{Spielman2000}, the measured linearly dispersing collective mode \cite{Spielman2001}, quantized Hall drag \cite{Kellogg2002,Kellogg2003}, counterflow transport \cite{Kellogg2004,Tutuc2004}, activated transport in the separate layers \cite{Wiersma2004}, and exciton transport and perfect Coulomb drag \cite{Finck2011,Nandi2012}.  These experiments establish the charged and neutral responses of the coherent bilayer; broad accounts are given in Refs.~\cite{Eisenstein2004,Eisenstein2014}.

Everything derived in this section and in Appendices~\ref{app:qh-stiffness}--\ref{app:qh-ed} is a Hartree-Fock statement, apart from the exact result at $d=0$ and the finite-sphere data of Appendix~\ref{app:qh-ed}.  Nothing is known in a controlled way beyond HF in this problem.  Our own expectations for the real system are collected, and labeled as such, in Sec.~\ref{qh:sec:beyond}.

\subsection{Hamiltonian and uniform Hartree-Fock saddle}
\label{sec:qh-model}

The LLL-projected bilayer Hamiltonian is
\begin{equation}
H=\frac{1}{2A}\sum_{\bm q}\sum_{\alpha\beta}V_{\alpha\beta}(q):\rho_\alpha(-\bm q)\rho_\beta(\bm q):,
\label{qh:eq:H}
\end{equation}
where $\alpha,\beta=T,B$ and
\begin{equation}
V_{TT}(q)=\frac{2\pi e^2}{\epsilon q},\qquad V_{TB}(q)=\frac{2\pi e^2}{\epsilon q}e^{-qd}.
\label{qh:eq:V}
\end{equation}
These are $V_q$ and $U_q$ of Eq.~\eqref{eq:VU}; the layers are labeled $T$ and $B$, and the interlayer interaction is repulsive. The LLL-projected density carries the Gaussian form factor $e^{-q^2\ell^2/4}$, where $\ell$ is the magnetic length. Thus each interaction term in Eq.~(\ref{qh:eq:H}) contains the Gaussian factor $e^{-q^2\ell^2/2}$, which appears explicitly in the exchange integral. Since there is no interlayer tunneling, the relative charge is therefore conserved and the Hamiltonian has an exact $U(1)$ symmetry.

At total filling one, the coherent state fills one common pseudospinor at every guiding center,
\begin{equation}
|\chi_\phi\rangle=\frac{1}{\sqrt2}\left(|T\rangle+e^{i\phi}|B\rangle\right).
\label{qh:eq:spinor}
\end{equation}
Its density matrix is
\begin{equation}
\rho_{m\alpha,m'\beta}=\delta_{mm'}\frac12\begin{pmatrix}1&e^{-i\phi}\\e^{i\phi}&1\end{pmatrix}_{\alpha\beta},\qquad \rho^2=\rho.
\label{qh:eq:rho}
\end{equation}
The direct Hartree term of the balanced uniform state cancels the neutralizing backgrounds.  The Fock contraction generates the interlayer exchange field.  This establishes a self-consistent uniform Hartree-Fock saddle, but it does \emph{not} establish exactness of finite-$d$ Hartree-Fock. At $T=0$, the coherent spinor of Eq.~(48) forms the filled $+$ band, while its orthogonal partner forms the empty $-$ band. For $\phi=0$, these are the symmetric and antisymmetric HF levels, respectively.  Intraband $+\!\rightarrow\!+$ bubbles vanish, whereas interband $+\!\rightarrow\!-$ particle-hole excitations are allowed.  The latter generate polarization, vertex, and multiparticle correlation corrections beyond Hartree-Fock.

\begin{figure*}[t]
\includegraphics[width=\textwidth]{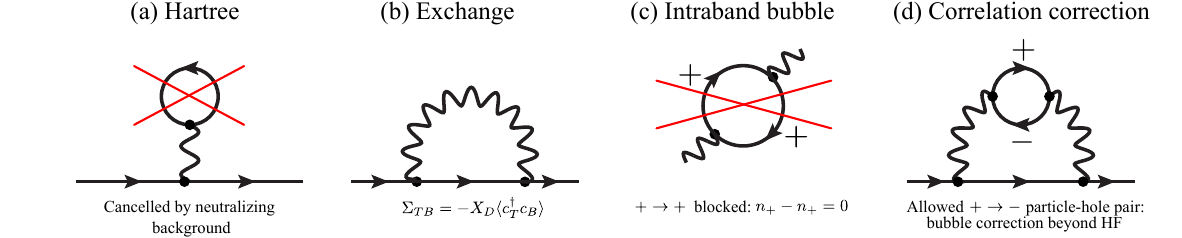}
\caption{Diagrammatics of the uniform LLL treatment.  (a) The direct uniform interaction is cancelled by the neutralizing backgrounds.  (b) Exchange is retained in the HF self-energy.  (c) An intraband bubble in the filled $+$ band vanishes.  (d) The physically relevant $+\!\rightarrow\!-$ pseudospin particle-hole pair is allowed and produces correlation omitted by HF.  The external momentum may be zero while the internal particle-hole momentum is finite.}
\label{qh:fig:diagrams}
\end{figure*}

Figure~\ref{qh:fig:diagrams} makes the distinction explicit.  The filled $+$ band removes only same-band particle-hole insertions.  It does not remove virtual transitions from the filled $+$ band to the empty $-$ band.  The LLL projection eliminates Landau-level mixing; it does not eliminate guiding-center or pseudospin correlation physics.  The uniform HF calculation retains the first-order exchange field and omits the higher-order diagrams, e.g., Fig.~\ref{qh:fig:diagrams}(d).  At $d=0$, $V_{TT}=V_{TB}$ and the Hamiltonian is SU(2) pseudospin invariant; the fully polarized $\nu_T=1$ ferromagnet is then an exact LLL ground state.  This special SU(2) statement does not extend to finite $d$, where $V_{TT}(q)-V_{TB}(q)\ne0$.

The finite-$d$ distinction can be displayed at the Hamiltonian level.  Writing $\rho=\rho_T+\rho_B$ and $S^z=(\rho_T-\rho_B)/2$, one obtains
\begin{equation}
\begin{aligned}
H=\frac{1}{2A}\sum_{\bm q}\bigl[&V_+(q):\rho(-\bm q)\rho(\bm q):\\
&+4V_-(q):S^z(-\bm q)S^z(\bm q):\bigr],
\end{aligned}
\label{qh:eq:splitH}
\end{equation}
where $V_\pm(q)=[V_{TT}(q)\pm V_{TB}(q)]/2$.  The $V_-$ term is identically zero only at $d=0$.  For $d>0$, it creates and scatters pseudospin particle-hole pairs about an in-plane polarized Slater determinant. Figure~\ref{qh:fig:diagrams}(d) illustrates one correlation correction generated by the $V_-$ term. It also makes clear why the total filling $\nu_T=1$ does not provide a finite-$d$ Pauli-blocking theorem: one pseudospin orbital is full, while its orthogonal pseudospin orbital remains available.

\subsection{Precise Hartree-Fock theorem}
\label{sec:qh-theorem}

The theorem established below is deliberately narrow and exact within its stated approximation.  For the spatially uniform HF variational manifold at $T=0$,
\begin{equation}
\mathcal M_{\rm uHF}=\left\{\prod_m\left[u c^\dagger_{mT}+v c^\dagger_{mB}\right]|0\rangle:\ |u|^2+|v|^2=1\right\},
\end{equation}
the interlayer Fock matrix element $X_D(d)$ [defined in Eq.~(\ref{qh:eq:XD})] is strictly positive for every finite $d$.  The balanced uniform stationary state therefore has nonzero transverse pseudospin.  Equivalently,
\begin{equation}
X_D(d)>0\quad (0\leq d<\infty)\ \Longrightarrow\ \langle S^+\rangle_{\rm uHF}\ne0.
\label{qh:eq:HFtheorem}
\end{equation}
In terms of the self-consistency condition, Eq.~\eqref{qh:eq:SC} below, the content of the statement is that a solution with nonzero polarization exists below a temperature $X_D/4k_B$ that is positive at every finite $d$, and that it has the positive HF gap $X_D$.  Equation~\eqref{qh:eq:HFtheorem} excludes a finite-$d$ transition from the uniform HF coherent saddle to a uniform HF incoherent saddle. It does not rule out a transition driven by correlations generated by the $V_-$ term in Eq.~\eqref{qh:eq:splitH}, since uniform HF neglects these effects. This statement is restricted to spatially uniform HF. It does not exclude a density-wave instability within unrestricted HF or a transition driven by correlations beyond HF. 

\subsection{Analytic exchange field, gap, and transition temperature}
\label{sec:qh-exchange}

Let $P=2|\langle c_{mT}^\dagger c_{mB}\rangle|$.  The two uniform HF levels are
\begin{equation}
\epsilon_S=-\frac{X_DP}{2},\qquad\epsilon_{AS}=+\frac{X_DP}{2},
\label{qh:eq:levels}
\end{equation}
where
\begin{align}
X_D(d)&=\int\frac{d^2q}{(2\pi)^2}V_{TB}(q)e^{-q^2\ell^2/2}\nonumber\\
&=\Ec\int_0^\infty dx\,e^{-x^2/2-(d/\ell)x}\nonumber\\
&=\Ec\sqrt{\frac\pi2}e^{d^2/(2\ell^2)}\operatorname{erfc}\!\left(\frac{d}{\sqrt2\ell}\right),
\label{qh:eq:XD}
\end{align}
with $\Ec=e^2/(\epsilon\ell)$. We denote the HF quasiparticle splitting between the antisymmetric and symmetric levels by $\Delta(T)\equiv\epsilon_{AS}-\epsilon_{S}=X_D(d)P(T)$. Since the integrand is positive, $X_D(d)>0$ for all finite $d$.  This is the $q=0$ interlayer exchange kernel of the earlier HF literature \cite{Fertig1989,Moon1995,JoglekarMacDonald2001}; the closed form appears as the interlayer exchange constant in Ref.~\cite{FilKravchenko2009}.

\begin{figure}[t]
\includegraphics[width=\linewidth]{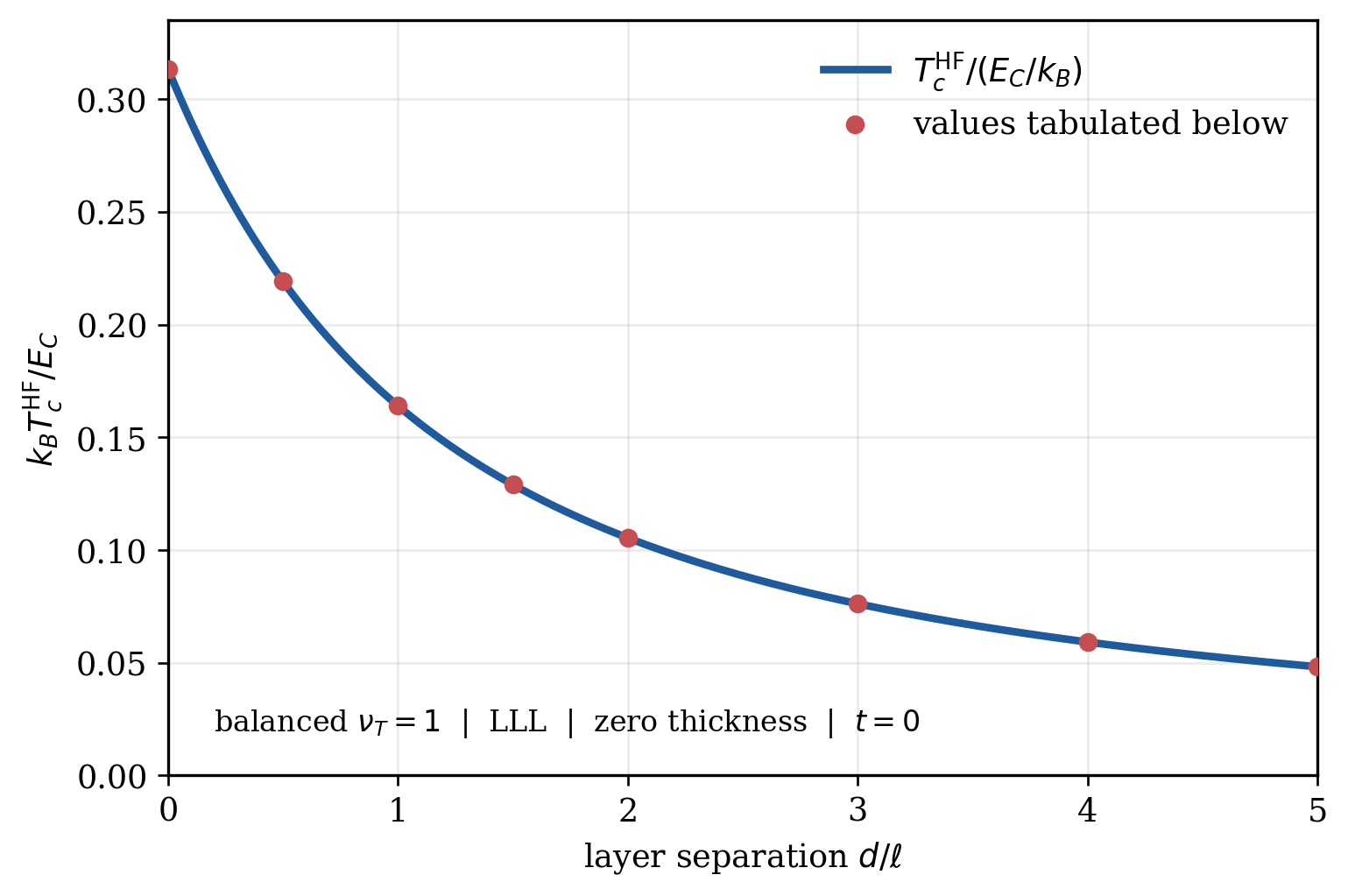}
\caption{Analytic HF transition temperature $k_BT_c^{\rm HF}/\Ec=X_D/(4\Ec)$. Dots: values of Eq.~\eqref{qh:eq:Tcexplicit}; a selection is listed in Table~\ref{qh:tab:scales}. In the figure $t=0$ denotes zero tunneling.}
\label{qh:fig:Tc}
\end{figure}

At filling one, the occupation difference of the two levels gives
\begin{equation}
P=\tanh\!\left(\frac{X_DP}{4k_BT}\right).
\label{qh:eq:SC}
\end{equation}
The uniform HF free energy per flux quantum is
\begin{equation}
{\cal F}(P)=\frac{X_D}{4}P^2-2k_BT\ln\left[2\cosh\left(\frac{X_DP}{4k_BT}\right)\right].
\label{qh:eq:F}
\end{equation}
The uniform HF solution follows immediately:
\begin{equation}
\begin{gathered}
P(T=0)=1,\qquad \Delta(T=0)=X_D,\\
k_BT_c^{\rm HF}=\frac{X_D}{4}.
\end{gathered}
\label{qh:eq:main}
\end{equation}
There is no numerical self-consistency problem and no fitting parameter.  The exchange gap and thermal scale are fixed by the same LLL integral.  At $d=0$ the thermal scale reduces to the HF transition temperature of the $\nu=1$ spin ferromagnet \cite{KasnerMacDonald1996}.  As there, $T_c^{\rm HF}$ is a mean-field amplitude scale and not the temperature of the physical transition (Appendix~\ref{qh:sec:finiteT}); it is shown in Fig.~\ref{qh:fig:Tc}.

\begin{figure}[t]
\includegraphics[width=\linewidth]{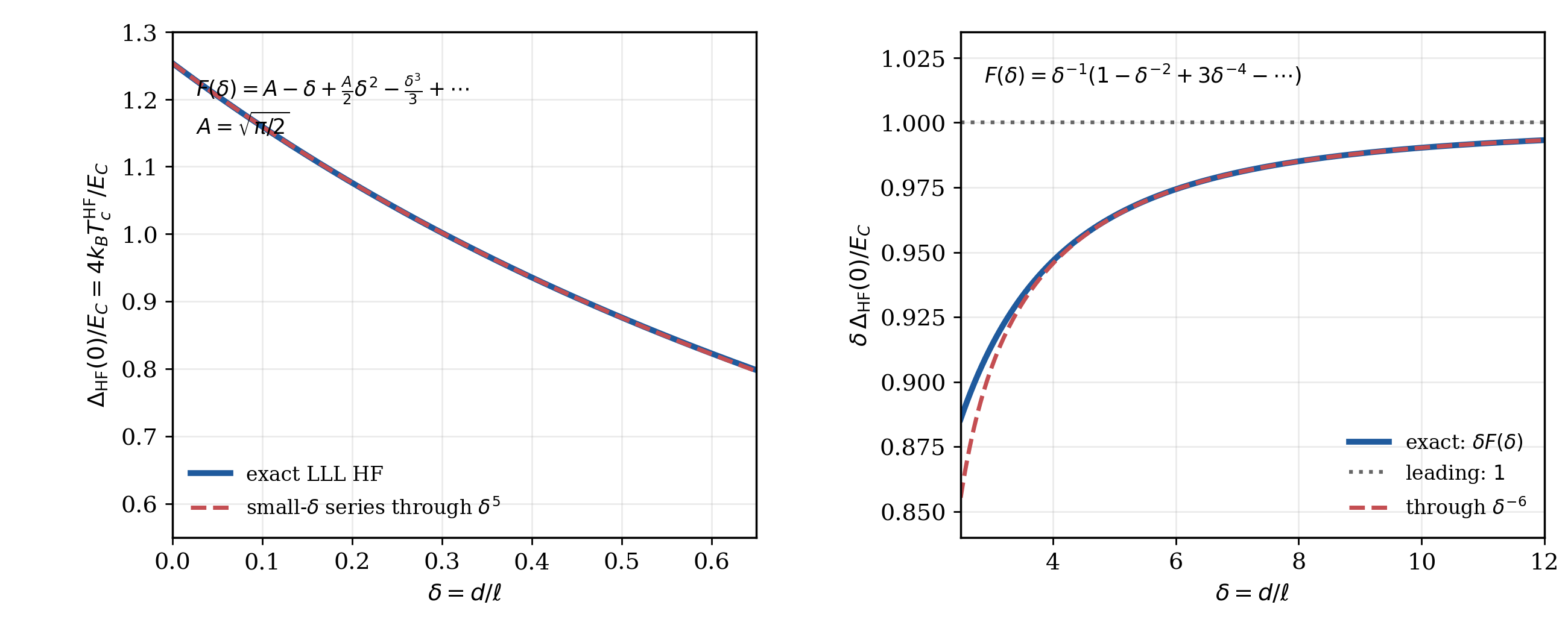}
\caption{Uniform-HF quasiparticle splitting $\Delta_{\rm HF}=X_D$ and transition scale $k_BT_c^{\rm HF}=X_D/4$, compared with the expansions of Eqs.~\eqref{qh:eq:small} (left) and \eqref{qh:eq:large} (right). In the figure $A=\sqrt{\pi/2}$.}
\label{qh:fig:gap}
\end{figure}

For $\delta=d/\ell$, the two asymptotic limits are
\begin{align}
\frac{X_D}{\Ec}&=\sqrt{\frac\pi2}-\delta+\frac12\sqrt{\frac\pi2}\delta^2-\frac{\delta^3}{3}+O(\delta^4),&&\delta\ll1,\label{qh:eq:small}\\
\frac{X_D}{\Ec}&=\frac1\delta\left(1-\frac1{\delta^2}+\frac3{\delta^4}-\frac{15}{\delta^6}+\cdots\right),&&\delta\gg1.
\label{qh:eq:large}
\end{align}
Thus
\begin{equation}
\Delta=4k_BT_c^{\rm HF}=\frac{e^2}{\epsilon d}\left(1-\frac{\ell^2}{d^2}+3\frac{\ell^4}{d^4}+\cdots\right).
\label{qh:eq:tail}
\end{equation}
The large-distance uniform-HF coherence scale is Coulombic.  This is a property of the HF exchange integral, which omits the intralayer correlations that dominate at large $d$.  The true large-$d$ scale is not known (Sec.~\ref{qh:sec:beyond}).  Figure~\ref{qh:fig:gap} compares the exact result with the two expansions.

\subsection{Goldstone mode and coherent density wave}
\label{sec:qh-mode}

\begin{figure}[t]
\includegraphics[width=\linewidth]{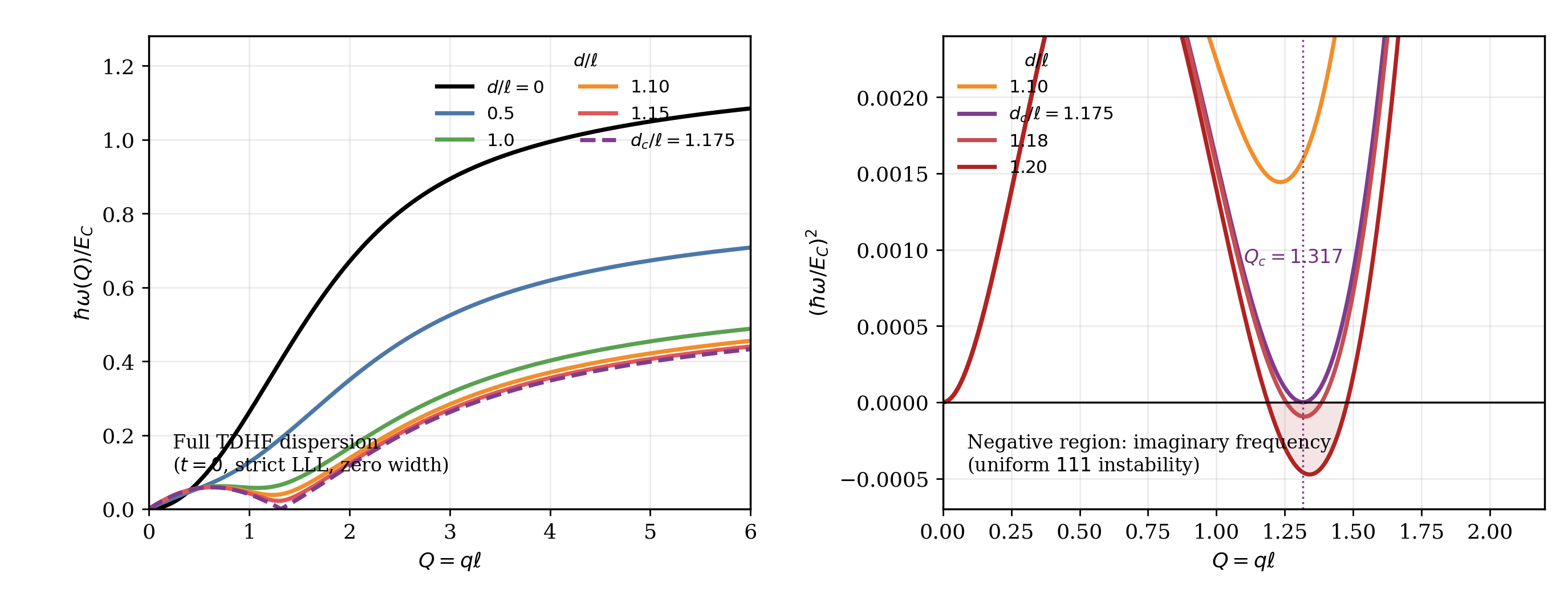}
\caption{Full projected-LLL collective-mode dispersion (left) and its square near the roton minimum (right); $t=0$ in the figure denotes zero tunneling.  The uniform state has a roton instability at $d/\ell\simeq1.175$ and $q\ell\simeq1.317$.}
\label{qh:fig:modefig}
\end{figure}

The coherent state breaks relative $U(1)$ and has a Goldstone mode.  The LLL time-dependent HF result \cite{Fertig1989,MacDonaldPlatzmanBoebinger1990} is
\begin{equation}
\hbar\omega(Q)=\sqrt{B_\delta(Q)K_\delta(Q)},\qquad Q=q\ell,
\label{qh:eq:mode}
\end{equation}
where
\begin{align}
\frac{B_\delta(Q)}{\Ec}&=\frac{\Gamma_D(0)-\Gamma_D(Q)}{\Ec},\nonumber\\
\frac{K_\delta(Q)}{\Ec}&=\frac{\Gamma_D(0)-\Gamma_S(Q)+D_\delta(Q)}{\Ec},\nonumber\\
\frac{\Gamma_D(Q)}{\Ec}&=\int_0^\infty dx\,e^{-x^2/2-\delta x}J_0(Qx),\nonumber\\
\frac{\Gamma_S(Q)}{\Ec}&=\sqrt{\frac\pi2}e^{-Q^2/4}I_0(Q^2/4),\nonumber\\
\frac{D_\delta(Q)}{\Ec}&=e^{-Q^2/2}\frac{1-e^{-\delta Q}}{Q}.
\label{qh:eq:kernels}
\end{align}
At long wavelength,
\begin{equation}
\hbar\omega(q)=\hbar vq+O(q^2),
\end{equation}
with
\begin{equation}
\frac{\hbar v}{\Ec\ell}=\frac12\sqrt{[(1+\delta^2)F(\delta)-\delta][F(\delta)-\sqrt{\pi/2}+\delta]},
\label{qh:eq:v}
\end{equation}
where $F=X_D/\Ec$.  At $d=0$ the SU(2)-symmetric ferromagnet has the expected quadratic mode.  The dispersion is shown in Fig.~\ref{qh:fig:modefig}.

The finite-$q$ zero has a definite symmetry meaning.  Define
\begin{equation}
\Psi=\langle S^+(\bm0)\rangle,\qquad\Phi_{\bm Q}=\langle S^z(\bm Q)\rangle.
\end{equation}
The uniform $(111)$ state has $\Psi\ne0$ and $\Phi_{\bm Q}=0$.  At the roton transition the $|\Phi_{\bm Q_0}|^2$ coefficient changes sign.  The $|\Psi|^2$ coefficient remains negative because $X_D$ remains positive.  Layer interchange combined with a relative-phase rotation takes $S^z\to-S^z$ at fixed $\Psi$, so no term cubic in $\Phi$ is allowed and the transition can be continuous.  The modulation is then small just beyond the instability, and the local transverse pseudospin $\sqrt{1-m_z^2}$ stays close to its uniform value.  The ordered state is therefore
\begin{equation}
\Psi\ne0,\qquad\delta n_T(\bm r)=-\delta n_B(\bm r)\propto\cos(\bm q_0\cdot\bm r+\theta).
\label{qh:eq:CDW}
\end{equation}
Here $\mathbf q_0=\mathbf Q_0/\ell$ is the physical modulation wave vector.
This is a coherent charge-density wave (CDW).  In exciton language it is a supersolid.  Unrestricted HF calculations confirm this structure \cite{BreyFertig2000}.  The finite-$q$ instability is not, within HF, a transition to an incoherent state.  It is a transition from a spatially uniform condensate to a spatially modulated condensate.

\subsection{Roton condensation as an excitonic supersolid}
\label{sec:qh-supersolid}

The finite-$q$ instability has an independent order parameter and an independent broken symmetry.  A minimal long-wavelength free energy for the uniform coherence $\Psi$ and the roton field $\Phi_{\bm Q_0}$ is
\begin{equation}
\begin{aligned}
{\cal E}={}&r_\Psi|\Psi|^2+u_\Psi|\Psi|^4+r_Q|\Phi_{\bm Q_0}|^2\\
&+u_Q|\Phi_{\bm Q_0}|^4+u_{\Psi Q}|\Psi|^2|\Phi_{\bm Q_0}|^2+\cdots.
\end{aligned}
\label{qh:eq:landau}
\end{equation}
Within the uniform-HF continuation, the coherent saddle has $r_\Psi<0$ and $r_Q>0$.  The roton transition is the point where $r_Q$ changes sign.  The positive uniform exchange integral keeps the \emph{uniform-HF} transverse field nonzero at every finite separation; it does not determine the renormalized coefficient after the higher-order correlation diagrams [e.g., Fig.~\ref{qh:fig:diagrams}(d)] are included.  The modulated HF phase has two independent broken symmetries: relative $U(1)$ from $\Psi\ne0$, and translation from $\Phi_{\bm Q_0}\ne0$.

The layer-density modulation in Eq.~\eqref{qh:eq:CDW} is out of phase between the two layers.  Consequently it is a pseudospin-density wave rather than a total-charge density wave.  This is precisely the modulation favored by the bilayer Coulomb interaction: the large Hartree cost of total charge modulation is avoided, while exchange energy is retained through the nonzero transverse pseudospin component.  The coherent CDW found in unrestricted HF calculations \cite{Brey1990,CoteBreyMacDonald1992} has this structure, as do the unidirectional coherent CDW of Brey and Fertig \cite{BreyFertig2000}, which in the lowest Landau level gives way to a coherent Wigner crystal near $d/\ell\simeq1.65$, and the excitonic CDW found for electron-hole layers by Chen and Quinn \cite{ChenQuinn1991}.  That the HF ground state beyond the instability is a coherent pseudospin-density wave was noted explicitly by Joglekar and MacDonald \cite{JoglekarMacDonald2001}.

The Hartree-Fock conclusion is straightforward.  A vanishing roton energy establishes an instability of the uniform saddle, not the vanishing of its uniform exchange integral.  The unrestricted-HF branch has nonzero $\Psi$ together with $\Phi_{\bm Q_0}$: a uniform $(111)$ condensate becomes a modulated coherent condensate.  In exciton language, it is an excitonic supersolid.  This coexistence of phase coherence and crystalline order is the defining supersolid structure emphasized for electron-hole bilayers by Joglekar, Balatsky, and Das Sarma \cite{Joglekar2006}; the present quantum-Hall problem supplies a microscopic pseudospin-density-wave realization.

The distinction resolves a simple apparent puzzle.  The roton collapses near $d/\ell\sim1$, while the interlayer exchange integral remains positive and has a long $1/d$ tail.  There is no conflict because these are different channels.  The roton is a finite-wave-vector density channel; $X_D$ is the uniform transverse-coherence channel.  The former determines the loss of translation invariance.  The latter determines the existence of interlayer phase coherence.

One argument beyond HF has been offered for the conventional identification.  A Gaussian fluctuation expansion about the uniform saddle finds the renormalized order parameter and stiffness driven toward zero as the roton softens \cite{JoglekarMacDonald2001}.  Such an expansion necessarily diverges at the instability of the saddle about which it is performed.  The divergence shows that the uniform saddle is the wrong starting point beyond the instability.  It cannot by itself decide whether coherence survives in the modulated state.

\subsection{Relation to earlier theory and experiment}
\label{sec:qh-relation}

The original theory of the $(111)$ state and its pseudospin representation established the basic structure of the problem \cite{Halperin1983,Fertig1989,MacDonaldPlatzmanBoebinger1990,Yang1994,Moon1995}.  The Chern--Simons formulation \cite{WenZee1992}, the theory of charged vortices and merons \cite{Moon1995,Moon1996}, and the Hartree-Fock density-wave solutions \cite{Brey1990,CoteBreyMacDonald1992} give complementary descriptions of the same coherent quantum Hall state.  The present account is more explicit.  The projected filled-band structure makes the uniform HF equation analytic at arbitrary separation.  The collective mode in Eq.~\eqref{qh:eq:mode} and its softening are those of Refs.~\cite{Fertig1989,MacDonaldPlatzmanBoebinger1990,Brey1990}.  The stiffness in Eq.~\eqref{qh:eq:rho_s} is that of Ref.~\cite{Moon1995} evaluated in closed form.  The exchange gap is the $q=0$ exchange kernel of the same literature, with the closed form given in Ref.~\cite{FilKravchenko2009}.  The closed formula for the HF amplitude scale at $d>0$, Eq.~\eqref{qh:eq:Tcexplicit}, has to our knowledge not been written down before; its $d=0$ limit is the result of Ref.~\cite{KasnerMacDonald1996}.  The closed formula in Eq.~\eqref{qh:eq:Tcexplicit}, the gap in Eq.~\eqref{qh:eq:gapT}, and the mode in Eq.~\eqref{qh:eq:mode} are determined by the same exchange integral.  This produces a complete one-parameter solution of the uniform theory.

The experimental record is equally direct.  The zero-bias tunneling anomaly measures the low-energy interlayer coherent response \cite{Spielman2000}.  The in-plane-field experiment measures the dispersing neutral collective mode \cite{Spielman2001}.  Hall drag demonstrates the charged quantum Hall response \cite{Kellogg2002,Kellogg2003}, while counterflow identifies the neutral superfluid channel \cite{Kellogg2004,Tutuc2004}.  Subsequent experiments demonstrated exciton transport and perfect Coulomb drag \cite{Finck2011,Nandi2012}.  These phenomena are naturally unified by the pseudospin-ferromagnet description rather than treated as separate effects.

The analytic $1/d$ exchange field gives a useful perspective on the rapid disappearance of strong coherence signatures with increasing effective layer separation.  Within uniform HF, an apparent boundary follows when $T_c^{\rm HF}$ reaches the experimental temperature or disorder scale.  This is distinct from the finite-$q$ instability of the uniform HF saddle.  Neither observation licenses an exact finite-$d$ conclusion for the interacting LLL bilayer.

\subsection{Phase structure}
\label{sec:qh-phase}

The zero-tunneling \emph{uniform-HF} problem consequently has a sharp theorem: its coherent saddle exists for every finite $d/\ell$, because $X_D(d)>0$.  An unrestricted-HF analysis adds a distinct finite-$q$ instability of that uniform saddle and produces a translation-breaking coherent density wave \cite{BreyFertig2000}.  The $d/\ell\rightarrow\infty$ limit has $X_D\rightarrow0$ and decoupled layers.  This is the  Hartree-Fock phase structure; it is not an exact phase diagram of the finite-$d$ LLL Hamiltonian.

The finite-temperature structure is equally clear within HF.  The interaction-generated amplitude scale is $T_c^{\rm HF}(d)$ in Eq.~\eqref{qh:eq:Tcexplicit}.  At a fixed temperature this produces an apparent critical separation.  The phase-mode stiffness and vortex dynamics determine the long-distance ordering behavior and the actual Kosterlitz-Thouless transition temperature, while the microscopic HF amplitude scale is set by the exchange integral. The HF amplitude scale and the Kosterlitz-Thouless transition temperature describe distinct aspects of coherence: exchange sets the local amplitude, while phase fluctuations determine long-distance order. Thus, they are not competing explanations.

\subsection{Analytic phase portrait}
\label{sec:qh-portrait}

The solution of the projected problem may be summarized by one positive function, $F(\delta)=X_D/\Ec$.  It decreases monotonically from $F(0)=\sqrt{\pi/2}$ to $F(\delta\gg1)=1/\delta+O(\delta^{-3})$.  Every uniform thermodynamic quantity follows from this function.  The exchange gap is $\Delta/\Ec=F$, the amplitude-transition temperature is $k_BT_c^{\rm HF}/\Ec=F/4$, the condensation energy is $F/4$ per flux quantum, and the sound velocity follows from Eq.~\eqref{qh:eq:v}.  The separation dependence of the coherent state is therefore analytic throughout the full range of $d/\ell$.

This HF structure is qualitatively different from a weak-coupling Cooper instability.  In BCS theory, a small gap is generated by a logarithmic instability and is exponentially sensitive to the interaction.  Here the HF gap is a direct Coulomb exchange matrix element in a filled Landau level.  The common phrase ``excitonic pairing'' is useful for identifying the broken relative-$U(1)$ symmetry, but it should not obscure this basic difference in scale.  The uniform-HF exchange gap has no BCS exponential.  Its large-$d$ behavior is fixed by electrostatics and by the LLL form factor.  This is the strong-coupling mean-field description continued to all $d$.  Whether it survives at large $d$, where each layer is separately correlated, is a question beyond HF (Sec.~\ref{qh:sec:beyond}).

The full HF phase portrait contains a uniform coherent region, a coherent modulated region, and the asymptotic decoupled limit.  In the uniform region, the only broken symmetry is relative $U(1)$.  In the modulated region, both relative $U(1)$ and translation are broken.  The two order parameters are experimentally distinguishable.  Interlayer tunneling and counterflow detect the transverse pseudospin order $\Psi$.  A density-sensitive probe detects the finite-$Q$ pseudospin component $\Phi_{\bm Q}$.  Within HF, a complete characterization of the roton-condensed state would therefore exhibit both a phase-coherent response and a density-wave Bragg component.

The same organization clarifies the meaning of the Hall response.  Within HF, the total charge sector of the uniform state remains incompressible and carries the integer Hall conductance.  The HF quasiparticle gap is $X_D\simeq0.60\Ec$ at the roton instability: the neutral mode softens, but the HF charge gap does not collapse.  The neutral sector supports the pseudospin Goldstone mode.  When the coherent density wave forms, the order parameter modulates the relative density, not the total charge, and near the onset, where the modulation is small, the HF gap remains close to $X_D$.  To that extent the quantum Hall charge response and the neutral supersolid response coexist within HF, and the state near the onset is the bilayer analog of a supersolid built from a neutral condensate embedded in a charged topological fluid.  The charge response of the fully developed modulated states, which have charged excitations of their own, is a separate question that this argument does not settle \cite{BreyFertig2000}.

The analytic result gives a sharp diagnostic \emph{inside uniform HF}.  A finite-separation loss of the uniform-HF coherence amplitude would require a zero of $X_D(d)$; Eq.~\eqref{qh:eq:XD} rules this out.  A finite-$q$ zero instead diagnoses the loss of spatial uniformity.  The unrestricted-HF density wave condenses in a coherent background \cite{BreyFertig2000}.  These statements do not constrain a correlation-driven loss of coherence beyond HF.

At large separation the distinction between a small scale and a vanished scale is essential.  The HF $1/d$ tail can easily fall below an experimental base temperature, giving the visual appearance of a transition at a finite $d/\ell$.  The same tail remains finite in the $T=0$ projected HF equation.  The apparent boundary and the zero-temperature analytic continuation are therefore both expected features of the same exchange physics.  One is controlled by a laboratory temperature; the other by the strict ground-state limit.

\section{The two problems compared}
\label{sec:compare}

Sections~\ref{sec:model}--\ref{sec:discussion} and Sec.~\ref{sec:qh} treat one phenomenon at the two ends of the range of magnetic field. This section compares them. It contains no new calculation. Its purpose is to state what the two problems have in common, where they differ, and what the results of one do and do not imply for the other. We return to the four questions posed in Sec.~\ref{sec:framework}~D: (1) Does mean-field theory predict interlayer coherence at every separation? (2) How does the coherence scale depend on $d$ at large separation, and does the mean-field result survive corrections? (3) Is the evolution from small to large $d$ a phase transition or a crossover from strong to weak coupling? (4) What is the relation between interlayer coherence and states that break translation symmetry?

\subsection{Mean-field theory: one diagram, two kinds of scale}
\label{sec:compare-mf}

Mean-field theory answers the first question in the same way in both problems, and for the same reason. The bare interlayer interaction has one sign at every wave vector, so the exchange self-energy that it generates cannot vanish. At zero field the kernel of the gap equation is positive and the instability exists at every $r_s$ and $d$ (Sec.~\ref{sec:gap}). At $\nu_T=1$ the exchange integral is positive at every finite $d$ [Eq.~\eqref{qh:eq:HFtheorem}]. In both, the scale falls off only algebraically with the separation,
\begin{align}
k_BT_c^{(0)}&\propto\frac{\hbar\vF}{d}&&(\text{zero field, fixed }r_s),\nonumber\\
k_BT_c^{\rm HF}&\to\frac{e^2}{4\epsilon d}&&(\nu_T=1),
\label{eq:twotails}
\end{align}
by Eqs.~\eqref{eq:Tc0} and \eqref{qh:eq:tail}. The origin of the power is also common. The interlayer interaction transfers wave vectors $q\lesssim1/d$, and nothing in mean-field theory screens it: at zero field this confines the pairing to within $\hbar\vF/d$ of the Fermi surface, and at $\nu_T=1$ it removes the form factor of the Landau level from the leading term of the exchange integral (Appendix~\ref{app:qh-exchange}).

The two scales are nevertheless of different kinds. At zero field $\hbar\vF/d$ is the prefactor of an exponential, $\exp[-\pi^{3/2}/(2\sqrt\alpha)]$, which is produced by the Cooper logarithm together with the logarithm of the unscreened kernel. At $\nu_T=1$ the gap is a Coulomb exchange energy with no exponential (Sec.~\ref{sec:qh-portrait}). In the language of the crossover, the mean-field theory at $\nu_T=1$ is a strong-coupling description continued to all $d$, whereas at zero field the same approximation contains the weak-coupling instability of the Fermi surface at one end and the two-body bound state at the other (Appendix~\ref{app:crossover}).

The zero-field counterpart of the calculation of Sec.~\ref{sec:qh} within mean-field theory is the Hartree-Fock theory of interlayer coherence in balanced electron-electron and electron-hole bilayers developed by Zhu and Das Sarma \cite{Zhu2024}.  That work obtains the zero-temperature coherence phase diagram and the finite-temperature onset by solving the self-consistent HF gap equation.  The common conclusion is structural: interlayer Coulomb exchange produces a spontaneous transverse pseudospin field and a finite HF coherence scale without interlayer tunneling.  The LLL calculation of Sec.~\ref{sec:qh} is the analytically soluble quantum-Hall specialization of this Hartree-Fock mechanism.  The distinction in scope is equally clear.  At zero field the kernel is set by the Fermi surface, whereas at $\nu_T=1$ it is fixed by the LLL form factor and guiding-center algebra.  In both cases the result is a Hartree-Fock statement, not an assertion that the corresponding finite-coupling many-body problem is exhausted by HF.  One difference is instructive.  At zero field the HF coherence of the electron-electron bilayer, unlike that of the electron-hole bilayer of Secs.~\ref{sec:model}--\ref{sec:discussion}, exists only for $r_s$ above a critical value that depends on $d$ \cite{Zhu2024}, whereas in the LLL the kinetic energy is quenched and the uniform-HF coherent saddle exists at every finite $d$.

\subsection{The fate of the mean-field tail at zero field}
\label{sec:compare-zf}

At zero field the second question has an answer. The layers are compressible, and their response screens the interlayer interaction. Static screening reduces the coupling to $\lams\propto r_s(\aB/d)^2$ [Eq.~\eqref{eq:lamsasym}], and the transition temperature becomes exponentially small in $d^2$ instead of algebraic in $d$. Within the ladder approximation dynamical screening appears to undo most of this, through the plasmon-induced coupling $\lampl\propto d^{-1/2}$. But that coupling is cancelled at second order by the crossed rung, and what is left of it is known [Eqs.~\eqref{eq:lamnetscaling} and \eqref{eq:lamnetlow}]. The result is rigorous for the leading exponent as $r_s\to0$ at fixed $d$ [Eq.~\eqref{eq:smallrs}], and for $d\to\infty$ at fixed $r_s$ it holds under the assumptions stated with Eq.~\eqref{eq:exactlarged}. In both limits the instability is present, and its scale decreases smoothly, as a Gaussian in $d$ for $d\gg\aB$. A first-order transition to a different state is not excluded by the theory (Appendix~\ref{app:rg}). Between the limits the existence of the instability at every $d$ is a property of the RPA ladder and is not established for the full problem (Sec.~\ref{sec:control}). The $1/d$ law of mean-field theory is an artifact of leaving out the response of the layers.

\subsection{Beyond Hartree-Fock at $\nu_T=1$: expectations}
\label{qh:sec:beyond}

Nothing in this problem is known in a controlled way beyond HF.  The field-theoretic treatments are uncontrolled, and the finite-size numerics \cite{NomuraYoshioka2002,ShengBalentsWang2003,ShibataYoshioka2006,ZhuFuSheng2017} are limited by size and geometry.  With that caveat we state our own expectation.  It is a conjecture, not a result of this paper.

We expect HF to be qualitatively right in its main conclusion: there is no zero-temperature transition at a finite $d/\ell$ at which interlayer coherence is lost.  The evolution from small to large separation is a crossover from strong to weak coupling, the analog of the crossover from Bose-Einstein condensation (BEC) to BCS pairing.  This is the picture suggested by experiments on graphene double layers \cite{Liu2022} and by the description of the large-$d$ state as a paired state of composite fermions connected to the exciton condensate \cite{Sodemann2017,RueggPRR2023,RueggPRL2024}.

We expect HF to be quantitatively wrong at large $d/\ell$.  The $1/d$ tail of Eq.~\eqref{qh:eq:tail} neglects intralayer correlations.  At large separation each layer is a composite-fermion liquid, and interlayer coherence is a weak-coupling pairing instability of these liquids with a scale far below $e^2/(\epsilon d)$.  We expect this scale to be exponentially small, so that the true coherence scale falls rapidly for $d/\ell\gtrsim1$ and mimics a transition.  The large-$d$ law is not settled: gauge-fluctuation treatments give an algebraic gap, $\Delta\propto1/d^2$ up to logarithms, strongly reduced by pair breaking \cite{BonesteelMcDonaldNayak1996,DengMendozaBonesteel2023}.  A crossover without a transition leaves no room for a phase of broken translation symmetry.  In this scenario the density-wave states of HF, the supersolid included, do not survive as distinct phases; that they are likely preempted in the lowest Landau level was already noted in Ref.~\cite{BreyFertig2000}.

The crossover is not the only possibility.  A sequence of distinct paired states separated by transitions was proposed in Ref.~\cite{Kim2001}, with an intermediate $(3,3,-1)$ state arising from $p_x-ip_y$ interlayer pairing of composite fermions.  That state has a Hall drag opposite in sign to that of the $(111)$ state.  No such sign reversal has been reported: the measured Hall drag rises from zero to the quantized $(111)$ value \cite{Kellogg2002,Kellogg2003}.  The $p_x+ip_y$ channel favored in trial-wave-function studies \cite{MollerSimonRezayi2008} is the one connected to the exciton condensate \cite{Sodemann2017,RueggPRL2024}.

\subsection{What the zero-field results imply for $\nu_T=1$, and what they do not}
\label{sec:compare-implications}

The expectation of Sec.~\ref{qh:sec:beyond} resembles the zero-field result of Sec.~\ref{sec:compare-zf} in its central element: a scale at large $d$ far below the mean-field one, with the instability still present. The zero-field theory is a case in which the failure of the mean-field tail is derived, rigorously for the leading exponent as $r_s\to0$ and conditionally for $d\to\infty$, and it identifies the mechanism. The $1/d$ tail is a property of the exchange integral with the bare interaction, and it does not survive once the carriers of each layer respond to the interlayer interaction. At $\nu_T=1$ the same ingredients are expected at large $d$: each layer is then expected to be a compressible composite-fermion liquid (Sec.~\ref{qh:sec:beyond}), and the interlayer interaction is confined to $q\lesssim1/d\ll1/\ell$. The zero-field result is consistent with the expectation that the HF tail of Eq.~\eqref{qh:eq:tail} is not the true law at large $d$. The other elements of the expectation, the absence of a transition and the crossover from strong coupling, are not derived at zero field either; there they rest on the estimates of Appendix~\ref{app:crossover}.

The zero-field result does not establish the expectation, and it does not determine the law that replaces the tail. Four differences prevent the argument from being carried over. First, neither controlled limit of the zero-field theory has an image in the lowest Landau level. By Eq.~\eqref{eq:dictionary} the projection sets $r_s=2\kappa_{\rm LL}$ to zero without making the interaction weak, and $d/\aB=\kappa_{\rm LL}\,d/\ell$ then vanishes as well, whereas the Gaussian law, $T_c\propto\exp[-\mathrm{const}\,(d/a_B)^2]$ at leading exponential order [Eq.~\eqref{eq:exactlarged}], requires $d\gg\aB$. Moreover the zero-field system at small $r_s$ is weakly paired at every $d$, while the $\nu_T=1$ state at small $d$ is strongly coupled. Second, the zero-field result at large $d$ rests on each layer being a Fermi liquid, whose static response is fixed by its compressibility and its quasiparticle parameters, and is itself conditional. The low-energy theory of the half-filled Landau level involves gauge fluctuations that have no counterpart at zero field, and the treatments that include them find an algebraic and not an exponential law \cite{BonesteelMcDonaldNayak1996,DengMendozaBonesteel2023}. Third, on the weak-pairing side at zero field the normal state from which the instability develops is the same pair of Fermi liquids at every $d$, whereas at $\nu_T=1$ the state at small $d$ is an incompressible quantum Hall state, whose HF description is exact at $d=0$, and the compressible layers appear only at large $d$. Fourth, the pairing at zero field is in the $s$-wave channel, while in the composite-fermion description of Refs.~\cite{Kim2001,MollerSimonRezayi2008} the paired states discussed in Sec.~\ref{qh:sec:beyond} are $p$-wave. The comparison is therefore a precedent and not a proof. It shows that one element of the scenario conjectured at $\nu_T=1$, the failure of the mean-field tail, occurs in the neighboring problem in the limits in which it can be checked. It is not evidence about the $\nu_T=1$ system itself.

\subsection{Broken translation symmetry and supersolids}
\label{sec:compare-ss}

In both problems states of broken translation symmetry appear, and in both they must be distinguished from the loss of coherence. At $\nu_T=1$ the statement is sharp within HF theory. The roton instability at $d/\ell\simeq1.175$ leads to a coherent pseudospin-density wave, which has both orders and is an excitonic supersolid (Sec.~\ref{sec:qh-supersolid}), and unrestricted HF theory finds a coherent Wigner crystal at larger separation \cite{BreyFertig2000}. Whether these phases exist beyond HF is not known; our expectation is that they do not (Sec.~\ref{qh:sec:beyond}). At zero field the crystalline phases lie at large $r_s$: the Wigner crystal of an isolated layer, and the crystal of dipolar excitons, which requires $r_s\gtrsim10$ and $d\gtrsim13\,\aB$ and gives way to the dilute superfluid at still lower density (Appendix~\ref{app:crossover}). The zero-field theory of this paper assumes a liquid and does not describe them (Sec.~\ref{sec:discussion}). That a crystalline state of the electron-hole bilayer can retain the exciton coherence, and is then a supersolid, was proposed in Ref.~\cite{Joglekar2006}.

The structural point is the same in both problems. Translation symmetry and the relative $U(1)$ symmetry are independent, their order parameters are independent [Eq.~\eqref{qh:eq:landau}], and an instability in the density channel is not by itself an instability of the coherence. At $\nu_T=1$ this is the content of result (vii) of the Introduction. At zero field the corresponding question, whether the crystalline phases of Fig.~\ref{fig:regimes} retain the exciton coherence, is outside the theory of this paper.

\subsection{The separation as the tuning parameter}
\label{sec:compare-crossover}

At zero field there are two crossovers between the condensation of tightly bound excitons and weak pairing, one as a function of $r_s$ at fixed $d$ and one as a function of $d$ at fixed $r_s$ (Appendix~\ref{app:crossover} and Fig.~\ref{fig:regimes}). For $r_s<(2/g)^{1/2}$ the binding energy is below the Fermi energy at every $d$ and there is no crossover. At $\nu_T=1$ the density and the field are tied by the filling factor, and only the crossover in $d$ can exist: $d/\ell=\kF d$ is varied at fixed $\nu_T$. The small-$d$ end is the strong-coupling one, where HF theory is exact at $d=0$. The large-$d$ end is the one at which we expect weak coupling (Sec.~\ref{qh:sec:beyond}); a crossover of this kind has been reported in graphene double layers \cite{Liu2022}.

In both problems the physical transition at finite temperature is of Kosterlitz-Thouless type and is bounded by the phase stiffness, and the mean-field scale is a different quantity. At zero field the bound is $k_BT_{\rm KT}\le g\EF/8$ [Eq.~\eqref{eq:TKT}], and the mean-field $T_c$ approaches the transition temperature only on the weak-pairing side. At $\nu_T=1$ the phase stiffness bound is $(\pi/2)\rho_s$, which is $0.0097\,\Ec$ at $d/\ell=1$, a factor of 17 below $k_BT_c^{\rm HF}$, and which falls off as $1/d^3$ within HF theory (Appendix~\ref{qh:sec:finiteT}), faster than the
$1/d$ decay of the zero-temperature exchange gap $X_D$. In a measurement at fixed temperature a scale that decreases smoothly with $d$ produces an apparent critical separation in either problem. This is a thermal effect and not a transition of the ground state.

\subsection{Experiment}

The experimental situation is the reverse of the theoretical one. At $\nu_T=1$ interlayer coherence, the collective mode, quantized Hall drag, and counterflow transport have all been observed (Sec.~\ref{sec:qh-relation}), and the theory beyond mean field is uncontrolled. At zero field the theory has controlled limits, but in them the transition temperature is far too small to be observed (Sec.~\ref{sec:discussion}). The interlayer coherence observed in the experiments is in the regime in which $E_B$ is comparable to $\EF$ or larger (Sec.~\ref{sec:discussion}), where neither limit applies.

\subsection{Status of the results}

Table~\ref{tab:status} lists the main results of the paper with the kind of statement that each of them is. The entries for the two problems should not be read across: a controlled result at zero field does not raise the status of the corresponding conjecture at $\nu_T=1$.

\begin{table*}[t]
\caption{Status of the main results. ``HF theorem'' and ``within HF'': statements about the stated Hartree-Fock approximation, not about the full problem. ``Exact'': an identity or a property that holds without approximation in the stated model. ``Controlled'': the leading term in a limit with a small parameter. ``Conditional'': follows from stated assumptions that are not proven. ``Approximation'': the output of a definite approximation with no estimate of its error. ``Conjecture'': an expectation, not derived.}
\label{tab:status}
\begin{ruledtabular}
\begin{tabular}{p{0.50\textwidth}p{0.20\textwidth}p{0.24\textwidth}}
Result & Where & Status\\
\hline
\multicolumn{3}{l}{\emph{Zero field}}\\
RPA interlayer interaction attractive at every $q$ and Matsubara frequency & Eq.~\eqref{eq:sign} & Exact within the RPA\\
Pairing instability at every $r_s$ and $d$ in the ladder with the RPA interaction & Sec.~\ref{sec:gap} & Exact within the RPA ladder\\
Mean-field $T_c$ of the ladder; $1/d$, $e^{-{\rm const}\,d^2}$, and $e^{-{\rm const}\sqrt d}$ laws & Secs.~\ref{sec:analytic}, \ref{sec:numerics} & Approximation (ladder)\\
Identity for the sum of the ladder iteration and the crossed rung & Eq.~\eqref{eq:Bnet} & Exact\\
Limits of $\lambda_{\rm net}$ for $r_s\to0$ and for $d\to\infty$ & Appendix~\ref{app:remainder} & Exact asymptotics of the second-order term\\
Cancellation at every order for linearized dispersion and Gaussian fluctuations & Sec.~\ref{sec:adiabatic} & Exact in that model\\
Leading exponent for $r_s\to0$ at fixed $d$, $\lams\ln(\EF/k_BT_c)\to1$ & Eq.~\eqref{eq:smallrs} & Controlled\\
$\ln T_c\simeq-(m^*/m)(16g^{3/2}/\pi r_s)(d/\aB)^2$ for $d\to\infty$ & Eq.~\eqref{eq:exactlarged} & Conditional\\
$T_c$ for $r_s\gtrsim1$ and $d\lesssim r_s\aB$ & Sec.~\ref{sec:control} & Approximation (ladder)\\
The two crossovers to Bose condensation; the maximum of $T_c$ & Appendix~\ref{app:crossover} & Estimates; crossover region not calculated\\
\hline
\multicolumn{3}{l}{\emph{$\nu_T=1$}}\\
$X_D(d)>0$: the uniform HF state is coherent at every finite $d$ & Eq.~\eqref{qh:eq:HFtheorem} & HF theorem\\
Closed forms for $\Delta_{\rm HF}$, $T_c^{\rm HF}$, the mode velocity, and the stiffness & Sec.~\ref{sec:qh}, Appendix~\ref{app:qh-stiffness} & Exact within uniform LLL HF and TDHF\\
Fully polarized pseudospin ferromagnet at $d=0$ & Sec.~\ref{sec:qh-model} & Exact\\
Roton instability leads to a coherent density wave, not to loss of coherence & Secs.~\ref{sec:qh-mode}, \ref{sec:qh-supersolid} & Within HF (TDHF; unrestricted HF of Ref.~\cite{BreyFertig2000})\\
No transition; crossover to weak coupling; HF tail does not survive & Sec.~\ref{qh:sec:beyond} & Conjecture\\
Density-wave phases of HF do not survive as distinct phases & Sec.~\ref{qh:sec:beyond} & Conjecture\\
\hline
\multicolumn{3}{l}{\emph{Comparison}}\\
Zero-field theory as a precedent for the expectation at $\nu_T=1$ & Sec.~\ref{sec:compare-implications} & Analogy, not a derivation\\
\end{tabular}
\end{ruledtabular}
\end{table*}

\section{Conclusion}
\label{sec:conclusion}

Exciton superfluidity in bilayers has been treated here at zero magnetic field and at $\nu_T=1$ as one problem in two settings. In both, mean-field theory gives interlayer coherence at every layer separation, with a scale that decays as $1/d$. The two parts of the paper establish what can be said beyond that statement in each case, and the answers are of different kinds because the problems differ in whether they have a small parameter.

\emph{Zero field.} Replacing the single electron layer by an electron-hole bilayer changes the problem of pairing by the dynamically screened Coulomb interaction in three ways. Within the RPA the interaction in the pairing channel is attractive at all Matsubara frequencies, so in the ladder approximation the instability exists without any balance between attraction and repulsion. The finite range $d$ of the interlayer interaction limits the momentum transfer to $1/d$, which confines the pairing to energies within $\hbar\vF/d$ of the Fermi surface and makes the frequency sum converge, as $1/\omega_c$ above the plasmon scale $\hbar\Ost=\sqrt{2g\EF E_d}$, so no cutoff has to be chosen. And the leading vertex correction (e.g., the crossed-rung correction), which in the single layer is unknown and of order unity, can be computed at large separation, where it is still of order unity.

The ladder theory has a simple analytic core. The unscreened interaction gives a transition temperature that decays as $1/d$, static screening one that decays as $\exp(-{\rm const}\,d^2)$, and dynamic screening one that decays as $\exp(-{\rm const}\sqrt d)$ through a plasmon-induced coupling $\lampl\simeq0.109\,r_s(E_d/\EF)^{1/2}$, and the full numerical solution confirms these forms. The plasmon-induced coupling, however, comes entirely from the region of frequency and momentum in which the carriers are slow compared with the plasmons, where Migdal's theorem fails and the adiabatic limit takes its place.

Beyond the ladder, the following is established. (i) At second order the crossed rung cancels the plasmon-induced coupling. The sum of the two diagrams is given by the identity of Eq.~\eqref{eq:Bnet}; it is less than \netfracmax\ of $\lampl$ in the whole range computed and about one percent of it or less for $\kF d>2$. (ii) The remainder is known in both limits. It is $C_3(d)\,r_s^3$ for $r_s\to0$ at fixed $d$, with $C_3$ in closed form [Eq.~\eqref{eq:C3}], and $J_w\sqrt g\,r_s^4(\aB/d)^{7/2}$ with $J_w=\Jw$ for $d\to\infty$ at fixed $r_s$, and the numerical results converge to both. The second-order dynamical contribution to $\ln T_c$, which in the ladder is $\lampl/\lams^2$, independent of $r_s$ and equal to $\ladExcessCoef\,(d/\aB)^{7/2}$ at large $d$, becomes $R(d)\,r_s$ at small $r_s$, with $R=\RlowOne$ at $d=\aB$ and $R\propto\aB/d$ at large $d$. (iii) For linearized dispersion and Gaussian density fluctuations the cancellation is exact at every order. (iv) For $r_s\to0$ at fixed $d$ the leading exponent is that of first-order perturbation theory in the statically screened interaction. The corrections to $\ln T_c$ are of order $\ln(1/r_s)$ and come from the static crossed rung and the intralayer terms.

What is not established is the cancellation of the dynamical terms beyond second order when the dispersion has curvature. The result for $d\to\infty$ at any $r_s$ in the liquid phase, $\ln T_c\simeq-(m^*/m)(16g^{3/2}/\pi r_s)(d/\aB)^2$, with $E_B/\EF$ as the small parameter when $r_s\gtrsim1$, is conditional on it. It requires that the terms not contained in the static quasiparticle interaction be small compared with $\lams$. At second order they are, by the factor $\lambda_w/\lams\propto r_s^3(\aB/d)^{3/2}$, and beyond second order we assume it. We consider the assumption very likely to be correct, but an evaluation of the third-order terms, or a proof, is needed before the large-$d$ result can be called exact, and the present work does not resolve the problem at all orders. The prefactor of $T_c$ has not been calculated in either limit. Between the limits, where the experiments are, the RPA ladder has an instability at every $r_s$ and $d$, but the enhancement of its scale by dynamical screening should not be relied upon.

\emph{High field.} The zero-tunneling $\nu_T=1$ bilayer has an analytic uniform LLL Hartree-Fock solution.  Its interlayer exchange field is nonzero at every finite separation.  The resulting uniform-HF theorem is that its $T=0$ saddle remains interlayer coherent, with $\Delta_{\rm HF}=X_D$ and $k_BT_c^{\rm HF}=X_D/4$.  The uniform-HF exchange scale is algebraic at large $d/\ell$.  The finite-wave-vector roton instability in unrestricted HF produces a coherent density wave and does not, by itself, set the uniform exchange field to zero.  The identification of this instability with the loss of interlayer coherence is not supported by the HF theory on which it rests.  Finite-$d$ LLL correlations are not eliminated by Pauli blocking; they are the physics omitted by HF and determine whether the Hartree-Fock phase structure survives beyond mean field.  Our expectation is that the absence of a zero-temperature transition survives, as a crossover from strong to weak coupling, while the corresponding coherence energy scale is expected to decay faster than the uniform-HF $1/d$ law at large separation.

\emph{The two together.} The persistence of the coherent mean-field solution at every separation is the part of mean-field theory that survives where it can be tested. It follows from the sign of the bare interlayer interaction, it is exact within HF in both problems, and at zero field it survives for $r_s\to0$ and, under the stated assumptions, for $d\to\infty$. At $\nu_T=1$ its survival beyond HF is a conjecture. The $1/d$ law of mean-field theory is not robust at zero field. In the controlled limits, conditionally for $d\to\infty$, it is replaced by a Gaussian once the response of the layers is included, and the larger scale that dynamical screening seems to restore within the ladder is removed by the vertex correction of the same order. At $\nu_T=1$ we expect the HF tail to fail for the same physical reason, but no controlled calculation shows it, and the law that replaces it is not known. In each problem one common inference is not supported. At zero field it is that dynamical screening greatly enhances the pairing scale: the enhancement is cancelled at second order and is absent in the controlled limits. At $\nu_T=1$ it is that the softening of the collective mode marks the loss of coherence: within the HF theory on which that inference rests, it marks the onset of a coherent density wave. What is missing differs in kind. At zero field it is the dynamical terms of third and higher order with curvature, the prefactor of $T_c$, and a controlled treatment of the intermediate regime and of the crossover to the Bose side. At $\nu_T=1$ it is any controlled theory of the coherence at large $d/\ell$, where each layer is separately correlated. The comparison made here suggests how the second problem might be approached, from the weak-pairing side. Whether the outcome there is a crossover and not a transition is our conjecture; the comparison does not test it.

Although not relevant to the current theoretical work where all impurity disorder effects are ignored focusing on the pristine bilayers, actual experimental systems invariably have disorder, which would suppress superfluidity in the same way as magnetic impurities suppress ordinary superconductivity since charge disorder in the two layers is unlikely to be identical.  Disorder effects have been considered in Refs.~\cite{Abergel2012,Abergel2013}.

\begin{acknowledgments}
This work is supported by the Laboratory for Physical Sciences.
This work made substantial use of artificial-intelligence tools, which is disclosed here in accordance with the policy of the American Physical Society. The author posed the problems, specified the approach, and directed the work at each stage. For the zero-field theory (Secs.~\ref{sec:model}--\ref{sec:discussion} and Appendices~\ref{app:pol}--\ref{app:remainder}) the analytic derivations, the numerical codes and calculations, the figures, and the text were produced with the AI system Claude (Anthropic; model identifier claude-opus-5-5) under that direction. Successive drafts of that part were criticized by a second AI system, Codex (OpenAI, GPT-6-Astra), acting as a referee, and were revised in response. The numerical results were checked against independent evaluations wherever these were available: the plasmon-induced coupling by two independent quadratures, the gap-equation solver against the BCS result, and the limiting forms of the net second-order coupling against the full quadrature (Appendices~\ref{app:num} and \ref{app:remainder}). The same AI system was used in revising the text of the $\nu_T=1$ theory (Sec.~\ref{sec:qh} and Appendices~\ref{app:qh-stiffness}--\ref{app:qh-ed}) and in writing the abstract, the introduction, the conclusion, and Secs.~\ref{sec:framework} and \ref{sec:compare}, which contain no calculation beyond those of the other sections. The authors are responsible for the content of the paper.

The data that support the findings of this article are not publicly available. The data are available from the authors upon reasonable request.
\end{acknowledgments}

\appendix

\section{Polarization of a two-dimensional layer}
\label{app:pol}

With $x=q/\kF$ and $z=\hbar\omega/\EF$, the retarded polarization at $T=0$ for ${\rm Im}\,z>0$ is
\begin{equation}
\frac{\Pi(q,z)}{N_0}=-1-\frac{a\sqrt{1-4x^2/a^2}-b\sqrt{1-4x^2/b^2}}{2x^2},
\end{equation}
with $a=z-x^2$, $b=z+x^2$, and the principal branch of the square root, which is analytic in the upper half plane \cite{Stern1967,GiulianiVignale}. On the imaginary axis, $z=i\nu$,
\begin{equation}
P(x,\nu)\equiv-\frac{\Pi(q,i\nu)}{N_0}=1-\frac{1}{x^2}\,{\rm Re}\sqrt{(x^2+i\nu)^2-4x^2},
\end{equation}
with the root of positive real part. It satisfies $0<P\le1$, decreases with $|\nu|$, and has the limits $P=1$ for $\nu=0$, $x\le2$; $P\simeq1-\nu/\sqrt{\nu^2+4x^2}$ for $x\ll1$; and $P\simeq2x^2/\nu^2$ for $\nu\gg2x+x^2$, that is, for $\hbar\nu$ large compared with both $\hbar\vF q$ and the recoil energy. We verified the implementation against the static limit, the high-frequency limit, the $f$-sum rule, and the long-wavelength plasmon.

\section{Exciton binding energy}
\label{app:exciton}

In units of $\aB$ and Ry the relative-motion Hamiltonian is $-2\nabla^2-2/\sqrt{r^2+d^2}$. For the $s$ state we use a logarithmic radial grid $r=e^t$, on which the equation is $-2\psi_{tt}+r^2(V-E)\psi=0$, symmetrize with $\phi=r\psi$, impose a regular boundary condition at the inner end, and diagonalize the resulting tridiagonal matrix. The result at $d=0$ is within $0.5\%$ of the exact $2$ Ry; for $0<d<0.05\,\aB$ it has a numerical scatter of about $2\%$, which is of no consequence at the separations considered here. For $d\gg\aB$ the potential is harmonic, $-2/d+r^2/d^3$, with zero-point energy $2\sqrt2/d^{3/2}$, which gives Eq.~\eqref{eq:EB}.

\section{Solution of Eq.~\eqref{eq:son}}
\label{app:son}

Replacing $\tanh(\xi'/2T)/\xi'$ by $1/\xi'$ with a lower limit $\xi_0=\pi T/2e^{\gamma}$, which reproduces $\int_0^\Omega d\xi\tanh(\xi/2T)/\xi=\ln(\Omega/\xi_0)$, and writing $s=\ln\xi$, Eq.~\eqref{eq:son} is $\Delta(s)=A\int ds'\,[s_b-\max(s,s')]\,\Delta(s')$. Two derivatives give $\Delta''=-A\Delta$ with $\Delta(s_b)=0$ and $\Delta'(s_0)=0$, so that $\Delta\propto\cos[\sqrt A(s-s_0)]$ and $\sqrt A\,\ln(\Omega_b/\xi_0)=\pi/2$. With $\Omega_b=2e^{-\gamma}\hbar\vF/d$ the factors of $e^{\gamma}$ cancel and $k_BT_c=(4/\pi)(\hbar\vF/d)\exp(-\pi/2\sqrt A)$. The replacement of $K_0$ by its logarithm and of the hyperbolic tangent by a cutoff are both correct to leading order in $\sqrt A$; the factor $\kappa_0=1+O(\sqrt A)$ in Eq.~\eqref{eq:Tc0} is determined numerically.

\section{Numerical details}
\label{app:num}

\emph{Angular average.} For $\theta<\pi/2$ we substitute $q=\delta\cosh t$ with $\delta=|k-p|$, which makes the integrand smooth for a $1/q$ interaction, and use 40 Gauss-Legendre points in $t$; for $\theta>\pi/2$ we use 16 points in $\theta$. The result agrees with adaptive quadrature to $10^{-10}$.

\emph{Momentum cells.} The cells are geometric in $|\xi|$ on each side of the Fermi surface. The kernel is averaged over each cell with four Gauss-Legendre points and over the diagonal cell, which contains the logarithmic singularity, with eight points on each half. The factor $1/(\omega^2+\xi^2)$ is integrated analytically, $\int d\xi/(\omega^2+\xi^2)=\omega^{-1}\arctan(\xi/\omega)$. The error of the eigenvalue is proportional to the square of the logarithmic cell size. We use 40 and 56 cells per side above the Fermi surface and extrapolate; for the frequency-independent kernels we use 120 cells. With 40, 56, 80, 120, and 170 cells the unscreened eigenvalue at $r_s=d/\aB=1$, $T=10^{-2}\EF$ is 2.6603, 2.6937, 2.7119, 2.7219, and 2.7259, and the extrapolations from successive pairs are 2.7285, 2.7295, and 2.7298.

\emph{Frequency blocks.} The first ten positive Matsubara frequencies are kept individually and the rest are grouped in blocks whose size grows by a factor 1.5, up to $10^5\EF$. The kernel is tabulated at 197 values of $\nu$ and averaged over each block with four points. The error is proportional to the square of the logarithm of the growth factor, and we extrapolate from growth factors 1.5 and 1.8. At $r_s=d/\aB=1$ and $T=10^{-2}\EF$ the extrapolations from the pairs of growth factors $(1.5,1.8)$ and $(1.25,1.5)$ are $0.7656$ and $0.7658$. A frequency cutoff $\omega_c$, where one is used, retains exactly the Matsubara frequencies below $\omega_c$.

\emph{Tests.} For a constant attraction $\lambda=0.25$ within $\pm0.2\,\EF$ of the Fermi surface the code gives $T_c$ within $0.2\%$ of the BCS value $1.134\,\Omega\,e^{-1/\lambda}$. For a frequency-independent kernel the Matsubara solver and the direct solution with $\tanh(\xi/2T)/2\xi$ agree to $0.15\%$ in $\eta$.

\emph{Crossed rung.} Equations~\eqref{eq:lamX} and \eqref{eq:gmb} are evaluated in the variables $\theta$, the angle between $\mathbf k$ and $\mathbf k'$, the polar coordinates $(q,\varphi)$ of $\mathbf q$, and $\nu$. The integrand is symmetric under $\mathbf q\to\bm\Delta-\mathbf q$, and we multiply it by $2|\bm\Delta-\mathbf q|^2/(q^2+|\bm\Delta-\mathbf q|^2)$, which leaves the integral unchanged and removes the singularity at $\mathbf q=\bm\Delta$. We use Gauss-Legendre nodes on geometric panels in $\theta$ and $q$, a geometric grid in $\nu$, and for $\varphi$ panels that are refined geometrically toward the zeros of $\xi_1$ and $\xi_2$ and toward the direction of $\bm\Delta$. The sum $B_{\rm lad}+B_{\rm cr}$ is evaluated from the closed form in Eq.~\eqref{eq:Bnet}, so the cancellation is not subject to rounding. Increasing the number of nodes in each variable by about 40\% changes $\lambda_{\rm lad}$ by about $10^{-3}$ or less in relative terms, and $\lambda_{\rm net}$ by less than $0.1\%$ at $\kF d>2$ and by up to $3\%$ at $\kF d<1$.

\emph{Low temperatures.} $T_c$ is located by decreasing $T$ in decades from $3\,\EF$ until $\eta>1$ and refining by the secant method in $\ln T$. The lowest temperature is about $10^{-12}\EF$. If $\eta<1$ there, $T_c$ is obtained by extrapolation and is marked as such: for the statically screened kernel $\eta$ is linear in $\ln T$ at low temperature, as in BCS theory, and it is extrapolated linearly from the last four temperatures. The slope of the numerical $\eta(T)$ agrees with the independently computed $\lams$ of Eq.~\eqref{eq:lams} to within \statSlopeErr\ at every point, and fits to the last three and the last six temperatures change $\ln T_c$ by at most \statWinErr, although the extrapolation extends over hundreds of decades. For the dynamically screened kernel the combination $\eta^2/(\lams\eta+\lampl)$, which is linear in $\ln T$ [Eq.~\eqref{eq:etaT} and Fig.~\ref{fig:scaling}(b)], is extrapolated linearly to its value at $\eta=1$.

\section{Diagrams through second order, the controlled limits, the two crossovers, and the renormalization group}
\label{app:long}

This appendix gives the complete set of pairing-channel diagrams through second order, both in the bare and in the screened interaction, and states for each one whether it is contained in our results and why. It then uses this classification to examine the two limits in which the theory is controlled: small $r_s$, where the leading exponent is that of leading-order perturbation theory, and large $d/\aB$, where the result follows from the large energy denominators of all processes that involve the plasmons, under assumptions that are stated. It finally discusses the two distinct crossovers between weak pairing and Bose condensation of excitons, as a function of $d$ at fixed $r_s$ and as a function of $r_s$ at fixed $d$, and what a renormalization-group (RG) treatment of the Polchinski-Shankar type \cite{Polchinski1992,Shankar1994} does and does not capture. The main conclusions are collected in Sec.~\ref{app:summary}.

\subsection{The object that is expanded}
\label{app:object}

The transition temperature is determined by the linearized equation for the anomalous self-energy,
\begin{align}
\phi(k)=-T\sum_{p}I(k,p)\,\mathcal G_e(p)\,\mathcal G_h(-p)\,\phi(p),
\label{eq:BS}
\end{align}
where $k=(\mathbf k,i\omega_n)$, $I$ is the vertex that is irreducible in the electron-hole pair channel, and $\mathcal G_{e,h}$ are the full normal propagators. A diagram of $I$ has an incoming electron and hole and an outgoing electron and hole, and it cannot be separated into two parts by cutting one electron line and one hole line. Equation~\eqref{eq:BS} is exact. Three things are expanded: the kernel $I$, the self-energy in $\mathcal G$, and the polarization inside the interaction lines. Charge conservation ties them together through Ward identities, so a truncation is consistent only if the three are truncated together.

The rules are the following. A solid line is the bare propagator $\mathcal G^0$. The electron line is in layer 1 and the hole line in layer 2. There is no interlayer tunneling, so the two external lines never join, and there is no exchange between an electron and a hole. Every closed loop is summed over the two layers. A wavy line is the bare interaction, $V_q$ between vertices in the same layer and $-U_q$ between vertices in different layers; a double wavy line is the RPA interaction of Sec.~\ref{sec:W}, $W_{11}$ or $W_{12}$. Diagrams that contain a tadpole are omitted: for the neutral balanced bilayer the Hartree terms add up to the capacitor shift $2\pi e^2nd/\epsilon$ of the chemical potential and its iterations, which are absorbed at fixed density.

\subsection{Expansion in the bare interaction}
\label{app:bare}

\begin{figure*}[t]
\centering
\includegraphics{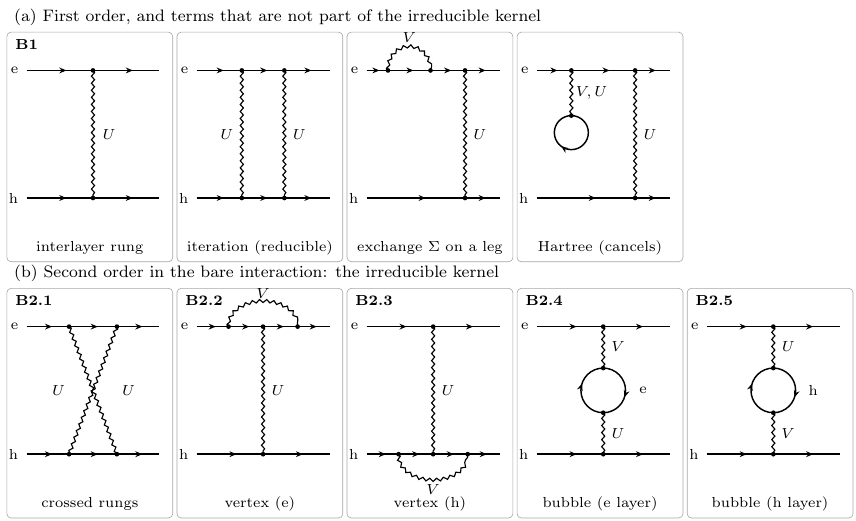}
\caption{Pairing-channel diagrams through second order in the bare interaction. Wavy lines are $V_q$ within a layer and $U_q$ between the layers. (a) The first-order kernel B1 and three terms that are not part of the irreducible kernel: the iteration of B1, which is generated by solving Eq.~\eqref{eq:BS}; the exchange self-energy, which belongs to the propagators; and the Hartree term, which is cancelled at fixed density. (b) The five second-order diagrams of the irreducible kernel.}
\label{fig:diagbare}
\end{figure*}

Figure~\ref{fig:diagbare} shows all diagrams through second order in the bare interaction.

\emph{First order.} The kernel has one diagram, the interlayer rung B1, $I^{(1)}=-U_q$. Solving Eq.~\eqref{eq:BS} with this kernel generates the ladder of all iterations, of which the first is shown in Fig.~\ref{fig:diagbare}(a). With bare propagators this is the unscreened theory of Sec.~\ref{sec:analytic}, Eq.~\eqref{eq:Tc0}. The HF theory of Ref.~\cite{Zhu2024} is the same ladder with the exchange self-energy included in the propagators.

\emph{Second order.} The irreducible kernel has five diagrams, Fig.~\ref{fig:diagbare}(b): the crossed rungs B2.1, the two vertex corrections B2.2 and B2.3, in which an intralayer line $V$ spans the interlayer vertex on the electron or on the hole line, and the two diagrams B2.4 and B2.5, in which the rung is interrupted by a bubble in the electron or in the hole layer, $-U_q\Pi V_q-V_q\Pi U_q$. At the same order the propagators acquire the exchange self-energy. There are no other diagrams: a second-order diagram has two interaction lines, and with at most one loop the possibilities are exhausted by these five and by the reducible iteration.

\emph{Why this is not a weak-coupling expansion.} The bare series is not an expansion in a small parameter, for two reasons that both come from small momentum transfer.

First, the bubble diagrams are larger than the rung they correct. Their ratio to B1 at $\nu\ll\vF q$ is
\begin{equation}
\frac{{\rm B2.4}+{\rm B2.5}}{{\rm B1}}=-2N_0V_q=-\frac{2\cG}{qd},\qquad\cG=\frac{gd}{\aB},
\label{eq:bubbleratio}
\end{equation}
and a rung with $n$ bubbles is of order $(\cG/qd)^n$. For $qd<\cG$ the series must be summed to all orders, which gives the RPA \cite{GellMannBrueckner1957}. The parameter $\cG$ is $d$ times the Thomas-Fermi wave vector $g/\aB$, which in two dimensions does not depend on the density. \emph{Screening of the interlayer interaction does not become weak when $r_s\to0$.} It is controlled by $d/\aB$ alone.

Second, the first-order term itself does not define a coupling constant. The Fermi-surface average of B1 is $(\alpha/\pi)\int dy\,e^{-y}/y$ with $y=qd$, which diverges logarithmically at small $y$. In the gap equation the divergence is cut off by the energy difference of the two states, Eq.~\eqref{eq:K0}, and the result is the double logarithm of Eq.~\eqref{eq:son} and $\ln T_c\propto-1/\sqrt\alpha$. This is not the behavior of the system at small $r_s$. The forward divergence is removed by the bubbles of Eq.~\eqref{eq:bubbleratio}, after which the coupling constant is finite and $\ln T_c\propto-1/\alpha$ (Sec.~\ref{app:smallrs}).

The expansion must therefore be reorganized in terms of the screened interaction.

\subsection{Expansion in the screened interaction}
\label{app:screened}

When the bubble chains are summed into $W$, counting interaction lines is no longer a consistent ordering, because $W\Pi=O(1)$: a closed loop compensates one screened line. The consistent count is
\begin{equation}
\text{order}=(\text{number of }W\text{ lines})-(\text{number of closed loops}),
\label{eq:order}
\end{equation}
which is the formal expansion in $1/N$ for $N$ fermion flavors, with $W\propto1/N$ and each loop $\propto N$. A bare bubble on a $W$ line is excluded, since it is already contained in $W$.

\begin{figure*}[t]
\centering
\includegraphics{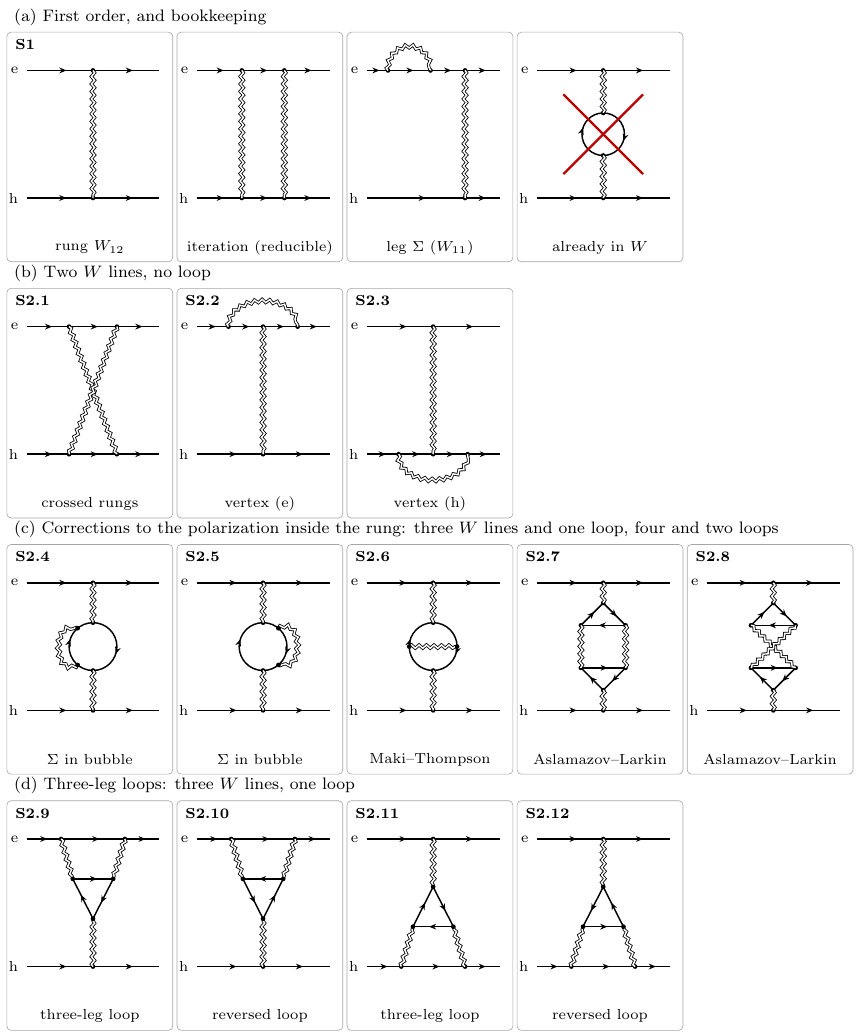}
\caption{Pairing-channel diagrams through second order in the screened interaction, with the order defined by Eq.~\eqref{eq:order}. Double wavy lines are $W_{12}$ between the layers and $W_{11}$ within a layer. (a) The first-order kernel S1; its iteration; the self-energy of the propagators, which is the Ward partner of S2.2 and S2.3; and the rung with a bare bubble, which is excluded because it is contained in $W$. (b)--(d) The twelve second-order diagrams of the irreducible kernel.}
\label{fig:diagscr}
\end{figure*}

\emph{First order.} One diagram, the screened rung S1, $I^{(1)}=W_{12}(q,i\nu)$ [Fig.~\ref{fig:diagscr}(a)].

\emph{Second order.} Twelve diagrams, shown in Fig.~\ref{fig:diagscr}(b)--(d). We obtained them by exhaustive computer enumeration of all pair-irreducible diagrams with two, three, and four $W$ lines: there are 3 with two lines and no loop, 7 with three lines and one loop, and 2 with four lines and two loops. They fall into three groups.

(i) Two $W$ lines and no loop: the crossed rungs S2.1, with two interlayer lines, and the vertex corrections S2.2 and S2.3, in which an intralayer line $W_{11}$ spans the interlayer vertex.

(ii) Corrections to the polarization inside the rung, $W\Pi^{(1)}W$: the two self-energy insertions S2.4 and S2.5, the Maki-Thompson diagram S2.6, and the two Aslamazov-Larkin diagrams S2.7 and S2.8. The loop is in either layer.

(iii) The three-leg loops S2.9--S2.12, in which a closed loop is attached by two lines to one external line and by one line to the other, in both orientations.

The first group is what line counting also gives. The other two groups are required by Eq.~\eqref{eq:order} and are absent from a naive count of screened lines.

\emph{Ward identities.} The vertex at which a $W$ line meets a fermion line is a density vertex, and charge conservation groups the diagrams into sets that are consistent only as a whole.

(a) S2.2 and S2.3 together with the self-energy $\Sigma=-\mathcal G W_{11}$ of the propagators. At $q=0$ and finite $\nu$ the vertex is given by Eq.~\eqref{eq:ward}. Where the self-energy is linear in frequency it equals $Z$ and cancels the factors $Z^{-1}$ of the propagators (Sec.~\ref{sec:ward}). In the opposite limit, $\nu\to0$ at fixed small $q$, the product of vertex and quasiparticle weight is $\chi/N^*$, the ratio of the compressibility to the quasiparticle density of states \cite{Nozieres}.

(b) S2.4, S2.5, and S2.6. The polarization must vanish at $q=0$ for $\nu\neq0$, and its high-frequency form $nq^2/m\nu^2$ must be unchanged ($f$-sum rule). The two self-energy insertions and the Maki-Thompson diagram satisfy this only as a sum.

(c) S2.7 and S2.8. The same identities. The two Aslamazov-Larkin diagrams cancel at $q=0$ only as a pair.

(d) S2.9 with S2.10, and S2.11 with S2.12. The two orientations of a three-leg loop cancel when one leg carries $q=0$ at $\nu\neq0$. When one leg carries $\nu=0$ and $q\to0$, their sum is the derivative of the two-leg loop with respect to the chemical potential, which is a part of the static density vertex of (a).

(e) S2.1 has no Ward partner. It is required by the order counting alone, and it is the only diagram with two interaction lines and no closed loop in which both lines are interlayer. With the loops summed over the layers, other topologies also have assignments with interlayer lines only, for example the Aslamazov-Larkin diagrams with their two loops in opposite layers.

\subsection{What our results contain}
\label{app:contain}

Table~\ref{tab:included} lists which diagrams enter each calculation.

\begin{table*}[t]
\caption{Diagrams of Figs.~\ref{fig:diagbare} and \ref{fig:diagscr} contained in the zero-field calculations of this paper and of Refs.~\cite{Zhu2024,Sreejith2024}. ``All orders'' means that the chain of bubbles is summed. The last column gives the section in which the diagram is evaluated or its effect is determined.}
\label{tab:included}
\begin{ruledtabular}
\begin{tabular}{lcccccl}
Diagram & Unscreened & Static & Dynamic & Eliashberg, & Crossed rung, & Where treated\\
 & ladder & ladder & ladder & Ref.~\cite{Sreejith2024} & Sec.~\ref{sec:vertex} & \\
\hline
B1; iterations & yes & yes & yes & yes & & Secs.~\ref{sec:analytic}, \ref{sec:numerics}\\
Bubble chains at $\nu=0$ (B2.4, B2.5, \dots) & no & all orders & all orders & all orders & & $\lams$, Eq.~\eqref{eq:lams}\\
Bubble chains, frequency dependence & no & no & all orders & all orders & & $\lampl$, Eq.~\eqref{eq:lampl}\\
Self-energy of the propagators & no\footnotemark[1] & no & no & yes & & Secs.~\ref{sec:ward}, \ref{app:larged}\\
Vertex corrections B2.2, B2.3; S2.2, S2.3 & no & no & no & no & & Secs.~\ref{sec:ward}, \ref{app:larged}\\
Crossed rungs B2.1; S2.1 & no & no & no & no & yes\footnotemark[2] & $\lambda_{\rm cr}$, $\lambda_{\rm GMB}$\\
Polarization corrections S2.4--S2.8 & no & no & no & no & & Sec.~\ref{app:larged}\\
Three-leg loops S2.9--S2.12 & no & no & no & no & & Sec.~\ref{app:larged}\\
\end{tabular}
\end{ruledtabular}
\footnotetext[1]{The HF theory of Ref.~\cite{Zhu2024} includes the unscreened exchange self-energy.}
\footnotetext[2]{Between states at the Fermi surface and at zero external frequency.}
\end{table*}

\emph{The ladder calculations} (Secs.~\ref{sec:analytic} and \ref{sec:numerics}) contain S1 and all its iterations, with bare propagators. The statically screened version keeps S1 at $\nu=0$, the dynamically screened version keeps its full frequency dependence, and the unscreened version replaces it by B1. In each case the eigenvalue problem is solved exactly, with no cutoff.

\emph{The self-energy and the vertex corrections S2.2, S2.3} are left out together. The reason is the Ward identity (a): they form one set, and keeping the self-energy alone, as in Eliashberg theory, produces an error of order unity in the region $\nu\gg\vF q$ from which the dynamical part of the ladder result comes (a factor as large as $Z^{-4}$ in $\lampl$, in the limit described in Sec.~\ref{sec:ward}). Their combined effect is small in neither limit of the identity. At large $d$ it is known in both: it cancels for $\nu\gg\vF q$ at frequencies at which the self-energy is linear, and for $\nu\ll\vF q$ it is expressed by Fermi-liquid parameters (Sec.~\ref{app:larged}). At general $r_s$ and $d$ it is not known, and leaving the set out is the statement that the calculation is of first order.

\emph{The crossed rungs S2.1} are evaluated in Sec.~\ref{sec:vertex}, with the full dynamically screened interaction on both lines, between states at the Fermi surface. This is the quantity that enters the exponent of a weak-coupling $T_c$. The result is $\lambda_{\rm cr}\simeq-\lampl$ from the frequency-dependent part and $\lambda_{\rm GMB}$ from the static part. We have not solved Eq.~\eqref{eq:BS} with S2.1 in the kernel at all frequencies.

\emph{The polarization corrections S2.4--S2.8 and the three-leg loops S2.9--S2.12} are not evaluated. The RPA uses the bare bubble. These diagrams are first order in the intralayer coupling relative to S1. Their effect is subleading at small $r_s$ and is absorbed in Fermi-liquid parameters at large $d$, as discussed next.

Table~\ref{tab:fate} summarizes the size of each contribution in the two limits.

\begin{table*}[t]
\caption{Contribution of each class of diagrams to the pairing coupling of the low-energy sector, and its fate in the two controlled limits. ``Subleading'' means that the term changes $\ln T_c$ by an amount that is small compared with the leading term in the limit, and therefore does not affect the leading asymptotic form; it need not be bounded. The entries for $d\to\infty$ assume that the cancellation found at second order holds at higher orders (Sec.~\ref{app:larged}).}
\label{tab:fate}
\begin{ruledtabular}
\begin{tabular}{llll}
Class & Coupling & $r_s\to0$ at fixed $d$ & $d\to\infty$ at fixed $r_s$\\
\hline
S1 at $\nu=0$ & $\lams=(\alpha/\pi)\,\cS(\cG)/\cG$ & leading term, $O(r_s)$ & leading term, $\propto r_s(\aB/d)^2$\\
S1 at $\nu\neq0$, iterated & $\lampl$ & $O(r_s^2)$; cancelled by S2.1 & $\propto d^{-1/2}$; cancelled by S2.1\\
S2.1, dynamical part & $\lambda_{\rm cr}=-\lampl+\lambda_{\rm net}$ & $\lambda_{\rm net}\to C_3(d)\,r_s^3$, Eq.~\eqref{eq:C3} & $\lambda_{\rm net}\to J_w\sqrt g\,r_s^4(\aB/d)^{7/2}$, Eq.~\eqref{eq:lamw}\\
S2.1, static part & $\lambda_{\rm GMB}\simeq-\lams^2[\ln\kF d+c]$ & subleading: $\hbar\vF/d\to\hbar^2/md^2$ & subleading: $\hbar\vF/d\to\hbar^2/md^2$\\
Self-energy, S2.2, S2.3 & $\lams\times O(\alpha\ln\alpha)$ & subleading (power of $r_s$) & cancel for $\nu\gg\vF q$; static part gives $m/m^*$\\
S2.4--S2.8 & $\lams\times O(\alpha\ln\alpha)$ & subleading (power of $r_s$) & change $\chi$, which drops out\\
S2.9--S2.12 & $\lams\times O(\alpha\ln\alpha)$ & subleading (power of $r_s$) & part of the static vertex; drops out with $\chi$\\
\end{tabular}
\end{ruledtabular}
\end{table*}

\subsection{The small-$r_s$ limit: leading-order perturbation theory}
\label{app:smallrs}

\emph{Statement.} For $r_s\to0$ at any fixed $d/\aB>0$,
\begin{equation}
\lim_{r_s\to0}\ \lams\,\ln\frac{\EF}{k_BT_c}=1,\qquad\lams=\frac{\alpha}{\pi}\,s(\cG),\quad s(\cG)=\frac{\cS(\cG)}{\cG},
\label{eq:smallrsthm}
\end{equation}
with $\alpha=\sqrt g\,r_s/2$ and $\cG=gd/\aB$. The function $s$ depends on the separation only. It interpolates between
\begin{equation}
s(\cG)\simeq\ln\frac{1}{2\cG}-\gamma\quad(\cG\ll1),\qquad s(\cG)\simeq\frac{\pi^2}{8\cG^2}\quad(\cG\gg1),
\label{eq:slimits}
\end{equation}
where $\gamma$ is Euler's constant; $s(1)=0.248$. This is the result of leading-order perturbation theory in the statically screened interaction: the first-order kernel S1 at $\nu=0$, iterated. The statement concerns the leading exponent. The form $\lams=(\alpha/\pi)s(\cG)$ requires $\kF d\gg1$, which holds in the limit at any fixed positive $d$.

\emph{Why the leading order is S1 at zero frequency.} By Sec.~\ref{app:bare} the bubbles must be summed, so the leading kernel is S1 and not B1. Its frequency dependence does not enter the leading logarithm. The logarithm of Eq.~\eqref{eq:BS} comes from states with $|\xi_p|$ and $|\omega_m|$ between $T_c$ and the cutoff, while the momentum transfer between two such states is fixed by the angle between them and is of order $1/d$. For all but a vanishing fraction of the logarithmic range the frequency transfer is small compared with $\vF q$, and S1 can be evaluated at $\nu=0$. Its Fermi-surface average is $\lams$.

\emph{Why the corrections do not change the limit.} Every other term is of higher order in $r_s$ (Table~\ref{tab:fate}). A correction $\delta\lambda$ changes $\ln T_c$ by $\delta\lambda/\lams^2$. The intralayer terms (the self-energy, S2.2, S2.3, S2.4--S2.12) are of relative order $\alpha\ln(1/\alpha)$, the logarithm being the one that appears in the effective mass of the two-dimensional electron gas at high density \cite{Galitski2004}. Then $\delta\lambda/\lams^2\sim\ln(1/\alpha)\,\pi/s(\cG)$, which grows logarithmically. The static crossed rung gives $\delta\lambda/\lams^2=-[\ln\kF d+c]$, which at fixed $d$ also grows as $\ln(1/r_s)$. Neither of these is bounded by a constant, but each is small compared with the leading term $1/\lams\propto1/r_s$. The net dynamical term is $C_3(d)\,r_s^3$ at fixed $d$, with the coefficient of Eq.~\eqref{eq:C3}, and gives $\lambda_{\rm net}/\lams^2=R(d)\,r_s$ [Eq.~\eqref{eq:Rd}], which vanishes in the limit. Equation~\eqref{eq:smallrsthm} follows. The remainder in $\ln T_c$ is of order $\ln(1/r_s)$: the prefactor of $T_c$ is a power of $r_s$.

\emph{Dependence on $d$, and the role of the crossed rung.} Two different statements must be kept apart.

At fixed $d$, the dynamically screened ladder has the same leading exponent as Eq.~\eqref{eq:smallrsthm}, because $\lampl/\lams\propto r_s\to0$. The crossed rung is not needed for the limit at fixed separation. The ladder is nevertheless not uniformly valid. It reaches the limit only for $r_s\ll0.64\,(\aB/d)^{3/2}$, and where it does it overestimates $T_c$ by $\exp(\lampl/\lams^2)$, a factor that does not depend on $r_s$ and is $\exp[\ladExcessCoef\,(d/\aB)^{7/2}]$ at large $d$ for $g=2$.

When $d$ is large as well, the ladder fails in the exponent: for $d>d_\times$ [Eq.~\eqref{eq:crossover}] its second-order term $\lampl$ exceeds its first-order term $\lams$, however small $r_s$ is, and it gives $\ln T_c\propto-\sqrt d/r_s^2$ where the first-order result is $\ln T_c\propto-d^2/r_s$. The crossed rung removes this: $\lambda_{\rm net}/\lams$, which at large $d$ is the sum of terms proportional to $r_s^3(\aB/d)^{3/2}$ and $r_s^2(\aB/d)^3$ with the coefficients of Appendix~\ref{app:remainder}, and $\lams\ln\kF d$ both vanish for $r_s\to0$ and for $d\to\infty$. With the crossed rung the first-order result holds along any path on which $r_s\to0$ and $\kF d\gg1$, and in this sense the statement that at small $r_s$ there is an exact result for every $d/\aB$ is correct. It refers to the statically screened theory, not to the HF theory.

The limit is not uniform as $d\to0$ in the form written. For $\kF d\lesssim1$ the coupling is the full Fermi-surface average of the static interaction [the first equality of Eq.~\eqref{eq:lams}], not $(\alpha/\pi)s(\cG)$. At $d=0$ it is $(\alpha/\pi)\ln(2/g\alpha)$, which behaves as $r_s\ln(1/r_s)$, whereas $s(\cG)$ diverges as $\ln(1/\cG)$. We expect Eq.~\eqref{eq:smallrsthm} to hold there with the full average, but the expressions of this paper for the second-order terms assume small-angle scattering and do not cover that regime.

\emph{The unscreened theory is not the small-$r_s$ limit.} The unscreened ladder gives $\ln T_c\simeq-\pi^{3/2}/2\sqrt\alpha$, which is larger than $-1/\lams$ by a factor that diverges as $r_s\to0$. It would be correct only if the forward singularity of the interaction survived down to momentum transfers of order $k_BT_c/\hbar\vF$, and static screening removes it at $q\lesssim\cG/d$, whatever the density. Dynamic screening restores the singularity for $\nu$ above the plasmon frequency, and the ladder converts this into $\lampl$, which the crossed rung cancels.

\emph{The prefactor.} Equation~\eqref{eq:smallrsthm} fixes the exponent. The prefactor requires every term of Table~\ref{tab:fate}, and we have computed two of them. One has a simple and exact structure. The static crossed rung is
\begin{equation}
\lambda_{\rm GMB}=-\lams^2\bigl[\ln\kF d+c\bigr],
\label{eq:gmblog}
\end{equation}
with a coefficient of the logarithm equal to unity. Our numerical results give a slope of \gmbslope\ when $\kF d$ is varied at fixed $d$, and $c$ between \cgmbA\ and \cgmbB\ for $5\le d/\aB\le20$. The logarithm has a direct interpretation. For scattering at small angle the crossed channel has the same logarithm as the Cooper channel, with the opposite sign, down to the energy at which the curvature of the Fermi surface distinguishes the two [Eq.~\eqref{eq:xisum}], which is
\begin{equation}
E_c=\frac{\hbar^2}{md^2}=\frac{\hbar\vF/d}{\kF d}.
\label{eq:Ec}
\end{equation}
Between $E_c$ and $\hbar\vF/d$ the two channels cancel, and the Cooper logarithm that determines $T_c$ extends only up to $E_c$. Equation~\eqref{eq:gmblog} expresses this: inserted in $1/(\lams+\lambda_{\rm GMB})\simeq1/\lams+\ln\kF d+c$, it replaces the cutoff $\hbar\vF/d$ of the statically screened ladder by $E_c$,
\begin{equation}
k_BT_c=C'\,\frac{\hbar^2}{md^2}\,\exp\Bigl[-\frac{1}{\lams}\Bigr].
\label{eq:Tcsmallrs}
\end{equation}
The crossed rung alone gives $C'=C_se^{-c}\approx0.2$ to $0.5$, where $C_s\approx0.7$ to $1.3$ is the prefactor of the statically screened ladder in units of $\hbar\vF/d$, obtained from the numerical solution. This is not the complete prefactor. The intralayer terms, which we have not evaluated, multiply $C'$ by a function of $d/\aB$ and, through the logarithm mentioned above, by a power of $r_s$. In units of $\EF$ the scale $\hbar^2/md^2$ is itself $2\EF/(\kF d)^2\propto r_s^2$.

\subsection{The large-$d$ limit: large energy denominators}
\label{app:larged}

\emph{Statement.} For $d\to\infty$ at any fixed $r_s$, under the two assumptions listed below,
\begin{equation}
\lim_{d\to\infty}\Bigl(\frac{\aB}{d}\Bigr)^{2}\ln\frac{\EF}{k_BT_c}=\frac{m^*}{m}\,\frac{16\,g^{3/2}}{\pi\,r_s},
\label{eq:largedthm}
\end{equation}
where $m^*$ is the quasiparticle effective mass of a single layer at the given $r_s$. The conditions are $\kF d\gg1$ and $d\gg\aB$. No smallness of $r_s$ is required. The assumptions are (A1) that each layer is a Fermi liquid, and (A2) that the cancellation between the Cooper channel and the crossed channel, which is exact at every order for linearized dispersion and which we have verified with curvature at second order, also holds for the curvature terms of third and higher order, in the precise sense that the sum $\delta\lambda$ of all contributions to the pair coupling beyond the static quasiparticle interaction satisfies $\delta\lambda/[(m/m^*)\lams]\to0$ as $d\to\infty$. Statement (1) below needs only (A1). Statements (2) and (3) are proven at second order and rest on (A2) beyond it.

\emph{The scales.} At large $d$ the problem has three energies below $\EF$, separated by the two large parameters,
\begin{equation}
\hbar\Ost:\frac{\hbar\vF}{d}:E_c=\sqrt{\cG}\,\kF d:\kF d:1,
\label{eq:hierarchy}
\end{equation}
with $\hbar\Ost/\EF=(g^2r_s^2\aB/d)^{1/2}$ and $k_BT_c$ far below all of them. The plasmon energy at $q=1/d$ is the largest, the range $\hbar\vF/d$ of the interlayer interaction in energy comes next, and the curvature energy $E_c$ of Eq.~\eqref{eq:Ec} is the smallest. The exciton binding energy is $E_d=\alpha\,\hbar\vF/d$. For $r_s=1$ and $d=8\aB$, in units of $\EF$, $\hbar\Ost=0.71$, $\hbar\vF/d=0.18$, $E_c=0.016$, and $E_d=0.13$.

\emph{The argument.} Three statements are needed.

(1) \emph{Below $E_c$ the pairing is that of the statically screened interaction.} For pair states within $E_c$ of the Fermi surface the frequency transfer is at most $E_c$, which is smaller than $\hbar\vF q$ at the relevant $q\sim1/d$ by the factor $1/\kF d$. The interaction is S1 at $\nu=0$, the crossed channel has no logarithm, and the Cooper channel gives $\ln(E_c/k_BT_c)=1/\lambda$ with $\lambda$ the Fermi-surface average of the static quasiparticle interaction.

(2) \emph{Processes between $E_c$ and $\hbar\vF/d$ do not renormalize $\lambda$.} All interlayer momentum transfers are $q\lesssim1/d\ll\kF$, so the scattering is at small angle. The electron of a pair and its hole move in opposite directions, so a small momentum transfer changes their energies by opposite amounts, and Eq.~\eqref{eq:Bnet} then gives $B_{\rm cr}=-B_{\rm lad}$ at every frequency, up to terms of relative order $E_c^2/(\nu^2+\xi^2)$. On these scales a patch of the Fermi surface and the opposite patch form a one-dimensional system of counterpropagating fermions with a density-density interaction, in which the Cooper logarithm is cancelled by the crossed channel at every order when the dispersion is linearized (Sec.~\ref{sec:adiabatic}) \cite{DzyaloshinskiiLarkin1974,Kopietz1997,Metzner1998}. Equation~\eqref{eq:gmblog} is the second-order term of this cancellation with the curvature included, and it is the term we have verified explicitly, including the momentum integration: the logarithm is regulated by the curvature at $E_c$ and its coefficient is unity. Beyond second order the statement rests on (A2).

\begin{figure}[t]
\includegraphics[width=\columnwidth]{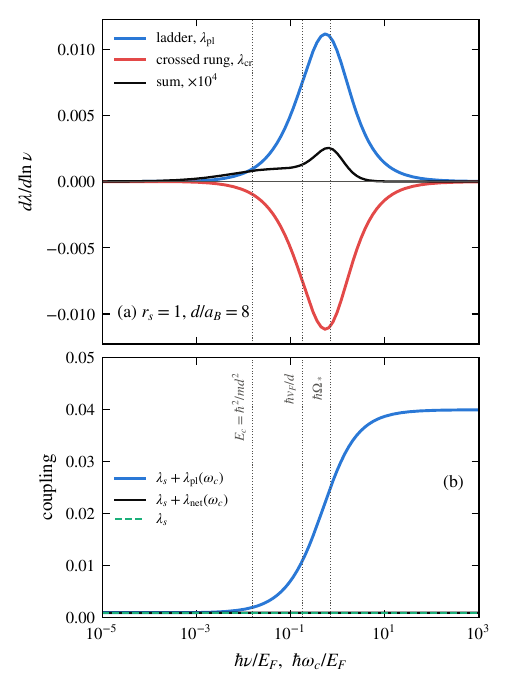}
\caption{Where the second-order couplings come from, for $r_s=1$ and $d=8\aB$. (a) Contribution per logarithmic interval of the intermediate frequency $\nu$ to the plasmon-induced coupling of the ladder, to the crossed rung, and to their sum, which is multiplied by $10^4$. (b) Coupling of the low-energy sector when the frequency integral is cut off at $\omega_c$, in the ladder and with the crossed rung. The vertical lines mark the three scales of Eq.~\eqref{eq:hierarchy}.}
\label{fig:dlam}
\end{figure}

(3) \emph{Processes at and above the plasmon energy do not renormalize $\lambda$.} These are the processes of the wedge, Eq.~\eqref{eq:wedge}, in which the pair is scattered into virtual states at frequencies $\nu\gtrsim\omega_+(q)$. Their energy denominators, $\hbar\nu\gtrsim\hbar\omega_+(q)$, are large compared with the energy $\hbar\vF q$ that the carriers can exchange at the same wave vector. The comparison is local in $q$, since the plasmon is gapless: the ratio $\omega_+(q)/\vF q$ is of order $\sqrt{\cG}$ at $q=1/d$ and increases as $q\to0$, so the condition holds uniformly at large $d$. On the time scale of such a process the carriers do not move, and for static carriers the dynamical part of the interaction changes the amplitude of the pair but not its energy, Eq.~\eqref{eq:static}. In the ladder alone these processes give $\lampl$; with the crossed diagrams they give $\lambda_{\rm net}$, which at second order is smaller by a factor of order $1/[\cG(\kF d)^2]$ (Fig.~\ref{fig:dlam}) and whose limit is given in Eq.~\eqref{eq:lamw}.

The same conclusion can be reached from the interaction itself. For $\nu\ll\omega_\pm(q)$ both dielectric functions are large, $\varepsilon_\pm\simeq(V_q\pm U_q)N_0P$, and Eq.~\eqref{eq:W12} becomes
\begin{equation}
W_{12}(q,i\nu)\simeq-\frac{1}{N_0^2P(q,\nu)^2}\,\frac{U_q}{V_q^2-U_q^2}=\frac{W_{12}(q,0)}{P(q,\nu)^2}.
\label{eq:Wslow}
\end{equation}
The plasmon frequencies have dropped out. What remains is the capacitive interaction between the layers divided by the square of the polarization of the particle-hole continuum, which is unity for $\nu\ll\vF q$. The plasmons have been eliminated, and their only trace is the static screening they provide.

Statements (1)--(3) leave the static quasiparticle interaction as the only input. Because $q\ll\kF$ it is given by Fermi-liquid theory, Sec.~\ref{sec:control}: the self-energy and the vertex corrections S2.2 and S2.3 combine into the static density vertex $\chi/N^*$, the three-leg loops are part of the same vertex, the polarization corrections S2.4--S2.8 replace $N_0$ by the exact compressibility $\chi$ in the dielectric functions, and for $(V_q\pm U_q)|\chi|\gg1$ the compressibility cancels between the vertices and the screening. The result is $\lambda=(m/m^*)\lams$ and Eq.~\eqref{eq:largedthm}. Including the prefactor from statement (1),
\begin{equation}
k_BT_c=C''\,\frac{\hbar^2}{md^2}\,\exp\Bigl[-\frac{m^*}{m}\,\frac{16\,g^{3/2}}{\pi\,r_s}\Bigl(\frac{d}{\aB}\Bigr)^2\bigl(1+\delta\bigr)\Bigr],
\label{eq:Tclarged}
\end{equation}
where $\delta\to0$ as $d\to\infty$ under assumption (A2); the corrections to $\cS(\cG)\simeq\pi^2/8\cG$ and to the Fermi-surface average contribute $O(\aB/d)+O(1/\kF d)$ to it. Since $\delta$ multiplies a large exponent it is not a correction to the prefactor, and Eq.~\eqref{eq:largedthm} is the precise form of the statement.

\emph{What is and is not established.} The identity of Eq.~\eqref{eq:Bnet}, the numerical cancellation of $\lampl$ by the crossed rung, and the logarithm of Eq.~\eqref{eq:gmblog} are results. The reduction of the intralayer terms to $m^*$, with the compressibility cancelling, follows from (A1) and the Ward identities at $q\ll\kF$. That the terms of third and higher order leave a contribution to the pair coupling that is small compared with $\lams\propto d^{-2}$, and not merely one that decays no more slowly, is (A2); a remainder proportional to $d^{-2}$ would change the coefficient in Eq.~\eqref{eq:largedthm}. It is supported by the exact solution for linearized dispersion and by the second-order results, and we know of no mechanism that would violate it, but we have not computed a third-order term. Equation~\eqref{eq:largedthm} is exact if (A2) holds.

\emph{Relation to Migdal's theorem.} The large energy denominators are the reason the limit is controlled, and it is worth stating precisely which theorem they give. There are two ways in which the intermediate states of a vertex correction can be far off the energy shell.

In Migdal's theorem the large denominator is electronic. The boson is slow, $\omega\ll\vF q$, the intermediate electron is off shell by $\hbar\vF q$, and the vertex correction is small, of order $\omega/\vF q\sim\omega_D/\EF$ \cite{Migdal1958}. The interaction remains retarded, and Eliashberg theory, the ladder without vertex corrections, is the leading-order theory.

Here the large denominator is bosonic. The boson is fast, $\omega_\pm\gg\vF q$ by Eq.~\eqref{eq:fast}, and the intermediate states are off shell by the plasmon energy. A process with a large bosonic denominator is instantaneous on the time scale of the carriers. That makes the interaction static, which is the result. It does not make the vertex corrections small. In the expansion in the dynamically screened interaction they are of order unity, $\lambda_{\rm cr}=-\lampl$, and the large denominators are what allows them to be summed. The corresponding statement for phonons is the canonical transformation that eliminates a fast boson and leaves an instantaneous interaction \cite{LangFirsov1963}, and the known breakdown of Migdal's theorem at $q<\omega/\vF$ \cite{Grimaldi1995} is the same physics.

The two theorems are mirror images, and both are realized in bilayers: Migdal's in the all-electron bilayer with a large ratio of layer masses \cite{Wang2026}, where the acoustic plasmon of the heavy layer is slow for the light carriers, and the present one in the equal-mass electron-hole bilayer, where both plasmons are fast. Both rest on a ratio of the boson frequency to $\vF q$, small in one and large in the other, and each fixes the leading exponent. They differ in the leading-order theory they select: Eliashberg theory for Migdal's limit, and BCS theory with the statically screened interaction for the present one. The smallness of $\hbar\Ost/\EF$, that is, of $E_B/\EF$, is what places the whole problem inside the domain of Fermi-liquid theory and guarantees $\kF d\gg1$ and $d\gg\aB$ when $r_s\gtrsim1$. In that sense $E_B/\EF\ll1$ is the condition for the controlled theory. It is not the condition $\omega\ll\vF q$ of Migdal's theorem, which is violated by the factor $\sqrt{\cG}$.

\emph{How fast the limit is approached.} The cancellation of the dynamical terms is complete to within about one percent of $\lampl$ already at $\kF d=2$ (Table~\ref{tab:vertex}). The static coupling reaches its asymptotic form more slowly: $\cS(\cG)\cG$ is within $10\%$ of $\pi^2/8$ only for $\cG\gtrsim\cGten$, that is, $d\gtrsim17\,\aB$ for $g=2$, and within $27\%$ at $\cG=10$. The effective mass in Eq.~\eqref{eq:largedthm} is a property of the single layer and is close to unity for $r_s\lesssim2$.

\subsection{The two crossovers between weak pairing and Bose condensation}
\label{app:crossover}

The linearized zero-field theory of this paper applies when the pairs are large and weakly bound. In the opposite regime the electron and the hole form an exciton, and the system is a gas of dipolar bosons. The two regimes are connected without a symmetry change \cite{Keldysh1965,Comte1982,Leggett1980,Nozieres1985}, and in the $(r_s,d/\aB)$ plane the connection can be made along two different axes. The two crossovers are physically distinct, and we discuss them in turn.

\subsubsection{Two-body scales and many-body parameters}

An isolated pair has three properties that matter (Sec.~\ref{sec:model} and Appendix~\ref{app:exciton}). Its binding energy $E_B(d)$ falls from $2\,{\rm Ry}$ at $d=0$ to $E_d=e^2/\epsilon d$ at large $d$. Its in-plane size $a_X=\langle r^2\rangle^{1/2}$, which we obtain from the same numerical solution, grows from $1.22\,\aB$ at $d=0$ to $(2\aB d^3)^{1/4}$ at large $d$, where the relative motion is harmonic with level spacing $\hbar\omega_0=2\sqrt2\,{\rm Ry}\,(\aB/d)^{3/2}$. And it carries a dipole moment $ed$, so two excitons repel at large distance as $e^2d^2/\epsilon r^3$, with a dipolar length $r_0=2d^2/\aB$ for bosons of mass $2m$.

At density $n$ these combine into three dimensionless parameters,
\begin{equation}
\frac{E_B}{\EF}=\frac{g\,r_s^2}{4}\,\frac{E_B}{\rm Ry},\qquad \pi na_X^2=\frac{a_X^2}{r_s^2\aB^2},\qquad nr_0^2=\frac{4d^4}{\pi r_s^2\aB^4}.
\label{eq:becparams}
\end{equation}
The first compares the binding of a pair with the kinetic energy that Pauli exclusion imposes on its constituents. The second measures the overlap of neighboring pairs. In the harmonic regime $\pi na_X^2=g\,\EF/\hbar\omega_0$, so it also compares the Fermi energy with the internal excitation energy of an exciton. The third is the strength of the interaction between excitons.

\begin{figure*}[t]
\includegraphics[width=0.96\textwidth]{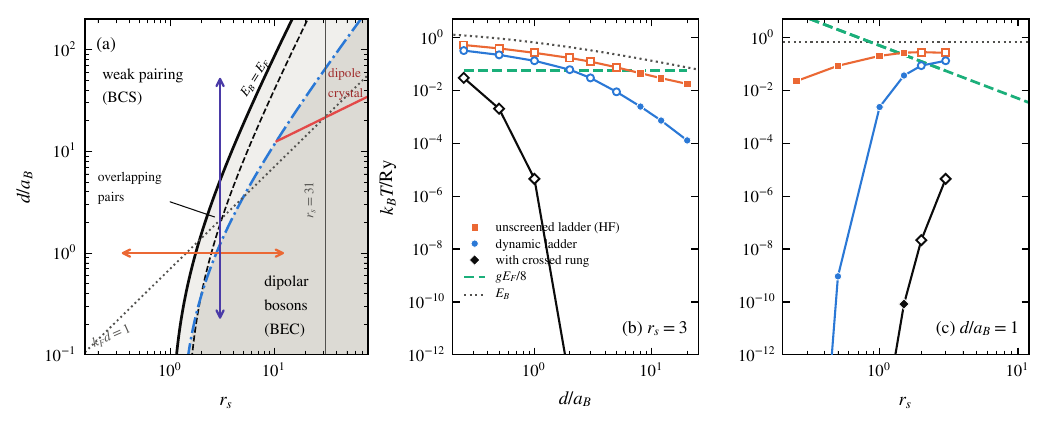}
\caption{(a) Regimes of the balanced electron-hole bilayer for $g=2$. Thick solid line: $E_B=\EF$. Dashed: $E_B=2\EF$, where the chemical potential changes sign in the short-range model. Dash-dotted: $\pi na_X^2=1$, with $a_X$ the size of the isolated exciton. Dotted: $\kF d=1$. Red line: $nr_0^2=290$, above which the dipolar bosons crystallize \cite{Astrakharchik2007}. Thin vertical line: Wigner crystallization of an isolated layer \cite{TanatarCeperley1989,DrummondNeeds2009}. The arrows are the two crossovers discussed in the text. (b), (c) Temperature scales along the two crossovers, in units of ${\rm Ry}$: (b) versus $d$ at $r_s=3$, (c) versus $r_s$ at $d=\aB$. Squares and circles: mean-field $T_c$ of the unscreened and of the dynamically screened ladder. Diamonds: weak-coupling estimate with the crossed rung, $0.5\,(\hbar\vF/d)\exp(-1/\lambda_{\rm eff})$ with $\lambda_{\rm eff}$ of Eq.~\eqref{eq:lamnet}. Open symbols: $E_B>\EF$, where the linearized theory at fixed chemical potential is not valid. Dashed line: upper bound $g\EF/8$ on the Kosterlitz-Thouless temperature. Dotted line: $E_B$. These are scales, not a calculation of $T_c$ in the crossover.}
\label{fig:regimes}
\end{figure*}

Figure~\ref{fig:regimes}(a) shows the lines on which these parameters are of order unity. They define three regimes.

(I) \emph{Weak pairing}, $E_B\ll\EF$. The Fermi surfaces exist, the pairs are much larger than their spacing, and the zero-field theory of this paper applies. At large $d$ the boundary $E_B=\EF$ is $d=(g/2)\,r_s^2\,\aB$.

(II) \emph{Overlapping pairs}, $E_B\gtrsim\EF$ but $\pi na_X^2\gtrsim1$. The binding energy exceeds the Fermi energy, yet the pairs overlap and their internal level spacing is below the Fermi energy. Neither description is quantitative. At large $d$ this regime lies between $d\simeq0.8\,r_s^{4/3}\aB$ and $d=(g/2)r_s^2\aB$, and it widens with increasing $r_s$.

(III) \emph{Dipolar bosons}, $\pi na_X^2\ll1$. The excitons are rigid, and the system is a two-dimensional Bose gas with dipolar repulsion. For $nr_0^2\gtrsim290$ the bosons crystallize \cite{Astrakharchik2007}; within regime III this requires $r_s\gtrsim10$ and $d\gtrsim13\,\aB$, and the crystal then lies between the line $d/\aB=3.9\sqrt{r_s}$ and the boundary of regime III [Fig.~\ref{fig:regimes}(a)]. For $d$ below about $0.9\,\aB$ for equal masses the repulsion is too weak to prevent the formation of biexcitons \cite{Meyertholen2008}, and the dilute phase is then a gas of biexcitons.

For a short-range attraction in two dimensions the mean-field crossover is governed by $E_B/\EF$ alone, with the chemical potential $\mu=\EF-E_B/2$ changing sign at $E_B=2\EF$ \cite{Miyake1983,Randeria1989}. For the long-range interaction considered here the size of the pair is an independent length, and the intermediate regime II has no counterpart in the short-range problem.

\subsubsection{What is known at the two ends}

\emph{Weak-pairing end.} $T_c$ is given by Eq.~\eqref{eq:Tcsmallrs} at small $r_s$ and by Eq.~\eqref{eq:Tclarged} at large $d$. The mean-field and Kosterlitz-Thouless temperatures coincide up to corrections of relative order $k_BT_c/\EF$.

\emph{Bose end.} The mean-field temperature at which pairs form is of order $E_B$ and has nothing to do with the transition. Superfluidity is lost by the unbinding of vortices, at
\begin{equation}
k_BT_{\rm KT}=\frac{\pi}{2}\,\frac{\hbar^2n_s}{2m}\le\frac{g}{8}\,\EF,
\label{eq:TKT}
\end{equation}
where $n_s\le n$ is the superfluid density of excitons of mass $2m$ at the transition \cite{NelsonKosterlitz1977}. Equation~\eqref{eq:TKT} is an upper bound and not an estimate. For a weakly interacting Bose gas $k_BT_{\rm KT}=(2\pi\hbar^2n/M)/\ln(\xi/\tilde g)$ with $\xi\simeq380$ and $\tilde g$ the dimensionless coupling \cite{Prokofev2001}, so the ratio of $T_{\rm KT}$ to the bound is $4/\ln(\xi/\tilde g)$. In the dilute limit $\tilde g$ itself vanishes logarithmically with the density, and the ratio tends to zero as the inverse of a double logarithm \cite{FisherHohenberg1988}. At $\tilde g=1$ it is about $0.7$, and it approaches unity for stronger interaction, where the weak-coupling formula no longer applies. The bound refers to the phase of the pair field; the larger order-parameter manifold that exists for $g>1$ can only lower the transition.

\emph{The interaction itself crosses over.} In regime III the excitons are neutral and there are no free carriers, so the interlayer attraction that binds a pair is unscreened. The unscreened ladder is then the Schr\"odinger equation of the pair, and HF theory gives the exact two-body binding energy. This does not make HF theory a controlled calculation of the transition temperature of the exciton gas, which is governed by Eq.~\eqref{eq:TKT} and by the interaction between excitons. In regime I the attraction is screened, and by Secs.~\ref{app:smallrs} and \ref{app:larged} the screening is static. Between the two limits the appropriate interaction changes from one to the other, because the polarization of the layers is suppressed at energies below the gap once a gap opens \cite{Perali2013,LopezRios2018}. This is the origin of the sensitivity of the energetics to the treatment of screening found in Ref.~\cite{Tu2024}. The unscreened HF theory and the statically screened theory are therefore not competing approximations to one problem. Each describes the pairing interaction correctly at one end, the unscreened one for the formation of pairs on the Bose side and the statically screened one on the weak-pairing side. The dynamically screened ladder interpolates too (Fig.~\ref{fig:tcd}), but for a reason, $\lampl$, that the crossed rung removes on the weak-pairing side.

\subsubsection{Crossover as a function of $d$ at fixed $r_s$}

Here the density is fixed and the two-body problem changes: $E_B$, $a_X$, and $r_0$ all depend on $d$, while $\EF$ and the bound of Eq.~\eqref{eq:TKT} do not.

For $r_s<(2/g)^{1/2}$ there is no crossover. Even at $d=0$ the binding energy $2\,{\rm Ry}$ is below $\EF=4\,{\rm Ry}/gr_s^2$, and the system is in the weak-pairing regime for every $d$. As $d$ decreases the coupling grows only logarithmically, Eq.~\eqref{eq:slimits}, and saturates when $\kF d\sim1$.

For $r_s>(2/g)^{1/2}$ the binding energy exceeds $\EF$ at small $d$, and regimes II and I are traversed with increasing $d$. Regime III is reached at small $d$ only if in addition the pairs do not overlap there, $\pi na_X^2<1$, which at $d=0$ requires $r_s>1.22$; for $g=2$ and $1<r_s<1.22$ the sequence is II, I. For larger $r_s$ the order is III, II, I [vertical arrow in Fig.~\ref{fig:regimes}(a)]. On the Bose side, where the exciton gas is a superfluid, $T_c$ is bounded by $g\EF/8$, is a sizable fraction of it, and depends on $d$ only through the interaction between excitons. On the weak-pairing side it is given by Eq.~\eqref{eq:Tclarged} and falls as a Gaussian in $d$. The fall sets in when $E_B$ drops below $\EF$, at $d\simeq(g/2)r_s^2\aB$, and it is very steep: at $r_s=3$ the weak-coupling exponent $1/\lambda_{\rm eff}$ is already $355$ at $d=8\aB$ (Table~\ref{tab:vertex}), where $E_B/\EF=0.72$. Figure~\ref{fig:regimes}(b) shows the scales. As a function of $d$ at fixed density we therefore expect the transition temperature to have a plateau followed by a collapse over a narrow range of $d$. This is a physical estimate. Neither the height of the plateau nor the form of the collapse through regime II is calculated here, and the picture assumes that no other phase intervenes. In an experiment at finite temperature and with finite disorder such a collapse is indistinguishable from a transition at a critical separation, although within the present theory it is a crossover and the ground state is paired on both sides.

Two features are specific to this crossover. The weak-pairing end is controlled for every $r_s$ in the liquid phase, by the large energy denominators of Sec.~\ref{app:larged}, and not only for small $r_s$. And the Bose end is not a simple limit. For $d\lesssim0.9\,\aB$ the dipolar repulsion is too weak, the excitons attract and form biexcitons \cite{Meyertholen2008}, and at $d=0$ the model is a single electron-hole plasma. For $r_s\gtrsim10$ the path also crosses the region of the dipole crystal, between $d/\aB\simeq3.9\sqrt{r_s}$ and the boundary of regime III.

\subsubsection{Crossover as a function of $r_s$ at fixed $d$}

Here the two-body problem is fixed, $E_B$, $a_X$, and $r_0$ are constants, and the density changes [horizontal arrow in Fig.~\ref{fig:regimes}(a)]. This is the crossover familiar from cold atomic gases, with $E_B/\EF\propto r_s^2$ as the tuning parameter.

Both ends are controlled, for $d$ above the biexciton threshold of about $0.9\,\aB$. For $r_s\to0$ the result is Eq.~\eqref{eq:Tcsmallrs}, with an essential singularity $\ln T_c\propto-1/r_s$. For $r_s\to\infty$ all three parameters of Eq.~\eqref{eq:becparams} behave as required for a dilute Bose gas, $\pi na_X^2\to0$ and $nr_0^2\to0$, and $T_c$ lies below the bound of Eq.~\eqref{eq:TKT}, which is ${\rm Ry}/2r_s^2$ for any $g$, by the slowly varying factor given there. In units of the Rydberg the transition temperature thus rises as $\exp(-{\rm const}/r_s)$ at high density and falls as $r_s^{-2}$, up to the double logarithm, at low density. It must therefore have a maximum in between, and we expect it where $E_B\sim\EF$ [Fig.~\ref{fig:regimes}(c)]. Since it is bounded by $(g/8)\EF$, our estimate for the maximum is a fraction of the exciton binding energy, of order $0.1\,E_B(d)$, which at large $d$ is of order $0.1\,e^2/\epsilon d$ near $r_s\approx(2d/g\aB)^{1/2}$. Neither the position nor the height of the maximum is determined by the calculations of this paper.

The path is not free of other phases at every $d$. By Eq.~\eqref{eq:becparams} the dipolar bosons are on the crystalline side of $nr_0^2=290$ for $r_s<2(d/\aB)^2/\sqrt{290\pi}=0.066\,(d/\aB)^2$. For $d\gtrsim13\,\aB$ this exceeds the value of $r_s$ at which regime III begins, and the path at fixed $d$ then passes through the crystal before it reaches the dilute superfluid at larger $r_s$. For example, at $d=100\,\aB$ and $r_s=500$ one has $\pi na_X^2\simeq0.006$ and $nr_0^2\simeq510$. For $d\lesssim13\,\aB$ no crystal of excitons intervenes.

Three features are specific to this crossover. The screening parameter $\cG=gd/\aB$ does not change along it, so on the weak-pairing side the interaction does not become unscreened as the density increases; only the coupling $\alpha$ decreases. The dimensionless separation $\kF d\propto1/r_s$ does change, so the weak-pairing end is always in the regime of small-angle scattering. And on the Bose side the exciton gas becomes more dilute and more weakly interacting as $r_s$ increases, so that the limit $r_s\to\infty$ is a superfluid, although a crystal can intervene on the way, as just described.

\subsubsection{Comparison}

The two crossovers differ in what is tuned, in which ends are controlled, and in the functional form of the weak-pairing side:
\begin{align}
&d\ \text{at fixed}\ r_s:\quad \ln\frac{k_BT_c}{\EF}\simeq-\frac{m^*}{m}\frac{16g^{3/2}}{\pi r_s}\Bigl(\frac{d}{\aB}\Bigr)^2,\nonumber\\
&r_s\ \text{at fixed}\ d:\quad \ln\frac{k_BT_c}{\EF}\simeq-\frac{2\pi}{\sqrt g\,s(\cG)}\,\frac1{r_s}.
\label{eq:twocross}
\end{align}
In both, $T_c$ in the controlled weak-pairing regime is many orders of magnitude below $\EF$. Where in the crossover it becomes observable is not determined by our calculation; the numbers of Table~\ref{tab:vertex} suggest that this requires $E_B$ comparable to $\EF$ or larger. This is consistent with the conclusion of quantum Monte Carlo calculations that the condensate is appreciable only at low density and small separation \cite{DePalo2002,LopezRios2018}. The present theory does not describe the crossover region itself, where the suppression of screening by the gap must be treated self-consistently \cite{Perali2013}; what it provides is the asymptotic form of the weak-pairing side of both crossovers.

\subsection{What a renormalization-group treatment captures}
\label{app:rg}

The RG of Polchinski and Shankar \cite{Polchinski1992,Shankar1994} is the natural framework for the statement that an attractive interaction guarantees pairing, and it is useful to say precisely which parts of the present problem it covers.

\subsubsection{The framework and what it captures}

The RG is formulated for fermions within a shell of width $\Lambda\ll\EF$ around the Fermi surface, and it follows the couplings as $\Lambda$ is reduced. For the bilayer the fermions are the electrons of layer 1 and the holes of layer 2, with identical Fermi surfaces. At tree level the only interactions that are not irrelevant are those between particles whose momenta are constrained to the Fermi surface in one of two ways: forward scattering, within a layer and between the layers, and the scattering of an electron-hole pair of zero total momentum into another such pair. The latter defines the pair couplings
\begin{equation}
\lambda_l=-\frac{N^*}{g}\int_0^{2\pi}\!\frac{d\theta}{2\pi}\,W^{\rm qp}_{12}\Bigl(2\kF\sin\frac\theta2\Bigr)\cos l\theta,
\end{equation}
one for each angular momentum $l$. At one loop the forward-scattering functions do not flow, and
\begin{equation}
\frac{d\lambda_l}{dt}=\lambda_l^2,\qquad t=\ln\frac{\Lambda_0}{\Lambda},
\label{eq:rgflow}
\end{equation}
so that an attractive coupling, $\lambda_l>0$ in our convention, grows and diverges at $k_BT_c\simeq\Lambda_0\exp[-1/\lambda_l(\Lambda_0)]$, while a repulsive one flows to zero. All other one-loop contributions to the flow, including the crossed channel, are suppressed by $\Lambda/\EF$.

The framework captures the following.

(a) \emph{The inevitability of pairing at weak coupling.} If the pair coupling at the scale $\Lambda_0$ is attractive in any channel, the normal state is unstable, however small the coupling. This is the rigorous content of the argument that an attractive electron-hole interaction guarantees superfluidity. It requires the two Fermi surfaces to coincide. A mismatch $\delta\mu$ stops the flow at $\Lambda\sim\delta\mu$.

(b) \emph{The form of $T_c$.} The exponent $-1/\lambda(\Lambda_0)$ and the fact that only the Fermi-surface average of the interaction in the given channel enters it.

(c) \emph{The asymptotic irrelevance of everything else.} Below a sufficiently small $\Lambda$ the flow is that of Eq.~\eqref{eq:rgflow} exactly. Vertex corrections, crossed diagrams, and the frequency dependence of the couplings affect only the initial condition.

(d) \emph{The contrast with the single layer.} For a repulsive bare interaction all $\lambda_l(\Lambda_0)$ are negative at leading order and flow to zero. Pairing can then come only from terms of second order in the interaction in the initial condition, as in the mechanism of Kohn and Luttinger \cite{KohnLuttinger1965}, or from the dynamics of screening, as in the plasmon mechanism, and neither is part of the leading-order RG. In the bilayer the attraction is present at leading order.

\subsubsection{What it does not capture}

(a) \emph{The initial condition.} The flow of Eq.~\eqref{eq:rgflow} begins at a scale $\Lambda_0$ at which the couplings must be supplied. They are the result of eliminating all modes between $\EF$ and $\Lambda_0$, and this elimination is the entire content of the zero-field part of this paper: the screening, the plasmons, the crossed rung, the Fermi-liquid renormalizations. The RG does not perform it. In particular, the sign of $\lambda(\Lambda_0)$ at general $r_s$ is not determined by the sign of the bare interaction; it is determined by the bare interaction only when the elimination is controlled. It is controlled, and the result is attractive, at small $r_s$, where $\lambda(\Lambda_0)=\lams$ to leading order. At large $d$ for any $r_s$ we have argued that $\lambda(\Lambda_0)=(m/m^*)\lams$, under the assumptions stated in Sec.~\ref{app:larged}; this result does not depend on the sign of the compressibility of the layers.

(b) \emph{The long range of the Coulomb interaction.} The tree-level analysis assumes coupling functions that are smooth on the Fermi surface. The bare interlayer interaction is singular in the forward direction, and the first-order pair coupling does not exist (Sec.~\ref{app:bare}). If the singular interaction is kept unscreened, the flow equation acquires a source term,
\begin{equation}
\frac{d\lambda}{dt}=\lambda^2+A,\qquad A=\frac\alpha\pi,
\end{equation}
as for the color-magnetic interaction in quark matter \cite{Son1999}. Its solution $\lambda=\sqrt A\tan(\sqrt A\,t)$ diverges at $\sqrt A\,t=\pi/2$, which is the unscreened result of Eq.~\eqref{eq:Tc0}. Screening must be included before the RG is set up, and the plasmons, which are collective modes of the same fermions at energies well above the shell, are not among its degrees of freedom.

(c) \emph{The scale at which the flow begins.} The suppression of the crossed channel by $\Lambda/\EF$ presupposes couplings that vary on the scale of $\kF$. At large $d$ the pair coupling varies over the angle $1/\kF d$, and the suppression factor is $\Lambda/E_c$ with $E_c=\hbar^2/md^2$. For $E_c<\Lambda<\hbar\vF/d$ the crossed channel is not suppressed and cancels the Cooper channel, so the coupling does not flow [statement (2) of Sec.~\ref{app:larged}]. The flow of Eq.~\eqref{eq:rgflow} begins at $\Lambda_0\sim E_c$ and not at $\hbar\vF/d$. An RG that starts the Cooper flow at $\hbar\vF/d$ overestimates $T_c$ by a factor of order $\kF d$.

(d) \emph{Retardation by a fast boson.} For phonons the RG reproduces Eliashberg theory, because the coupling becomes marginal only below the Debye energy and Migdal's theorem controls the elimination above it \cite{Polchinski1992,Tsai2005}. A one-loop flow in the Cooper channel alone, with the frequency-dependent coupling $W_{12}(q,i\nu)$ and started at $\EF$, is the RG form of the dynamically screened ladder. It generates $\lampl$ as $\Lambda$ passes through the plasmon energy [upper curve in Fig.~\ref{fig:dlam}(b)]. At those scales the crossed channel is of the same size, and the flow of the two together generates nothing [lower curve]. The correct elimination of the plasmons is the one described in Sec.~\ref{app:larged}.

(e) \emph{The Bose regime and the crossovers.} The RG is an expansion around the Fermi-surface fixed point. When $E_B\gtrsim\EF$ the coupling is of order unity at the scale $\EF$, there is no range of scales in which Eq.~\eqref{eq:rgflow} applies, and the composite bosons, their dipolar interaction, and the crossover of Sec.~\ref{app:crossover} are outside the framework.

(f) \emph{Fluctuations of the order parameter and first-order transitions.} The divergence of the coupling gives the mean-field instability. The Kosterlitz-Thouless physics is beyond it, although at weak coupling this is a small correction. A first-order transition to a different state, such as crystallization, is not signaled by the flow, and its absence is an assumption.

\subsubsection{The sequence of scales at large $d$}

The elimination of modes at large $d$ can be summarized as a sequence, which Fig.~\ref{fig:dlam} illustrates.

From $\EF$ to $\hbar\Ost$: the carriers of each layer become Landau quasiparticles with mass $m^*$; the interlayer pair coupling is the bare one in the wedge and the screened one outside it.

Through $\hbar\Ost$: the plasmons are eliminated. The Cooper channel and the crossed channel each generate a coupling of magnitude $\lampl$, with opposite signs. The interaction between the quasiparticles becomes Eq.~\eqref{eq:Wslow}.

From $\hbar\vF/d$ to $E_c$: the interaction is static, the scattering is at small angle, and the pair coupling does not flow.

Below $E_c$: Eq.~\eqref{eq:rgflow}, with $\lambda(E_c)=(m/m^*)\lams$, up to $k_BT_c=C''E_c\exp[-1/\lambda(E_c)]$.

\subsubsection{Bearing on the existence of superfluidity at all $r_s$ and $d$}

The physical argument that the bilayer is superfluid at zero temperature for every $r_s$ and $d$, because the bare electron-hole interaction is attractive at every wave vector, can now be placed. The RG shows that it is sufficient for the renormalized pair coupling at low energy to be attractive. We find it attractive in every regime in which it can be computed: at small $r_s$ at fixed $d$; at large $d$ for all $r_s$ in the liquid phase, under the assumptions of Sec.~\ref{app:larged}, where the result holds whatever the compressibility; and on the Bose side, where the pairs are bound. Within the RPA the first-order kernel is attractive at all $r_s$ and $d$, by Eq.~\eqref{eq:sign}, which is a statement about that kernel and not about the complete irreducible vertex. No mechanism that would reverse the sign at intermediate parameters has appeared, but its absence is not proven. The argument is therefore supported at every point where it can be tested, with three qualifications that the analysis makes explicit. It does not exclude a first-order transition to a state of different symmetry. The scale of the superfluidity on the weak-pairing side is set by the statically screened interaction, which makes it unobservably small once $E_B$ is well below $\EF$. And because that scale falls as a Gaussian in $d$, at sufficiently large $d$ it drops below the scale of any instability that a single layer has on its own, such as pairing of the Kohn-Luttinger type \cite{KohnLuttinger1965}, which does not depend on $d$; in the ideal model the two would then have to be compared, at energies far below anything accessible.

\subsection{Summary of this appendix}
\label{app:summary}

(1) Through second order the irreducible pairing vertex has one plus five diagrams in the bare interaction and one plus twelve in the screened interaction (Figs.~\ref{fig:diagbare} and \ref{fig:diagscr}). The bare series is not a weak-coupling expansion, because screening is governed by $gd/\aB$ and not by $r_s$.

(2) Our numerical results contain the first-order screened rung and all its iterations. The crossed rung is evaluated at the Fermi surface. The remaining eleven second-order diagrams and the self-energy are not evaluated, and Tables~\ref{tab:included} and \ref{tab:fate} state what they do in each limit.

(3) At small $r_s$ and fixed $d>0$ the leading exponent is that of first-order perturbation theory in the statically screened interaction: $\lams\ln(\EF/k_BT_c)\to1$ with $\lams=(\alpha/\pi)\,s(gd/\aB)$. The static crossed rung and the intralayer terms of second order change $\ln T_c$ by amounts of order $\ln(1/r_s)$, and the energy scale of the prefactor is $\hbar^2/md^2$; the net dynamical interlayer term changes $\ln T_c$ by $R(d)\,r_s$, which vanishes in the limit (Appendix~\ref{app:remainder}). The dynamically screened ladder has the same limit at fixed $d$ but not when $d$ is large as well; the crossed rung removes that defect.

(4) At large $d$, for every $r_s$ in the liquid phase, every process that involves the plasmons has a large energy denominator, and the result is $\ln T_c\to-(m^*/m)(16g^{3/2}/\pi r_s)(d/\aB)^2$. This is the fast-boson counterpart of Migdal's theorem and selects the statically screened interaction. The cancellation on which it rests is established at second order, where the remainder is $J_w\sqrt g\,r_s^4(\aB/d)^{7/2}$ and is smaller than $\lams$ by a factor proportional to $r_s^3(\aB/d)^{3/2}$ (Appendix~\ref{app:remainder}), and at every order for linearized dispersion; that the remaining terms are small compared with $\lams$ is an assumption that we consider very likely to hold, and the result is conditional on it.

(5) Weak pairing and Bose condensation of excitons are connected by two distinct crossovers. As a function of $r_s$ at fixed $d\gtrsim0.9\,\aB$ both ends are controlled, and we estimate a maximum of $T_c$ of order $0.1\,E_B(d)$; for $d\gtrsim13\,\aB$ a crystal of excitons intervenes. As a function of $d$ at fixed $r_s$ large enough for dilute excitons to exist at small $d$ we expect a plateau bounded by $g\EF/8$ followed by a collapse that is Gaussian in $d$. The pairing interaction is unscreened on the Bose side and statically screened on the weak-pairing side. The crossover region itself is not calculated.

(6) An RG of the Polchinski-Shankar type captures the inevitability of pairing for an attractive low-energy coupling and the form of $T_c$. It does not supply the coupling or the scale at which the flow begins, and these are what the diagrams determine.
\section{Asymptotic evaluation of the net second-order coupling}
\label{app:remainder}

This appendix evaluates the net dynamical coupling of second order, $\lambda_{\rm net}=\lampl+\lambda_{\rm cr}$, in its two limits. By Eqs.~\eqref{eq:Bnet} and \eqref{eq:lamX},
\begin{align}
\lambda_{\rm net}={}&\frac{N_0}{g}\Bigl\langle\int\!\frac{d\nu}{2\pi}\!\int\!\frac{d^2q}{(2\pi)^2}\,\mathcal W(\mathbf q,\bm\Delta,\nu)\nonumber\\
&\times\frac{(\xi_1+\xi_2)^2}{2(\nu^2+\xi_1^2)(\nu^2+\xi_2^2)}\Bigr\rangle_{\mathbf k'},
\label{eq:netdef}
\end{align}
where $\mathcal W=W_{12}(\mathbf q,i\nu)W_{12}(\bm\Delta-\mathbf q,i\nu)-W_{12}(\mathbf q,0)W_{12}(\bm\Delta-\mathbf q,0)$. Within the RPA $W_{12}(q,i\nu)\le W_{12}(q,0)<0$, so $\mathcal W\ge0$ and the integrand is nonnegative everywhere. Two regions of the integration contribute at leading order, and they are well separated in frequency when $\kF d\gg1$: the region $\hbar|\nu|\sim\hbar^2/md^2\ll\hbar\vF q$, and the wedge $|\nu|\gg\vF q$. In this appendix $\hbar=m=e^2/\epsilon=1$, so that lengths are in units of $\aB$, energies in units of $2\,{\rm Ry}$, $\kF=2/(\sqrt g\,r_s)$, and $N_0/g=1/2\pi$. We take $\mathbf k=(\kF,0)$ and denote by $\theta$ the angle between $\mathbf k'$ and $\mathbf k$.

\subsection{$r_s\to0$ at fixed $d$}

The momentum transfers are of order $\min(1,1/d)$ and are small compared with $\kF$. With $\theta=t/\kF$ and $\mathbf q=(q_x,q_y)$, to leading order in $1/\kF$,
\begin{align}
\xi_1&=\kF q_x+\tfrac12q_y^2,\qquad \xi_2=-\kF q_x-q_yt+\tfrac12q_y^2,\nonumber\\
\varepsilon&\equiv\xi_1+\xi_2=q_y(q_y-t),
\end{align}
in agreement with Eq.~\eqref{eq:xisum}. The integrand of Eq.~\eqref{eq:netdef} is concentrated where $\xi_1$, $\xi_2$, and $\nu$ are of the order of the curvature energy $\varepsilon$. There $q_x\sim\varepsilon/\kF\ll q_y$, so the two interaction lines carry the momenta $q=|q_y|$ and $q'=|q_y-t|$, and $|\nu|\ll\kF q$. In this range the polarization of a layer is $\Pi(q,i\nu)=-N_0[1-|\nu|/\kF q]$ up to terms of higher order in $\nu/\kF q$ and $q/\kF$ (Appendix~\ref{app:pol}); the term linear in $|\nu|$ is the Landau damping of the particle-hole continuum, continued to imaginary frequency. Expanding Eq.~\eqref{eq:W12} to first order in it,
\begin{align}
W_{12}(q,i\nu)&=W_0(q)\Bigl[1+D(q)\,\frac{|\nu|}{\kF q}\Bigr],\nonumber\\
W_0(q)&=-\frac{2\pi e^{-qd}}{q\,\varepsilon_+\varepsilon_-},\qquad D(q)=2-\frac{1}{\varepsilon_+}-\frac{1}{\varepsilon_-},
\label{eq:Wlandau}
\end{align}
with the static dielectric functions of the two modes, $\varepsilon_\pm=1+(g/q)(1\pm e^{-qd})$. Hence $\mathcal W=W_0(q)W_0(q')\,(|\nu|/\kF)\,[D(q)/q+D(q')/q']$. The integral over $q_x$ is the convolution of two Lorentzians,
\begin{equation}
\int\!dq_x\,\frac{1}{(\nu^2+\xi_1^2)(\nu^2+\xi_2^2)}=\frac{1}{\kF}\,\frac{2\pi}{|\nu|\,(\varepsilon^2+4\nu^2)},
\end{equation}
and its factor $1/|\nu|$ cancels the $|\nu|$ of the Landau damping. The frequency integral is then $\int d\nu\,(\varepsilon^2+4\nu^2)^{-1}=\pi/2|\varepsilon|$, and since $|\varepsilon|=qq'$ the remaining integrals over $q_y$ and $q_y'=q_y-t$ separate. The result is
\begin{equation}
\lambda_{\rm net}=C_3(d)\,r_s^3\,\bigl[1+O\bigl(r_s\ln r_s^{-1}\bigr)\bigr],\qquad C_3=\frac{g^{3/2}}{64\pi^3}\,I_1I_2,
\label{eq:C3}
\end{equation}
with
\begin{equation}
I_1=\int_0^\infty\!dq\,|W_0(q)|\,D(q),\qquad I_2=\int_0^\infty\!dq\,q\,|W_0(q)|.
\label{eq:I12}
\end{equation}
Equation~\eqref{eq:C3} is the exact leading term of $\lambda_{\rm net}$ at small $r_s$ for any fixed $d>0$; the form of its correction is discussed below. The frequencies that contribute are $\hbar|\nu|\sim\hbar^2/md^2$ for $d\gtrsim\aB$. In the same units the static coupling of Eq.~\eqref{eq:lams} is $\lams=\sqrt g\,r_sI_0/4\pi^2$ with $I_0=\int_0^\infty dq\,|W_0(q)|$, so that
\begin{equation}
\frac{\lambda_{\rm net}}{\lams^2}\ \to\ R(d)\,r_s,\qquad R=\frac{\pi\sqrt g}{4}\,\frac{I_1I_2}{I_0^2}.
\label{eq:Rd}
\end{equation}
For $gd\gg\aB$, $W_0\to-\pi q/(g^2\sinh qd)$ and $D\to2$, which give $I_0=\pi^3/4g^2d^2$, $I_1=2I_0$, $I_2=7\pi\zeta(3)/2g^2d^3$, and
\begin{align}
C_3&\to\frac{7\pi\zeta(3)}{256\,g^{5/2}}\Bigl(\frac{\aB}{d}\Bigr)^{5}=\frac{\ClowCoef}{g^{5/2}}\Bigl(\frac{\aB}{d}\Bigr)^{5},\nonumber\\
R&\to\frac{7\zeta(3)\sqrt g}{\pi}\,\frac{\aB}{d}.
\label{eq:C3asym}
\end{align}
The approach to these forms is slow, with corrections of relative order $\aB/gd$ (Table~\ref{tab:remainder}).

\subsection{$d\to\infty$ at fixed $r_s$}

In the wedge $|\nu|\gg\vF q$ one has $\Pi=-N_0(\vF q)^2/2\nu^2$, the interaction is the product of the bare one and the two plasmon poles,
\begin{equation}
W_{12}(q,i\nu)=-U_q\,\frac{\nu^4}{[\nu^2+\omega_+^2(q)][\nu^2+\omega_-^2(q)]},
\end{equation}
with $\omega_\pm^2(q)=\tfrac12\Ost^2\,qd\,(1\pm e^{-qd})$, and the bracket of Eq.~\eqref{eq:netdef} is $(\xi_1+\xi_2)^2/2\nu^4$ with $\xi_1+\xi_2=\mathbf q\cdot(\mathbf q-\bm\Delta)$. The static product in $\mathcal W$ is smaller than the dynamical one by a factor of order $\cG^{-2}$ and is dropped. In the variables $\mathbf y=\mathbf qd$, $\tau=\kF\theta d$, and $w=\nu/\Ost$ the integral becomes a pure number,
\begin{equation}
\lambda_w=J_w\,\sqrt g\,r_s^{4}\Bigl(\frac{\aB}{d}\Bigr)^{7/2},
\label{eq:lamw}
\end{equation}
\begin{align}
J_w&=\frac{1}{256\pi^3}\int_{-\infty}^{\infty}\!\!d\tau\!\int\!d^2y\;\frac{e^{-y-y'}}{yy'}\,\bigl[\mathbf y\cdot(\mathbf y-\bm\tau)\bigr]^2F,\nonumber\\
F&=\int_{-\infty}^{\infty}\!\!\frac{w^4\,dw}{(w^2+a_+)(w^2+a_-)(w^2+a_+')(w^2+a_-')},
\label{eq:Jw}
\end{align}
where $\bm\tau=(0,\tau)$, $y'=|\mathbf y-\bm\tau|$, $a_\pm=\tfrac12y(1\pm e^{-y})$, and $a'_\pm$ is the same function of $y'$; the quantities $\Ost^2a_\pm$ are the squared frequencies of the two plasmons. The value is $J_w=\Jw$. The integral converges at both ends of the frequency range: the integrand vanishes as $w^4$ at small $w$, because both plasmon factors do, and as $w^{-4}$ at large $w$. It is dominated by $w\sim1$ and $y\sim1$, that is, by $\nu\sim\Ost$ and $q\sim1/d$, where $\nu/\vF q\sim\sqrt{\cG}\gg1$, so the use of the wedge forms is consistent for $\cG\gg1$ and $\kF d\gg1$. Numerical quadrature of Eq.~\eqref{eq:Jw} at two resolutions gives the same value to five figures.

Since $\lambda_w\propto d^{-7/2}$ and the contribution of Eq.~\eqref{eq:C3} falls off as $d^{-5}$, the wedge is the leading term of $\lambda_{\rm net}$ for $d\to\infty$ at fixed $r_s$. In their asymptotic forms the two are equal at $r_s(d/\aB)^{3/2}=7\pi\zeta(3)/(256\,g^3J_w)\approx300/g^3$.

\subsection{Comparison with the numerical evaluation}

\begin{figure*}[t]
\includegraphics[width=0.9\textwidth]{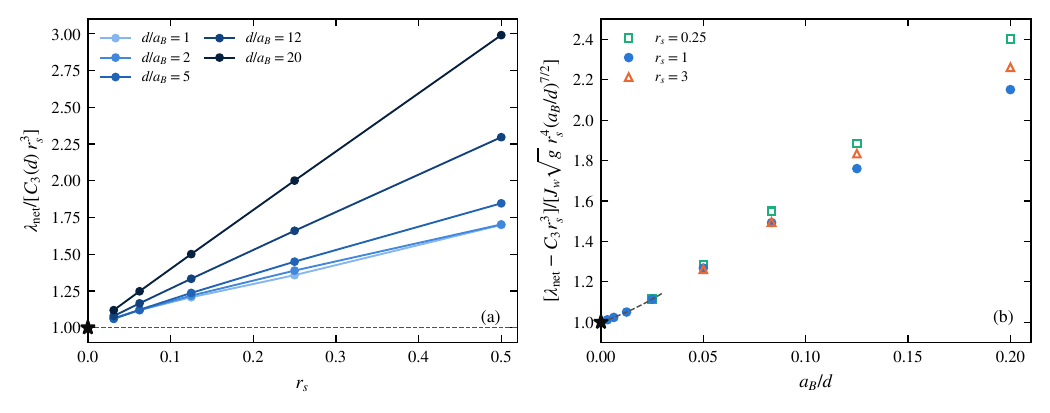}
\caption{Tests of the asymptotic forms of the net second-order coupling against the full RPA quadrature ($g=2$). (a) $\lambda_{\rm net}$ divided by the leading term $C_3(d)r_s^3$ of Eq.~\eqref{eq:C3}, which contains no adjustable quantity. The star marks the exact limit. (b) The remainder $\lambda_{\rm net}-C_3r_s^3$ divided by the wedge term of Eqs.~\eqref{eq:lamw} and \eqref{eq:Jw}, versus $\aB/d$, for $5\le d/\aB\le320$. Dashed line: quadratic fit to the four points with $d\ge40\aB$. The star marks the exact limit.}
\label{fig:remainder}
\end{figure*}

The two results are tests of the four-dimensional quadrature of Sec.~\ref{sec:vertex}, and conversely. Figure~\ref{fig:remainder}(a) shows $\lambda_{\rm net}/C_3r_s^3$ from the full RPA calculation for $r_s$ down to $1/32$. It tends to 1 linearly in $r_s$ for every $d$ examined: at $r_s=1/32$ it is between \lowRatA\ and \lowRatB\ for $d\le20\aB$, and the linear extrapolation to $r_s=0$ from the two smallest values of $r_s$ differs from 1 by at most \lowExtrapErr. Figure~\ref{fig:remainder}(b) shows the remainder, $\lambda_{\rm net}-C_3r_s^3$, divided by $\lambda_w$ of Eq.~\eqref{eq:lamw}. For $d\ge5\aB$ this ratio depends on $d$ and only weakly on $r_s$: it varies by at most \CfourSpread\ over $0.25\le r_s\le3$. For $r_s=1$ it is \wedgeRatA, \wedgeRatB, \wedgeRatC, and \wedgeRatD\ at $d/\aB=40$, 80, 160, and 320, and a fit quadratic in $\aB/d$ to these four values extrapolates to \wedgeIntercept\ at $d=\infty$. For $d\ge5\aB$ the computed values are therefore described by
\begin{equation}
\lambda_{\rm net}\simeq C_3(d)\,r_s^3+C_4(d)\,r_s^4,\qquad C_4\to J_w\sqrt g\,\Bigl(\frac{\aB}{d}\Bigr)^{7/2},
\label{eq:twoterm}
\end{equation}
with $C_3$ given by Eq.~\eqref{eq:C3} and $C_4/[J_w\sqrt g(\aB/d)^{7/2}]\approx1+\wedgeSlope\,\aB/d$ at large $d$ for $g=2$. We have not derived $C_4(d)$ at finite $d$. It is the coefficient of the region $|\nu|\sim\vF q$, of which the wedge is the part that survives at large $d$, and its integral, taken alone, diverges at low frequency, where it joins the region of Eq.~\eqref{eq:C3}. The leading divergence is the term of order $r_s^3$; a subleading logarithmic one would give a term of order $r_s^4\ln(1/r_s)$, which is therefore possible. The numerical values of the remainder for $d\le2\aB$ increase slowly as $r_s$ decreases from $1/4$ to $1/32$, which is consistent with such a term, and this is the reason for the form of the correction in Eq.~\eqref{eq:C3}.

\begin{table}[tb]
\caption{Net second-order coupling at small $r_s$ and at large $d$ for $g=2$. $C_3$: leading coefficient, Eq.~\eqref{eq:C3}. $C_3/C_3^{(\infty)}$: its ratio to the large-$d$ form of Eq.~\eqref{eq:C3asym}. $R$: Eq.~\eqref{eq:Rd}. $\rho$: the ratio $\lambda_{\rm net}/C_3r_s^3$ from the full quadrature at $r_s=1/32$. $c_4$: the remainder $\lambda_{\rm net}-C_3r_s^3$ at $r_s=1$ divided by the wedge term $\lambda_w$ of Eq.~\eqref{eq:lamw}.}
\label{tab:remainder}
\begin{ruledtabular}
\begin{tabular}{cccccc}
$d/\aB$ & $C_3$ & $C_3/C_3^{(\infty)}$ & $R$ & $\rho$ & $c_4$\\
\hline
1 & $9.27\times10^{-4}$ & 0.051 & 1.563 & 1.066 & ---\\
2 & $8.79\times10^{-5}$ & 0.154 & 1.055 & 1.062 & ---\\
5 & $2.33\times10^{-6}$ & 0.399 & 0.559 & 1.060 & 2.15\\
8 & $3.01\times10^{-7}$ & 0.540 & 0.385 & 1.065 & 1.76\\
12 & $4.77\times10^{-8}$ & 0.651 & 0.273 & 1.079 & 1.49\\
20 & $4.36\times10^{-9}$ & 0.765 & 0.173 & 1.119 & 1.27\\
40 & $1.55\times10^{-10}$ & 0.871 & 0.090 & 1.259 & 1.11\\
\end{tabular}
\end{ruledtabular}
\end{table}

\subsection{Consequence for the transition temperature}

A weak addition $\delta\lambda$ to the coupling changes $\ln T_c$ by $\delta\lambda/\lams^2$. For the dynamical terms of second order this is now known in both limits. For $r_s\to0$ at fixed $d$, by Eq.~\eqref{eq:Rd},
\begin{equation}
\delta\ln T_c\big|_{\rm net}=R(d)\,r_s\ \to\ 0,
\end{equation}
with $R$ of order one at $d\sim\aB$ and decreasing as $\aB/d$ (Table~\ref{tab:remainder}). The dynamical part of the interlayer interaction therefore does not contribute to the prefactor of $T_c$ in this limit at second order; the terms that do are $\lambda_{\rm GMB}$ and the intralayer corrections. In the ladder approximation the corresponding quantity is $\lampl/\lams^2$, which does not vanish with $r_s$ and is $\ladExcessCoef\,(d/\aB)^{7/2}$ at large $d$. For $d\to\infty$ at fixed $r_s$, by Eqs.~\eqref{eq:lamw} and \eqref{eq:lamsasym},
\begin{equation}
\delta\ln T_c\big|_{\rm net}=\frac{256\,J_w}{\pi^2}\,g^{7/2}\,r_s^{2}\Bigl(\frac{d}{\aB}\Bigr)^{1/2}=\JwT\,r_s^2\Bigl(\frac{d}{\aB}\Bigr)^{1/2}
\label{eq:dlnTcw}
\end{equation}
for $g=2$. This increases with $d$, but as $\sqrt d$, while the leading term $1/\lams$ increases as $d^2$: the ratio is $\lambda_{\rm net}/\lams\propto r_s^3(\aB/d)^{3/2}$. At second order the condition $\delta\lambda/\lams\to0$ of Sec.~\ref{sec:control} is thus satisfied with a known coefficient. These statements concern the second-order term only. The dynamical terms of third and higher order with curvature have not been evaluated, and the assumption about them stated in Sec.~\ref{sec:control} remains an assumption.

\section{Stiffness and response functions at $\nu_T=1$}
\label{app:qh-stiffness}

The two factors in Eq.~\eqref{qh:eq:mode} are the phase and density sectors of the coherent state.  A phase twist changes the exchange energy by $B_\delta(Q)$, and a relative-density fluctuation costs $K_\delta(Q)$.  The projected equations of motion are
\begin{align}
\hbar\partial_tS^y(\bm q)&=K_\delta(q\ell)S^z(\bm q),\nonumber\\
\hbar\partial_tS^z(\bm q)&=-B_\delta(q\ell)S^y(\bm q),
\label{qh:eq:eom}
\end{align}
for a condensate along $x$, with $S^y$ the phase variable.  They give Eq.~\eqref{qh:eq:mode} without any long-wavelength approximation.  In particular, $B_\delta(0)=0$ is the microscopic statement of the Goldstone theorem: a uniform change of the spontaneously chosen interlayer phase costs no energy.

The phase stiffness follows by expanding the first factor at small $Q$,
\begin{equation}
\begin{aligned}
B_\delta(Q)&=4\pi\rho_sQ^2+O(Q^4),\\
\frac{\rho_s}{\Ec}&=\frac{1}{16\pi}\left[(1+\delta^2)F(\delta)-\delta\right],
\end{aligned}
\label{qh:eq:rho_s}
\end{equation}
where $F(\delta)=X_D/\Ec$.  Here $\rho_s$ is the conventional stiffness, defined by the energy density $\rho_s(\nabla\phi)^2/2$.  It is the interlayer stiffness $\rho_E$ of Ref.~\cite{Moon1995} in closed form and equals $\Ec/(16\sqrt{2\pi})$ at $d=0$.  The second factor at $Q=0$ is the inverse relative-density susceptibility.  The sound velocity is the geometric mean of these two microscopic quantities.  This result identifies the charged and neutral responses of the $(111)$ state.  Symmetric electric fields couple to the total density and produce the integer quantum Hall response.  Antisymmetric fields couple to the neutral phase-density variables in Eq.~\eqref{qh:eq:eom}.  A quantized Hall plateau and a gapless neutral mode are therefore simultaneous and complementary signatures of the same state.

The stiffness also makes the connection with counterflow immediate.  A counterflow current is a gradient of the interlayer phase.  The energy density of a long wavelength twist is $\rho_s(\nabla\phi)^2/2$.  The same $\rho_s$ sets the in-plane-field tunneling response and the long-distance phase dynamics.  In the present problem the stiffness is not an independent phenomenological parameter: it is an explicit analytic function of $d/\ell$ through Eq.~\eqref{qh:eq:rho_s}.

The exchange gap has an equally direct response interpretation.  The symmetric and antisymmetric bands in Eq.~\eqref{qh:eq:levels} are split by $X_DP$.  The coherent state therefore has a neutral collective mode together with a finite quasiparticle splitting,
\begin{equation}
\Delta(T)=X_DP(T).
\label{qh:eq:gapT}
\end{equation}
The gap is maximal at $T=0$, where $P=1$.  It vanishes at the thermal HF transition together with the coherence amplitude.  The ratio $\Delta(0)/(k_BT_c^{\rm HF})=4$ is universal in the projected uniform theory.

\section{Finite temperature and the separation-dependent coherence scale at $\nu_T=1$}
\label{qh:sec:finiteT}

The finite-temperature solution is contained in one equation, Eq.~\eqref{qh:eq:SC}.  It has the usual bifurcation from $P=0$ at
\begin{equation}
k_BT_c^{\rm HF}(d)=\frac{e^2}{4\epsilon\ell}\sqrt{\frac\pi2}
e^{d^2/(2\ell^2)}\operatorname{erfc}\!\left(\frac{d}{\sqrt2\ell}\right).
\label{qh:eq:Tcexplicit}
\end{equation}
Near the transition, writing $t=T/T_c^{\rm HF}$, one obtains
\begin{equation}
P^2=3(1-t)-\frac{12}{5}(1-t)^2+O[(1-t)^3].
\label{qh:eq:criticalP}
\end{equation}
Thus the gap closes as $\Delta(T)=X_D\sqrt{3(1-T/T_c^{\rm HF})}+\cdots$.  At low temperature $P$ is saturated and the condensation energy per flux quantum is $X_D/4$.

The experimental implication must be stated with care.  $T_c^{\rm HF}$ is the mean-field amplitude scale at which the HF gap closes.  It is not the temperature of the physical transition, which in two dimensions is of Kosterlitz-Thouless type and bounded by the stiffness, $k_BT_{\rm KT}\le(\pi/2)\rho_s$ \cite{KT1973,Moon1995}.  From Eq.~\eqref{qh:eq:rho_s} this bound is $0.0097\Ec$ at $d/\ell=1$, a factor of $17$ below $T_c^{\rm HF}$.  The same overestimate is well known for the $\nu=1$ spin ferromagnet \cite{KasnerMacDonald1996}.  The qualitative point survives.  At a fixed laboratory temperature a coherence scale that decreases smoothly with $d/\ell$ produces an apparent separation boundary.  Since $T_c^{\rm HF}$ falls as $e^2/(4\epsilon d)$, and the stiffness bound falls faster, as $1/d^3$, increasing $d/\ell$ eventually drives the sample above its coherence scale.  This is a thermal loss of observable coherence.  It does not imply that the $T=0$ exchange field has vanished at that separation.  The distinction is particularly important at large $d/\ell$, where the uniform-HF scale is small but nonzero.

The no-tunneling limit is also the cleanest setting for the Goldstone mode.  A small symmetric--antisymmetric splitting adds
\begin{equation}
H_t=-\Delta_{\rm SAS}\sum_mS_m^x,
\label{qh:eq:tunnel}
\end{equation}
after choosing the condensate along $x$.  It pins the phase and produces a Josephson gap in the collective mode.  It does not generate the exchange field.  In the interaction-dominated regime $\Delta_{\rm SAS}\ll X_D$, tunneling probes a spontaneously coherent state already present in the zero-tunneling Hamiltonian.

An in-plane field is an especially direct probe.  It supplies the tunneling wave vector
\begin{equation}
q_B=\frac{edB_\parallel}{\hbar},\qquad eV_{\rm res}=\hbar\omega(q_B\ell).
\label{qh:eq:resonance}
\end{equation}
The experimentally measured dispersing tunneling resonance is therefore a direct measurement of Eq.~\eqref{qh:eq:mode} \cite{Spielman2001}.  Its low-$Q$ slope measures Eq.~\eqref{qh:eq:v}; its finite-$Q$ evolution measures the approach to the roton instability.

\begin{table}[t]
\caption{Uniform LLL HF scales in units of $\Ec=e^2/(\epsilon\ell)$.}
\label{qh:tab:scales}
\begin{ruledtabular}
\begin{tabular}{ccc}
$d/\ell$ & $\Delta/\Ec$ & $k_BT_c^{\rm HF}/\Ec$\\\hline
0 & 1.253314 & 0.313329\\
0.5 & 0.876364 & 0.219091\\
1.0 & 0.655680 & 0.163920\\
2.0 & 0.421369 & 0.105342\\
4.0 & 0.236652 & 0.059163
\end{tabular}
\end{ruledtabular}
\end{table}

\section{Quantitative scales and controlled perturbations at $\nu_T=1$}
\label{app:qh-scales}

The numbers in Table~\ref{qh:tab:scales} display the full strength of the analytic result.  At $d/\ell=1$, the zero-temperature exchange gap is $0.656\Ec$ and the HF transition scale is $0.164\Ec$.  Even at $d/\ell=4$, where the coherent scale is small in comparison with the $d=0$ value, the analytic exchange field remains $0.237\Ec$.  The decrease is rapid but smooth.  No fitting form, numerical interpolation, or finite-size extrapolation is involved.

For GaAs, the conversion to Kelvin is immediate after specifying $B_\perp$ and the dielectric constant.  The Coulomb unit scales as $\Ec\propto\sqrt{B_\perp}/\epsilon$, while $d/\ell\propto d\sqrt{B_\perp}$.  Changing density or field therefore changes both the absolute exchange scale and the dimensionless separation.  This is the microscopic origin of the strong field dependence of interlayer-coherence phenomena.  The formula in Eq.~\eqref{qh:eq:Tcexplicit} gives the entire HF dependence without introducing a material-specific coherence parameter.

Layer imbalance provides another controlled test.  A voltage bias couples to $S^z$,
\begin{equation}
H_V=-eV\sum_mS_m^z.
\label{qh:eq:bias}
\end{equation}
It tilts the pseudospin out of the easy plane.  If $m_z$ denotes the resulting normalized layer polarization, the transverse coherent component is $P_\perp=\sqrt{1-m_z^2}$.  The transverse exchange field becomes $X_DP_\perp$, while the HF quasiparticle splitting, which is the magnitude of the self-consistent field along the tilted pseudospin, remains $X_D$ at $T=0$.  Thus imbalance suppresses tunneling coherence by geometrically reducing the transverse pseudospin component while retaining the same interaction-generated exchange field.  This is a direct pseudospin-ferromagnet consequence and is distinct from a loss of coherence induced by increasing $d/\ell$.

Finite layer width modifies the interaction form factors in Eq.~\eqref{qh:eq:V}.  The calculation proceeds identically after replacing the strict-2D Coulomb kernels by the appropriate subband form factors.  The analytic structure remains the same: the interlayer exchange integral determines the gap, amplitude scale, and the collective kernels.  The zero-width result is therefore the universal reference curve against which finite-width samples may be compared.

Disorder and density inhomogeneity broaden experimental crossovers but do not alter the basic organizing principle.  The coherent phase is controlled microscopically by the positive interlayer exchange field.  The roton transition is controlled by the finite-$Q$ stiffness.  These two quantities are different functions of the same projected interaction, and they lead to different physical boundaries.  The analytic separation of these mechanisms is useful precisely because experimental phase diagrams often show a single apparent boundary while the microscopic theory contains distinct thermal and translational phenomena.

\section{Exchange integral at $\nu_T=1$}
\label{app:qh-exchange}

The exchange integral is
\begin{equation}
\frac{X_D}{\Ec}=\int_0^\infty dx\,e^{-x^2/2-\delta x}=e^{\delta^2/2}\int_\delta^\infty du\,e^{-u^2/2}.
\end{equation}
Completing the square gives the last expression.  The short-distance series follows by expanding the complementary error function.  The large-distance expansion follows more transparently by observing that the integral is dominated by $x\sim1/\delta$:
\begin{align}
\frac{X_D}{\Ec}&=\int_0^\infty dx\,e^{-\delta x}\left(1-\frac{x^2}{2}+\frac{x^4}{8}-\cdots\right)\nonumber\\
&=\frac1\delta-\frac1{\delta^3}+\frac3{\delta^5}-\frac{15}{\delta^7}+\cdots.
\end{align}
The LLL form factor has disappeared from the leading term because the exchange process is carried by $q\sim1/d$.  This gives a simple physical explanation for the $e^2/(\epsilon d)$ scale.  At large separation the HF exchange field is long ranged.

\section{Uniform density matrix and the exchange self-energy}
\label{app:qh-density}

The uniform density matrix can be written in pseudospin form as
\begin{equation}
\rho_m=\frac12\left[1+P(\cos\phi\,\sigma_x+\sin\phi\,\sigma_y)\right].
\label{qh:eq:pauli}
\end{equation}
The exchange self-energy is obtained by contracting one density operator in Eq.~\eqref{qh:eq:H},
\begin{equation}
\Sigma_{\alpha\beta}=-\int\frac{d^2q}{(2\pi)^2}V_{\alpha\beta}(q)e^{-q^2\ell^2/2}\rho_{\alpha\beta}.
\label{qh:eq:selfenergy}
\end{equation}
For a balanced state the diagonal intralayer terms shift both HF bands equally.  The off-diagonal interlayer term is
\begin{equation}
\Sigma_{TB}=-X_D\langle c_{mT}^\dagger c_{mB}\rangle,
\label{qh:eq:offdiag}
\end{equation}
which immediately gives Eq.~\eqref{qh:eq:levels}.  The resulting Green function is a $2\times2$ matrix with poles at $\epsilon_S$ and $\epsilon_{AS}$.  Taking the difference of their Fermi functions yields Eq.~\eqref{qh:eq:SC}.

The diagrammatic content of the approximation is most transparent in the $\{+,-\}$ basis defined by Eq.~\eqref{qh:eq:spinor}.  The HF Green functions are
\begin{equation}
G_s(i\omega_n)=\frac{1}{i\omega_n-\epsilon_s},\qquad s=+,-,
\end{equation}
with $n_+=1$ and $n_-=0$ at $T=0$.  The same-band bubble is Pauli blocked,
\begin{equation}
\Pi_{++}(i\Omega_n)\propto\frac{n_+-n_+}{i\Omega_n}=0.
\label{qh:eq:blocked}
\end{equation}
The interband bubble is instead nonzero,
\begin{equation}
\Pi_{+-}(\bm q,i\Omega_n)=\frac{n_+-n_-}{i\Omega_n+\epsilon_+-\epsilon_-}\,\mathcal F_{+-}(\bm q),
\label{qh:eq:interbubble}
\end{equation}
where $\mathcal F_{+-}(\bm q)$ is the LLL guiding-center form factor.  Equation~\eqref{qh:eq:interbubble} is the microscopic particle-hole propagator of the pseudospin mode.  It enters the TDHF ladder that produces Eq.~\eqref{qh:eq:mode}, and it also enters self-energy and vertex corrections omitted by static HF.  A zero external momentum does not set the internal momentum in Eq.~\eqref{qh:eq:interbubble} to zero.

Consequently, $\rho^2=\rho$ establishes only that the \emph{one-body} HF density matrix is a projector.  It is not an identity eliminating the connected two-particle correlator.  The finite-$d$ interaction contains the anisotropy
\begin{equation}
V_-(q)=\frac{V_{TT}(q)-V_{TB}(q)}{2}>0,
\end{equation}
which couples pseudospin-density fluctuations and has nonzero matrix elements from the coherent Slater determinant to states containing $+\!\rightarrow\!-$ particle-hole pairs.  The first-order Fock self-energy is therefore the uniform HF self-energy, not the full finite-$d$ projected self-energy.  The unique exact point of this argument is $d=0$, where $V_-(q)=0$ and global pseudospin rotations map the state to a completely filled one-component LLL.

At finite temperature the same equation has a transparent physical interpretation.  The exchange field is $X_DP$, while thermal occupation of the upper pseudospin band reduces $P$.  The self-consistency condition is exactly the magnetization equation of a pseudospin-$1/2$ ferromagnet in its self-generated in-plane field.  The ordered solution is selected for $T<T_c^{\rm HF}$; its phase remains arbitrary because tunneling is absent.

\section{Derivation of the collective-mode kernels}
\label{app:qh-kernels}

Introduce the projected pseudospin densities
\begin{equation}
S^a(\bm q)=\frac12\sum_{mm'}\langle m|e^{-i\bm q\cdot\bm R}|m'\rangle c_{m\alpha}^\dagger\sigma^a_{\alpha\beta}c_{m'\beta}.
\label{qh:eq:spin-density}
\end{equation}
Here $\mathbf q$ is the physical wave vector, and $Q=|\mathbf q|\ell$ is its dimensionless magnitude, with $\ell$ the magnetic length.
Their LLL commutator algebra contains the magnetic-translation phase $\sin[(\bm q\times\bm k)\ell^2/2]$.  Linearization about the state in Eq.~\eqref{qh:eq:spinor} couples $S^y(\bm q)$ and $S^z(\bm q)$ and produces the two equations in Eq.~\eqref{qh:eq:eom}.  The interlayer exchange kernel is the Bessel transform $\Gamma_D(Q)$ in Eq.~\eqref{qh:eq:kernels}; the intralayer exchange kernel is $\Gamma_S(Q)$; and $D_\delta(Q)$ is the direct capacitive contribution.

The long-wave expansion is obtained from $J_0(Qx)=1-Q^2x^2/4+O(Q^4)$.  It gives the phase-stiffness factor in Eq.~\eqref{qh:eq:rho_s}.  The direct term has a finite $Q\rightarrow0$ limit because
\begin{equation}
\lim_{Q\rightarrow0}\frac{1-e^{-\delta Q}}{Q}=\delta.
\end{equation}
This is the capacitive energy of a uniform relative-density fluctuation.  Combining the two terms gives the sound velocity in Eq.~\eqref{qh:eq:v}.  The complete formula, rather than only the gradient expansion, is needed to follow the dispersion to the roton wave vector.

The roton minimum is an exchange effect. At momenta of order $1/\ell$, the LLL form factor and the noncommutative guiding-center algebra make the phase and density kernels strongly momentum dependent. The direct term stabilizes the long-wavelength density sector, whereas exchange produces the finite-$Q$ competition. At the instability, the collective mode vanishes at $Q_0=q_0\ell\simeq1.317$, identifying the physical wave number $q_0$ of the emergent coherent density wave.

\section{Finite-sphere data at $\nu_T=1$}
\label{app:qh-ed}

\begin{figure}[tb]
\includegraphics[width=\linewidth]{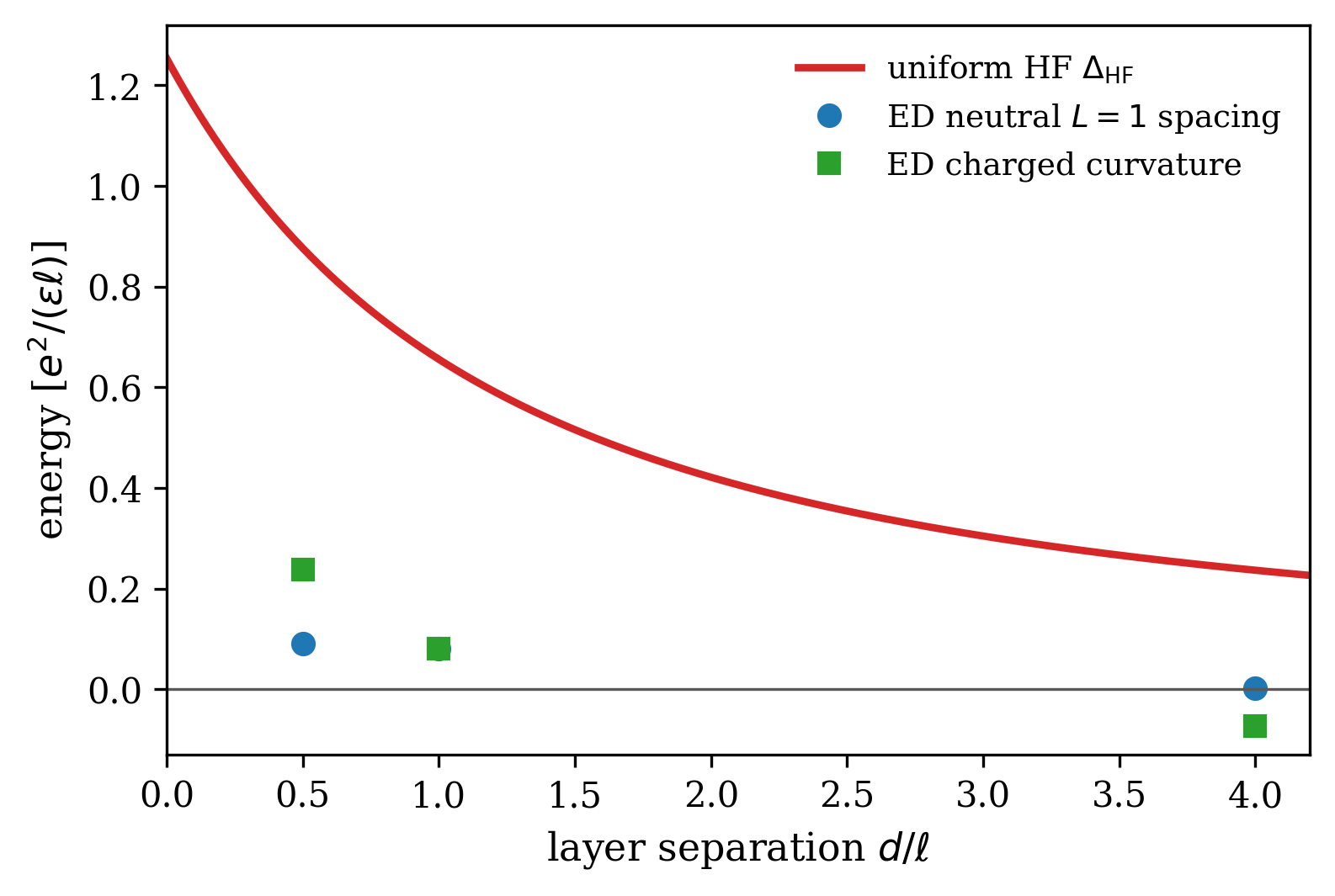}
\caption{Finite-sphere $N=8$ spectrum diagnostics and the analytic uniform exchange gap.}
\label{qh:fig:ed}
\end{figure}

The $N=8$, $N_\phi=7$ finite-sphere spectrum is summarized in Table~\ref{qh:tab:ed} and Fig.~\ref{qh:fig:ed}.  The neutral $L=1$ spacing and charged curvature are finite-size spectral diagnostics, whereas $X_D$ is the thermodynamic uniform-HF one-particle splitting.  They are not the same observable and should not be identified.
\begin{table}[tb]
\caption{Finite-sphere data in units of $e^2/(\epsilon\ell)$.}
\label{qh:tab:ed}
\begin{ruledtabular}
\begin{tabular}{cccc}
$d/\ell$ & $\Delta_n$ & $\Delta_c$ & $X_D$\\\hline
0.5 & 0.090652 & 0.238798 & 0.876364\\
1.0 & 0.080985 & 0.080815 & 0.655680\\
4.0 & 0.002135 & $-0.072450$ & 0.236652
\end{tabular}
\end{ruledtabular}
\end{table}

The comparison is useful in a limited and concrete sense.  At $d/\ell=0.5$, the finite-size charged curvature is positive and the spectrum is consistent with a robust coherent incompressible regime.  By $d/\ell=1$, both finite-size diagnostics have fallen sharply.  At $d/\ell=4$, the neutral spacing is nearly zero and the charged curvature is negative for this geometry.  This entry requires a caveat.  At $N_\phi=7$ each layer holds four electrons, and the composite fermions of a decoupled layer see an effective flux $2Q^*=N_\phi-2(N/2-1)=1$, which leaves the second composite-fermion shell half filled.  The vanishing $L=1$ spacing and the negative curvature at large $d$ therefore reflect the open-shell degeneracy imposed by the $(111)$ shift, and are not evidence about the thermodynamic state.  Thus the $N=8$ spectrum does \emph{not} support extrapolating the uniform-HF gap to an exact many-body charge gap at large $d/\ell$; it is consistent with strong finite-$d$ correlation physics absent from HF, but at this size it cannot establish it.  Conversely, it cannot locate a thermodynamic transition by itself.  This is the appropriate division of labor: HF supplies the analytic uniform saddle and its exchange scale, while finite-size diagonalization provides a nonperturbative but size-limited test of competing correlations.  Larger-scale numerical studies find nontrivial evolution away from the small-$d$ coherent regime \cite{NomuraYoshioka2002,ShengBalentsWang2003,ShibataYoshioka2006,ZhuFuSheng2017}.

\end{document}